%% file: seft.tex
\documentclass[11pt]{article}

\usepackage[letterpaper, margin=0.8in]{geometry}

\usepackage{amsmath, amsfonts, amssymb}
\usepackage{hyperref}
\usepackage{slashed,comment}
\usepackage{cite}
\usepackage{youngtab}
\usepackage{empheq}
\usepackage{booktabs}
\usepackage{breqn}
\usepackage{mathrsfs}
\usepackage{xcolor}

\usepackage{hyperref}
\hypersetup{
	unicode,	
	colorlinks,
	citecolor=blue,
	linkcolor=[rgb]{0,.3,.5},
	urlcolor=[rgb]{0,.3,.5}, 
	bookmarksopen=true,
	bookmarksopenlevel=\maxdimen,
        linktocpage,
}

\newcommand{\ua}{{\ul a}}
\newcommand{\ub}{{\ul b}}
\newcommand{\uc}{{\ul c}}

\newcommand{\um}{{\ul m}}
\newcommand{\un}{{\ul n}}
\newcommand{\up}{{\ul p}}
\newcommand{\uq}{{\ul q}}
\newcommand{\ur}{{\ul r}}
\newcommand{\us}{{\ul s}}
\newcommand{\ut}{{\ul t}}
\newcommand{\uk}{{\ul k}}

\newcommand{\CC}{\text{c.c.}}

\newcommand{\SU}{\rm SU}

\newcommand{\E}[1]{\ensuremath{{\rm E}_{#1(#1)}}}

\newcommand{\scB}{\mathscr{B}}
\newcommand{\scG}{\mathscr{G}}
\newcommand{\scH}{\mathscr{H}}

\newcommand{\w}[1]{\\[0.#1cm]}
\newcommand{\eq}[1]{(\ref{#1})}
 \def\det{{\rm det\,}}
\def\be{\begin{equation}}
\def\ee{\end{equation}}
\def\bea{\begin{eqnarray}}
\def\eea{\end{eqnarray}}
\def\ba{\begin{array}}
\def\ea{\end{array}}
\def\nn{\nonumber}

\include{sugracom_v3}
\newcommand{\rep}[1]{\mathbf{#1}}
\newcommand{\brep}[1]{\mathbf{\overline{#1}}}

\newcommand{\tK}{\mathtt K}
\newcommand{\tL}{\mathtt L}
\newcommand{\tM}{\mathtt M}
\newcommand{\tN}{\mathtt N}
\newcommand{\tP}{\mathtt P}

\newcommand{\tR}{\mathtt R}
\newcommand{\tS}{\mathtt S}

\newcommand{\balpha}{\boldsymbol \alpha}
\newcommand{\bbeta}{\boldsymbol \beta}
\newcommand{\bgamma}{\boldsymbol \gamma}

\makeatletter
\g@addto@macro\bfseries{\boldmath}
\makeatother

\numberwithin{equation}{section}

\begin{document}

\thispagestyle{empty}

\begin{flushright}\small
MI-TH-186    \\

\end{flushright}


\bigskip
\bigskip

\vskip 10mm

\begin{center}

{\Large{\bf E$_{7(7)}$ Exceptional Field Theory in Superspace}}

\end{center}


\vskip 6mm

\begin{center}

{\bf Daniel Butter$^\ast$, Henning Samtleben$^\dagger$ and Ergin Sezgin$^\ast$}

\vskip 4mm

$^\ast$\,{\em George P. and Cynthia W. Mitchell Institute \\for Fundamental
Physics and Astronomy \\
Texas A\&M University, College Station, TX 77843-4242, USA}\\
\vskip 4mm

$^\dagger$\,{\em Univ Lyon, Ens de Lyon, Univ Claude Bernard, CNRS,\\
Laboratoire de Physique, F-69342 Lyon, France} \\
\vskip 4mm

\end{center}

\vskip0.5cm

\begin{center} {\bf Abstract } \end{center}

\begin{quotation}\noindent

We formulate the locally supersymmetric E$_{7(7)}$ exceptional field theory in a $(4+56|32)$ dimensional superspace, corresponding to a 4D $N\!=\!8$ ``external'' superspace augmented with an ``internal'' 56-dimensional space. This entails the unification of external diffeomorphisms and local supersymmetry transformations into superdiffeomorphisms. The solutions to the superspace Bianchi identities lead to on-shell duality equations for the $p$-form field strengths for $p\leq 4$.
The reduction to component fields provides a complete description of the on-shell supersymmetric theory. 
As an application of our results, we perform a generalized Scherk-Schwarz reduction and obtain the superspace formulation of maximal gauged supergravity in four dimensions parametrized by an embedding tensor. 
\end{quotation}

\newpage

\tableofcontents

\section{Introduction}

It is well known that compactifications of 11-dimensional supergravity on an $n$-torus give rise to an enhancement of the manifest $SL(n,\mathbb R)$ symmetry to symmetries including the exceptional groups $\E{n}$ \cite{Cremmer:1979up,Julia:1980gr}, and that their suitable discrete subgroups are interpreted as the U-duality symmetries of M-theory \cite{Hull:1994ys}. The search for a manifest origin of these symmetries in 11-dimensions prior to any toroidal compactification and without any truncation, which started in \cite{deWit:1986mz}, has culminated in a series of papers \cite{Hohm:2013pua,Hohm:2013vpa,Hohm:2013uia,Hohm:2014fxa} where this was achieved in a framework called exceptional field theory (ExFT). It is based on a generalization to exceptional geometry \cite{Berman:2010is,Berman:2011cg,Coimbra:2011ky,Coimbra:2012af,Aldazabal:2013mya}
of the double field theories (DFT) that provide a manifest realization of the T-duality group $O(n,n)$ that arises in toroidal compactification of string theory \cite{Siegel:1993th,Hull:2009mi,Hohm:2010jy}. In that case the 10-dimensional spacetime coordinates are doubled and certain conditions on fields known as section constraints are imposed. The latter are required for the symmetries to form a closed algebra and, in effect, remove dependence on coordinates beyond ten dimensions. For a more detailed description of the ideas behind these theories, with several references to earlier works, see \cite{Hohm:2013pua}. 

Exceptional field theories are well motivated for a number of reasons. Firstly, they have made it possible to derive fully nonlinear and consistent reductions to gauged supergravities in lower dimensions. For example the long standing problem of finding the nonlinear and consistent reduction of Type IIB supergravity on $AdS_5 \times S^5$ was solved in this way \cite{Hohm:2014qga}. Second, exceptional field theory provides a convenient framework for taking into account the BPS states in the computation of loop corrections to the string low energy effective action \cite{Bossard:2015foa}. Furthermore, higher derivative corrections to the supergravity limit of string/M-theory may be powerfully tackled by employing the DFT/ExFT in which the U-duality symmetry is manifestly realized. For the  case of DFT, see \cite{Hohm:2015doa}
and several references therein for earlier work. Last but not least,  the generalized geometry underlying exceptional field theories may pave the way to the construction of effective actions that genuinely go beyond 11D supergravity, thereby shedding light on important aspects of M-theory.

In this paper, we shall focus on the exceptional field theory based on $\E7$ \cite{Hohm:2013uia}, and starting from its supersymmetric extension provided in \cite{Godazgar:2014nqa}, we formulate the theory in  $(4+56|32)$ dimensional superspace. One of our main motivations is the construction of actions for M-branes propagating in a target space described by the generalized geometry of exceptional field theory. This problem is still open, though progress has been made in the form of exceptional sigma models for string theory \cite{Arvanitakis:2018hfn}. The importance of a superspace formulation of target space supergravities becomes especially clear with the realization that all known actions for branes beyond strings are feasible only as sigma models in which the target is a superspace. Another motivation for the exploration of supergeometry in exceptional field theories is to find clues in the search for an extended geometrical framework which would unify the external (spacetime) and internal space diffeomorphisms.

Our approach to the superspace formulation of the supersymmetric $\E7$ ExFT is to elevate the 4-dimensional ``external'' spacetime to $(4|32)$ dimensional ``external'' superspace\footnote{Ungauged 4D $N=8$ superspace was constructed in \cite{Brink:1979nt, Howe:1981gz}, see also \cite{Bandos:2015ila, Howe:2015hpa}. Our construction will reduce to this upon discarding all dependence on internal coordinates.}
and to augment this with a 56-dimensional ``internal'' space. As such, the external diffeomorphisms and local supersymmetry transformations of ExFT are unified to external superdiffeomorphisms with structure group GL$(4|32)$, with $\E7$ internal diffeomorphisms treated separately. In particular, there are separate (super)vielbeins for the two spaces.
This is in contrast to early work involving so-called ``central charge superspace'' \cite{Howe:1980th} where the vielbeins were unified into a single sehrvielbein but
with all fields independent of the additional 56 coordinates,\footnote{For a discussion
of how to derive the $\E7$ section condition from a superparticle
moving in central charge superspace, see \cite{Bandos:2015cha}.}
as well as more recent efforts in superspace double field theory
where a unified description is sought
(see e.g. \cite{Hatsuda:2014qqa, Polacek:2014cva, Bandos:2015cha,
Cederwall:2016ukd}).

Our approach turns out to require more than just a superspace lift of \cite{Godazgar:2014nqa}. We find that it is important to redefine a constrained two-form of the theory, so that it transforms inhomogeneously under Lorentz transformations. This allows one to eliminate the internal part of the Lorentz spin connection everywhere, with the constrained two-form now playing its role.
Another important step is the relaxation of the constraints imposed on the $\E7$ connection $\Gamma_{\um \un}{}^\up$ in \cite{Godazgar:2014nqa}. Recall that these constraints amounted to (i) the elimination of non-metricity of the internal generalized vielbein postulate; (ii) the vanishing of the $\E7$ torsion tensor; and (iii) requiring that the 4D volume form be covariantly constant, $\nabla_\um e = 0$.
Here we will find it convenient to relax all of them, and to take a completely generic internal $\E7$ connection. Naturally, this is consistent only if the undetermined pieces drop out of the supersymmetry transformations, which we will show.

We also probe further the sector of the theory that involves extra 3-form and 4-form potentials within the framework of the tensor hierarchy formalism. In particular, we show that the solutions to the superspace Bianchi identities lead to on-shell duality equations for the $p$-form field strengths for $p\leq 4$.
We also show that the reduction to component fields provides a complete description of the on-shell supersymmetric theory, including the higher order fermion terms.
As an application of our results, we perform a generalized Scherk-Schwarz reduction and obtain the superspace formulation of maximal gauged supergravity in four dimensions parametrized by an embedding tensor. 

The paper is organized as follows. In Section 2, we review the locally supersymmetric $\E7$ exceptional field theory in components. In Section 3 we lay the groundwork for the superspace formulation, in particular describing the required redefinition of a constrained 2-form potential, and its consequences. In Section 4, we describe the superspace formulation, including the superspace Bianchi identities and their solutions. In Section 5, we present the component results, establishing that they agree with the component formulation of \cite{Godazgar:2014nqa} subject to the redefinition of the 2-form potential. In Section 6, we perform a generalized Scherk-Schwarz reduction and obtain the superspace formulation of maximal gauged supergravity in four dimensions parametrized by an embedding tensor. In Section 7 we comment further on our results and point out future directions. In Appendix A, we give some details of our conventions. Appendix B contains some technical details of the algebra of external and internal covariant derivatives that we found useful in explicit computations.

\section{ Supersymmetric $\E7$ exceptional field theory in components}
\label{S:Sec2}

Let us begin by reviewing the structure of the E$_{7(7)}$-covariant ExFT,
first in its original bosonic formulation \cite{Hohm:2013uia} and then its
supersymmetrized extension \cite{Godazgar:2014nqa}.
The bosonic field content is given by 
\bea
\left\{ e_m{}^a\ ,\cV_\um{}^\ua\ ,A_m{}^\um\ , B_{mn\,\balpha}\ , B_{mn\,\um} \right\}\ ~.
\label{fieldcontent}
\eea
The vierbein $e_m{}^a$ describes the geometry of external 4D spacetime, while the
56-bein ${\cal V}_{\um}{}^{\ua}$, parametrizing the coset $\E7/\SU(8)$, describes
the internal geometry.
The 1-form $A_m{}^{\um}$ gauges internal diffeomorphisms on external spacetime and
lies in the fundamental ($\rep{56}$) of $\E7$. Requiring closure of internal
diffeomorphisms on the 1-form requires the existence of 2-forms $B_{mn\, \balpha}$ 
and $B_{mn\, \um}$ valued respectively in the adjoint ($\rep{133}$) and fundamental 
($\rep{56}$) representations.
The internal tangent space index $\ua$ on the $56$-bein decomposes under 
$\SU(8)$ as $\rep{28} + \brep{28}$,
\be
\cV_\um{}^\ua = \{ \cV_\um{}^{ij}, \cV_\um{}_{ij}\}\ ,
\ee
satisfying $\cV_{\um ij} = (\cV_\um{}^{ij})^*$ with $\SU(8)$ indices $i,j, \dots=1, \dots, 8$.

All fields in the theory, including the symmetry transformation parameters that will be encountered below, depend on both external ($x^m$) and internal ($y^\um$) coordinates, with the dependence
on the latter subject to the section conditions. We write these as
\begin{align}
(t_\balpha)^{\um\un}\,\partial_\um \otimes \partial_\un \ = \ 0 \ ,\qquad
  \Omega^{\um\un}\,\partial_\um \otimes \partial_\un = 0
\end{align}
where the derivatives are understood to act on any two (or the same) fields or parameters.
Here $(t_\balpha)_\um{}^{\un}$ are the E$_{7(7)}$ generators in the fundamental representation,
$\Omega^{\um \un}$ is the invariant symplectic form of $\E7 \subset {\rm Sp}(56)$, and we employ
the usual (NW-SE) conventions for raising and lowering $\rep{56}$ indices, e.g.
$(t_\balpha)^{\um \un} = \Omega^{\ul{mp}} \,(t_\balpha)_\up{}^\un$ and
$(t_\balpha)_{\um \un} = (t_\balpha)_\um{}^\up \,\Omega_{\ul{pn}}$.

In addition, the field $B_{mn\,\um}$ is constrained on its internal index so that it obeys the
section condition with respect to both $\pa_\um$ and itself, i.e.
\begin{gather}
(t_\balpha)^{\um\un}\,B_\um \otimes \partial_\un \ = \ 0 \ ,\qquad
\Omega^{\um\un}\,B_\um  \otimes \partial_\un = 0~, \eol
(t_\balpha)^{\um\un}\,B_\um B_\un \ = \ 0 \ ,\qquad
\Omega^{\um\un}\,B_\um  B_\un = 0~.
\label{eq:ConstrainedTensor}
\end{gather}
In the first set of equations, the derivative may act on another field or on
$B_{mn\, \um}$ itself.

In principle, 3-forms and 4-forms are also required for a complete description of the
tensor hierarchy, but these drop out of the action, and so one can usually avoid any
explicit discussion of their properties. Nevertheless, we will find it useful to
discuss them briefly in a few places. The 3-forms are $C_{mnp}{}^\um{}_\balpha$ and
$C_{mnp}{\,}_\um{}^\un$, with the former valued in the $\rep{912}$ and the latter
constrained on its lower index. The unconstrained 4-forms are 
$D_{mnpq}{\,}_\balpha$ and $D_{mnpq}{\,}^{\um\un}{}_\balpha$, respectively in the
$\rep{133}$ and in the $\rep{8645}$, while there appear to be as many as
three constrained 4-forms
$D_{mnpq}{\,}_\um$, $D_{mnpq}{\,}_\um{}^\balpha$, and $D_{mnpq}{\,}_\um{}^{\ul{np}}$,
each obeying the section condition on their lower index $\um$, with the last
field constrained in the $\rep{1539}$ in its upper indices.

\subsection{Generalized vielbein postulates}

For later purposes, we record the generalized vielbein postulates (GVP) satisfied 
by external and internal vielbeins:\footnote{Our convention for the spin
connection matches that of \cite{deWit:2007kvg}, which differs in sign from
that used in the previous $\E7$ papers \cite{Hohm:2013uia, Godazgar:2014nqa}.}
\bea
0 &=& \partial_{m} e_n{}^a - A_m{}^\un\partial_\un e_n{}^a
-\frac12\,\partial_\un A_m{}^\un \,e_n{}^a - \omega_m{}^{ab} e_{nb}
- \Gamma_{mn}{}^p  e_p{}^a
\nn\\
 &=& D_m e_n{}^a  - \omega_m{}^{ab} e_{nb} - \Gamma_{mn}{}^p  e_p{}^a\ ,
\label{vp1}\w2
0&=& \partial_\um e_n{}^a 
-\frac13\,\Gamma_{\underline{km}}{}^\uk\,e_n{}^a - \omega_\um{}^{ab} e_{nb}
+ \pi_\um{}^{ab} e_{nb}\ ,
\label{vp2}\w2
0&=&
D_m {\cal V}_\un{}^\ua
-  {\cal V}_{\un}{}^{\ub}  {\cal Q}_m{}_{\underline b}{}^{\underline{a}}
-  {\cal V}_{\un}{}^{\ub}  {\cal P}_m{}_{\underline b}{}^{\underline{a}}
\label{vp3}\w2
0 &=& 
\partial_\um \cV_\un{}^\ua 
- \cV_{\un}{}^\ub \cQ_\um{}_\ub{}^\ua - \Gamma_{\underline{mn}}{}^\up\, \cV_\up{}^\ua\ .
\label{vp4}
\eea
The connections $\Gamma_{m n}{}^p$ and $\pi_\um{}^{ab}= \pi_\um{}^{(ab)}$ are
defined by \eqref{vp1} and \eqref{vp2}, and ${\cal Q}$ and ${\cal P}$ live in \SU(8) and its orthogonal complement in $\E7$, respectively, so that
\begin{align}
\cQ_m{}_{\ul b}{}^{\ul a} =
\begin{pmatrix}
\delta_{[k}{}^{[i} \cQ_m{}_{l]}{}^{j]} & 0 \\
0 & -\delta_{[i}{}^{[k} \cQ_m{}_{j]}{}^{l]}
\end{pmatrix}~, \qquad
\cP_m{}_{\ul b}{}^{\ul a} = 
\begin{pmatrix}
0 & \cP_m{}^{kl ij}\\
\cP_m{}_{kl ij} & 0
\end{pmatrix}
\end{align}
and similarly for $\cQ_\um$ and $\cP_\um$.
The $\E7$ covariant derivative $D_m$ is defined as 
\be
D_m := \partial_m - \mathbb{L}_{A_m}\ ,
\label{dm}
\ee
where the  generalized Lie derivative acts on a fundamental vector $V^\um$ of weight $\lambda(V^\um)$ as 
\be
\mathbb{L}_{\Lambda} V^\um \ := \ \Lambda^\un \partial_\un V^\um - 12 \left[\partial_\un \Lambda^\um\right]_{\rm adj}\,V^\un +\lambda(V)\,\partial_\un \Lambda^\un\,V^\um
\ee
where the second term is projected onto the adjoint of $\E7$, i.e.
\be
\left[\partial_\un \Lambda^\um\right]_{\rm adj} :=  \mathbb{P}^\um{}_\un{}^\ur{}_\us\ \partial_\ur \Lambda^\us\ ,
\ee
with the adjoint projector given by
\bea
\mathbb{P}^\um{}_\un{}^\ur{}_\us &=&  (t^\balpha)^\um{}_\un \, (t_\balpha)^\ur{}_\us
= \frac{1}{24} \delta^\um{}_\un\,\delta^\ur{}_\us +\frac{1}{12} \delta^\um{}_\us\, \delta^\ur{}_\un  + (t^\balpha)^{\um\ur}\, (t_\balpha)_{\un\us} -\frac{1}{24} \Omega^{\um\ur}\, \Omega_{\un\us}\ .
\label{tti}
\eea
We emphasize that the symplectic metric $\Omega^{\um\un}$ is used to raise and lower the $56$-plet indices as $V^\um = \Omega^{\um\un} V_\un$ and $V_\un=V^\um \Omega_{\um\un}$, and it is an invariant tensor of weight $0$, namely $\mathbb{L}_{\Lambda} \Omega^{\um\un}=0$.

Defining the generalized torsion tensors as
\bea
\Gamma_{[mn]}{}^p &=& \frac12 T_{mn}{}^p\ ,
\label{t1}\w2
\Gamma_{\underline{mn}}{}^\uk
-12\,\mathbb{P}^\uk{}_\un{}^\ur{}_\us\, \Gamma_{\underline{rm}}{}^\us
+4\,\mathbb{P}^\uk{}_\un{}^\ur{}_\um\, \Gamma_{\underline{sr}}{}^\us
&=& \cT_{\underline{mn}}{}^\uk\ ,
\label{t2}
\eea
the following constraints are imposed in \cite{Godazgar:2014nqa}
\be
T_{mn}{}^p = 0\ , \qquad 
\cT_{\underline{mn}}{}^\uk =0\ .
\label{t12}
\ee
The definition of generalized torsion  $\cT_{\underline{mn}}{}^\uk$ is motivated by the relation
\be
\label{r}
\mathbb L_V^\nabla W^\um - \mathbb L_V W^\um = \cT_{\un\uk}{}^{\um}\, V^\un W^\uk\ ,
\ee
for vectors $V,W$ of weight $\frac12$ where $\mathbb L_V^\nabla$ denotes the generalized Lie derivative with all partial derivatives replaced by covariant derivatives. Explicit evaluation of this relation gives the expression \eq{t2}. 

The connection $\Gamma_{\underline{mn}}{}^\up$ lives in the algebra $\mathfrak{e}_{7(7)}$ and as such we can write
\be
\Gamma_{\underline{mn}}{}^\up =  \Gamma_\um{}^\balpha (t_\balpha)_\un{}^\up
~\in~ \mathfrak{e}_{7(7)}\ .
\label{e7c}
\ee
Using this relation in \eq{t2}, one finds that
\be
\frac17 \cT_\um{}^\balpha = \mathbb{P}_{(912) \um}{}^{\balpha\,\un}{}_\bbeta \ \Gamma_\un{}^\bbeta\ ,
\label{ac}
\ee
where the projector onto the $\bf 912$ dimensional representation is given by \cite{deWit:2002vt}
\be
\mathbb{P}_{(912) \um}{}^{\balpha\,\un}{}_\bbeta := \frac17 \left(  \delta_\um{}^\un\,\delta^\balpha{}_\bbeta + 4 (t^\balpha t_\bbeta)_\um{}^\un -12 (t_\bbeta t^\balpha)_\um{}^\un \right)\ .
\ee
We have used the notation $(t_\balpha t_\bbeta)_\um{}^\un \equiv (t_\balpha)_\um{}^\uk\,(t_\bbeta)_\uk{}^\un$.

\subsection{Bosonic symmetries, duality equations and tensor hierarchy}

The full bosonic theory is invariant under the generalized diffeomorphisms, vector, and tensor gauge symmetries, and shift symmetries with parameters $(\xi^m,\  \Lambda^{\um}, \ \Xi_{m \,\balpha},\ \Xi_{m \,\um},\ \Omega_{mn\,\um}{}^{\un},\  \Omega_{mn}{}^{\un}{}_\balpha)$, respectively. These transformations are given by
\bea
\delta e_m{}^a &=& \xi^n D_n e_m{}^a
 + D_m\xi^n e_n{}^a +  \mathbb{L}_{\Lambda} e_m{}^a\ ,
\nn\w2
\delta \cV_\um{}^\ua &=&  \xi^m D_m \cV_\um{}^\ua  +  \mathbb{L}_{\Lambda} {\cal V}_\um{}^\ua\ , 
\nn\w2
\delta A_m{}^\um &=& \xi^n\,{\cal F}_{nm}{}^\um + {\cal M}^{\um\un}\,g_{mn} \,\partial_\un \xi^n + D_m \Lambda^\um  
   +  12\,(t^\balpha)^{\um\un}\,\partial_\un\Xi_{m\,\balpha}
   +\frac12\,\Omega^{\um\un}\,\Xi_{m\,\un} \ ,
\nn\w2
\Delta  B_{mn\,\balpha} &=& \xi^p\,H_{mnp\,\balpha}
+ (t_\balpha)_{\um\un}\, \Lambda^\um{\cal F}_{mn}{}^\un 
+ 2\,{ D}_{[m}\Xi_{n]\balpha} 
\nn\w2
&& 
+ \partial_\um \Omega_{mn}{}^\um{}_\balpha + (t_\balpha)_\um{}^\un \Omega_{mn}{}_\un{}^\um\ ,
\nn\w2
\Delta B_{mn\,\um} &=& \xi^p\,H_{mnp\,\um}
-2i\,{\varepsilon}_{mnpq} g^{qr} {D}^p\left(g_{rs}\partial_\um \xi^s\right) 
+ \cF_{mn}{}^\un\partial_\um \Lambda_\un - \partial_\um \cF_{mn}{}^\un   \Lambda_\un
\nn\w2
&& + 2 D_{[m}\Xi_{n]\um} 
+48\,(t^\balpha)_\un{}^\ur  \left(\partial_\ur\partial_\um A_{[m}{}^\un\right) \Xi_{n]\balpha}
\nn\w2
&& - \partial_\um \Omega_{mn}{}_\un{}^\un - 2 \,\partial_\un \Omega_{mn}{}_\um{}^\un\ ,
\label{bst}
\eea
where 
\bea
\Delta  B_{mn\,\balpha} &:=& \delta  B_{mn\,\balpha} + (t_\balpha)_{\um\un}\,A_{[m}{}^\um\, 
\delta A_{n]}{}^\un \ ,
\nn\w2
\Delta B_{mn\,\um} &:=& \delta B_{mn\,\um} +\left(A_{[m}{}^\un \partial_\um \delta A_{n]}{}_\un
- \partial_\um A_{[m}{}^\un \delta A_{n]}{}_\un\right)\ ,
\eea
and
\bea
\cM_{\um\un} &=& \cV_{\um\,ij} \cV_\un{}^{ij} + \cV_{\un\,ij} \cV_\um{}^{ij}\ ,
\label{VV}\w2
\cF_{mn}{}^\um &=& 2 \partial_{[m} A_{n]}{}^\um
-[A_m{}, A_n{}]^\um_{\rm E}
-12 \,  (t^\balpha)^{\um\un} \,\partial_\un B_{mn\,\balpha}
-\frac12\,\Omega^{\um\un}\,B_{mn\,\un}\ \;,
\label{cF}
\eea
with the E-bracket defined by 
\bea
{}[A_m{}, A_n{}]^\um_{\rm E} &:=&
2\,A_{[m}{}^\un \partial_\un A_{n]}{}^\um 
+\frac1{2}\left(24\, (t_\balpha)^{\um\un} (t^\balpha)_{{\ul p}\uq}
-\Omega^{\um\un}\Omega_{{\ul p}\uq}\right)\,A_{[m}{}^{\ul p}\,\partial_\un A_{n]}{}^\uq\ .
\eea

The field strengths $H_{mnp\,\balpha}$ and $H_{mn\, \up}$ are defined by the Bianchi identity
\be
D_{[m}\cF_{np]}{}^\um = -4 (t^\balpha)^{\um\un}\, \partial_\un H_{mnp\,\balpha} -\frac16 \Omega^{\um\un} H_{mnp\,\un}\ .
\label{fb}
\ee
Note that ${\cal F}, D_m {\cal F}$ and $H_\balpha$ transform covariantly under internal diffeomorphisms, while $H_\um$ does not, which is evident from the presence of the non-covariant term $\partial_\un H_\balpha$ term in \eqref{fb}. From this equation, upon using the definition of $\cal F$, one finds that 
\bea
H_{mnp\,\balpha} &:=&
3\,D_{[m} B_{np]\,\balpha}
-3\,(t_\balpha)_{\um\un}\,A_{[m}{}^\um\partial_{n\vphantom]} A_{p]}{}^\un
+(t_\balpha)_{\um\un}\,A_{[m}{}^\um [ A_n{}, A_{p]}{}]^\un_{\rm E}
\nonumber\\
&&{}
- \partial_\um C_{mnp}{}^\um{}_\balpha - (t_\balpha)_\um{}^\un C_{mnp}{\,}_\un{}^\um
\;,
\label{ha}\w2
H_{mnp\,\um} &:=&
\Big[3\,D_{m} B_{np\,\um}
-3\left(A_{m}{}^\un \partial_\um\partial_{n\vphantom]} A_{p}{}_\un -
 \partial_\um A_{m}{}^\un  \partial_{n\vphantom]} A_{p}{}_\un
\right) 
+72\, (t^\balpha)_{\uk}{}^{{\ul p}}\, \pa_\um \pa_{\ul p} A_{m}]^\uk \, B_{np\balpha}
\nonumber\\
&&{}
+\left(A_{m}{}^\un \partial_\um
 [ A_n{}, A_{p}{}]_{\rm E\,\un}
 -
 \partial_\um A_{m}{}^\un   [ A_n{}, A_{p}{}]_{\rm E\,\un}
\right)\Big]_{[mnp]} 
+ \partial_\um C_{mnp}{\,}_\un{\,}^\un + 2 \,\partial_\un C_{mnp}{\,}_\um{}^\un
\ .
\label{hm}
\eea
The three-forms $C_{mnp}{}^\um{}_\balpha$ and $C_{mnp}{\,}_\un{}^\um$ introduced in \eqref{ha} and \eqref{hm}
are projected out of the Bianchi identity \eqref{fb} using the section condition.\footnote{%
We use the identity \eq{tti} and $t^{\balpha(\um\un} C_{mnp}{}^{\up){}_\balpha}=0$. The latter identity follows from the fact that $C_{mnp}{}^\up{}_\balpha$ belongs to the $\rep{912}$ of $\E7$ while $(\rep{56}\times \rep{56}\times \rep{56})_S$ does not contain the $\rep{912}$.}
They may be thought of as parametrizing the part of the field strengths $H_{mnp\,\balpha}$, $H_{mnp\,\um}$,
which is left
undetermined by \eqref{fb} with their presence being necessary for invariance of the curvatures under 
the higher $p$-form gauge transformations.
The covariant derivatives read explicitly
\bea
D_m B_{np \,\balpha} &=&
\partial_m B_{np\,\balpha} - A_m{}^\um \,\partial_\um B_{np \,\balpha}
- 12\, (t_\bgamma)^\ur{}_\us\,f^{\bgamma\bbeta}{}_\balpha \partial_\ur A_m{}^\us\,B_{np\, \bbeta} - \partial_\un A_m{}^\un\,B_{np\,\balpha}\ ,
\nn\w2
 D_m B_{np\,\um} &=& \partial_m B_{np\,\um}- A_m{}^\un \,\partial_\un  B_{np\,\um}
-\partial_\um A_m{}^\un  B_{np\,\un} - \partial_\un A_m{}^\un\,B_{np\,\um}\ .
\eea
%


\begin{table}[tb]
\begin{center}
\begin{tabular}{|c||c|c|c|c|c|c||c|c|}\hline
{\rm field} & $e_m{}^a$ & $\xi^m$ & ${\cal V}_{\um}{}^{ij}$  & $A_m{}^\um$, $\Lambda^\um$ &
$B_{mn\,\balpha}$\,, $\Xi_{m\,\balpha}$ &
$B_{mn\,\um}$, $\Xi_{m\,\um}$ & $\chi_{ijk}$ & $\psi_m{}^i$\,, $\epsilon^i$
\\ \hline
$\lambda$ & $\frac12$ & 0 & 0 & $\frac12$ & $1$ & $\frac12$ & $-\frac14$ & $\frac14$ \\ \hline
\end{tabular}
\caption{$\lambda$-weights for the bosonic and fermionic fields and parameters.}
\label{tab:weights}
\end{center}
\end{table}


The 3-form field strengths in turn obey the Bianchi identities
\bea
4\,D_{[m} H_{npq] \,\balpha} &=& -3 \,(t_{\balpha})_{\um\un} \,
\cF_{[mn}{}^\um \cF_{pq]}{}^\un-\partial_\um G_{mnpq}{}^\um{}_\balpha 
- (t_{\balpha})_{\um}{}^\un\, G_{mnpq\,\un}{}^\um\ ,
\label{hb1}\w2
4\,D_{[m} H_{npq]\um} &=& -6\,{\cal F}_{[mn}{}^\un\,\partial_{|\um|} {\cal F}_{pq]\,\un}
- 24 (t^\balpha)_\up{}^\un \pa_\um \pa_\un A_{[m}{}^\up\,H_{npq]\,\balpha}
\nn\w2
&& + \partial_\um G_{mnpq}{\,}_\un{\,}^\un 
+ 2 \,\partial_\un G_{mnpq}{\,}_\um{}^\un\ ,
\label{hb2}
\eea
which serve to define the curvatures $G_{mnpq}{}^\um{}_\balpha$ and $G_{mnpq}{\,}_\um{}^\un$ associated with the 3-form potentials. This leads to the introduction of 4-form potentials in certain representations of $\E7$ and obeying certain constraints. The transformation rules for the 3-form and 4-form potentials can be determined from the requirement of the closure of the algebra. We will not need these transformation rules, except for the behavior of the 3-form potentials under external diffeomorphisms, which we shall derive below. We will also derive the duality equations obeyed by 4-form field strengths below, and we shall comment on the occurrence of particular 4-form potentials in their definitions in Section \ref{app:gc}.

The curvatures associated with the $p$-form potentials with $p=1,2,3$  obey duality equations given by \cite{Hohm:2013uia}\footnote{The first equation can also be written as $\cF^-_{mn\,ij} \equiv \tfrac{1}{2} \cF_{mn\,ij} -\frac{1}{4}\,  \varepsilon_{mnpq} \cF^{pq}{}_{ij}=0$. The third equation \eqref{hd2} had an overall sign mistake
in \cite{Hohm:2013uia, Godazgar:2014nqa} that is corrected below, keeping in mind the
change in sign of the spin connection.}
\bea
\cF_{mn}{}^\um &=& \frac{i}{2}\,  \varepsilon_{mnpq}\, \Omega^{\um\un}\,\cM_{\un\uk}\,\cF^{pq\,\uk}\ ,
\label{fd}\w2
\varepsilon^{mnpq} H_{npq\,\balpha} &=& \left(t_\balpha\right)_\un{}^\um \left( P^{m ijkl} \cV^\un{}_{ij} \cV_{\um kl}- P^m{}_{ijkl} \cV^{\un ij} \cV_\um{}^{kl}\right)\ , 
\nn\w2
&=& -\frac{i}{2} (t_\balpha)_\um{}^\un (D^m \cM^{\um\uk}) \cM_{\un\uk}\ ,
\label{hd1}\w2
\frac{1}{12} \varepsilon^{mnpq} H_{npq\,\um} &=&
    - 2i \,e_a{}^m e_b{}^n \left(\partial_\um \omega_n{}^{ab} - D_n \omega_\um{}^{ab}\right) 
    - \frac13 D^m \cV^{\un ij} \partial_\um \cV_{\un ij}\ ,
\label{hd2}
\eea
where
\be
\omega_\um{}^{ab} := -e^{n[a} \partial_\um e_n{}^{b]}\ .
\label{umab}
\ee
The first duality equation \eqref{fd} is required together with the second-order pseudo-action given in \cite{Hohm:2013uia}
in order to describe the correct vector field dynamics. 
The second order field equation for the vector fields can then be obtained by the external curl of \eqref{fd} 
together with the Bianchi identity \eqref{fb}. Comparing this second order equation to the one  obtained from variation
of the pseudo-action gives rise to the duality equations \eqref{hd1}, \eqref{hd2}. As in the
Bianchi identity \eqref{fb}, the duality equations \eqref{hd1} and \eq{hd2} 
only follow under projection with $(t^\balpha)^{\um\un}\, \partial_\un$, 
and their remaining parts may thus be taken as an equations for the three-form potentials
introduced in \eqref{ha} and \eq{hm}.\footnote{It is worth noting that the variation of the duality equation \eq{fd} yields \eq{hd1} but not \eq{hd2} for the constrained field. The latter involves two derivative terms on the right hand side, and these are derived in \cite{Hohm:2013uia} by employing a suitable action.} 
The variations of the 3-form potentials under all the bosonic symmetries can be determined from the requirement of the invariance of the duality equations above. For later purposes we shall in particular need their variations under the external diffeomorphisms. To determine them, we consider the invariance of the duality equation \eq{hd1} under the external diffeomorphisms. To this end, the following formula for the general variation of $H_\balpha$ is useful:
\bea
\delta H_{mnp\,\balpha}
&=& \Big[3\, D_{m} (\Delta B_{np\, \balpha})
-3\,(t_{\balpha})_{\um\un} \,\delta A_{m}{}^\um\,{\cal F}_{np}{}^\un
-\partial_\um \Delta C_{m n p}{}^\um{}_\balpha 
-(t_{\balpha})_\um{}^\un\,\Delta C_{mnp}{\,}_\un{}^\um 
\Big]_{[mnp]}
\;,
\eea
where
\bea
\Delta C_{mnp}{}^\um{}_\balpha &=& \Big[\delta C_{mnp}{}^\um{}_\balpha 
+ 21\,(\mathbb{P}_{912}){}_\balpha{}^{\um\,\bbeta}{}_\un\,
B_{mn\,\bbeta}\,\delta A_{p}{}^\un
-7\,(\mathbb{P}_{912}){}_\balpha{}^{\um\,\bbeta}{}_\un\,
(t_\bbeta)_{\ur\us}\,A_m{}^\un A_n{}^\ur \delta A_p{}^\us
\Big]_{[mnp]}
\;,
\nonumber\\
\Delta C_{mnp\,\un}{}^\um &=&
\Big[\delta C_{mnp\,\un}{}^\um - \frac32\,\delta A_m{}^\um \,B_{np\,\un}
+24\,\partial_\un\delta A_m{}^\ur\, B_{np \,\balpha} (t^\balpha)_\ur{}^\um
-12\,\delta A_m{}^\ur\,\partial_\un B_{np \,\balpha} (t^\balpha)_\ur{}^\um 
\nonumber\\
&&{}
-12\,(t_\balpha)_{\ur\us}(t^\balpha)_\uk{}^\um\,\partial_\un A_m{}^\uk A_n{}^\ur \delta A_{p}{}^\us
-\frac56\,\delta A_m{}^\um \partial_\un(A_{n \uk} A_p{}^\uk)
+\frac16\,\partial_\un \delta A_m{}^\um A_{n\uk} A_p{}^\uk
\nonumber\\
&&{}
-\frac23\,A_m{}^\um \partial_\un\delta A_{n\,\uk} A_p{}^\uk
-\frac16\,\partial_\un A_m{}^\um \delta A_{n\,\uk} A_p{}^\uk
+\frac13\,A_m{}^\um \delta A_{n\,\uk} \partial_\un A_p{}^\uk \Big]_{[mnp]}
\;,
\eea
and we have used the following identity:
\be
(t^\bbeta)^{\uk\us}\, (t_\bbeta)_{(\um\un}\,(t^\balpha)_{\ur)\us}=\frac13 (t_\bbeta)_{\us(\um}\, (t^\bbeta)_{\un\ur)}\, (t_\balpha)^{\uk\us} -\frac{1}{12} (t^\balpha)_{(\um\un}\,\delta_{\ur)}{}^\uk\ .
\ee
Under external diffeomorphisms
\bea
\Delta C_{mnp}{}^\um{}_\balpha 
&=& \xi^r\,G_{rmnp}{}^\um{}_\balpha 
\;,
\nonumber\\
\varepsilon_m{}^{npq}\,\Delta C_{npq\,\un}{}^\um &=&
\varepsilon_m{}^{npq}\,\xi^r G_{rnpq\,\un}{}^\um
+6\,i\,\partial_\un\xi^n\,F_{nm}{}^\um
+12\,i\,\partial_{(\un} \left(
{\cal M}^{\um\uk}\,\partial_{\uk)}\xi^n\,g_{mn}
\right)
\nonumber\\
&&{}
-8i\,\partial_\uk {\cal M}^{\uk\um}\,g_{mn}\partial_\un\xi^n\ ,
\eea
the duality equation of $H_\balpha$ transforms covariantly, provided that we also impose the following duality equation for the four-form field strength
\bea \label{eq:G912duality}
\varepsilon^{mnpq}\,G_{mnpq}{}^\um{}_\balpha
&=&
14i\,(\mathbb{P}_{912})^\um{}_{\balpha\,\un}{}^\bbeta\,
{\cal M}^{\un\ur} J_{\ur\bbeta}\ ,
\eea
with the current $J_{\ur\bbeta}$ defined by 
\be
{\cal M}^{\un\uk} \partial_\ur {\cal M}_{\uk\um}=J_{\ur\balpha}\,(t^\balpha)_\um{}^\un \ .
\ee
In this calculation the terms involving the field strength $G_{mnpq\,\um}{}^\un$ cancel, and consequently a duality equation for this field strength does not follow. However, $H_{mnp\,\balpha}$ determined from \eq{hd1}, substituted into the Bianchi identity \eq{hb1} gives
\be
\left(t_\balpha\right)_\um{}^\un \Big( e\,\varepsilon^{mnpq}  G_{mnpq\,\un}{}^\um +48i\cM_{\un\ur} \frac{\delta (e V) }{\delta \cM_{\um\ur}} 
+14i (t^\bbeta){}_\un{}^\um (\mathbb{P}_{912})^\uk{}_{\bbeta\,\ur}{}^\bgamma \partial_\uk \left( e \cM^{\ur\us} J_{\us\bgamma}\right) \Big)=0 \ ,
\label{Geq}
\ee
from which one can derive\footnote{In order to strip off the $(t_\balpha){}_\um{}^\un$, we make use
of the fact that $G_{mnpq\,\um}{}^\un$ is constrained on its $\um$ index. This condition means that
$G$ lies in a generic $\rep{56} \times \rep{56}$ representation, as any projection operation
would spoil the section condition. Since a generic $\rep{56} \times \rep{56}$ always contains 
the adjoint representation, there is no ambiguity in solving this equation.}
\begin{align} \label{eq:Gmn.duality}
    \frac{1}{24} e \, \veps^{mnpq} \,G_{mnpq}{\,}_\un{}^\um
    &= i \pa_\ur \Big( 
        e \,\pa_\un \cM^{\ur \um} \Big)
        + \frac{i}{3} \pa_\un \Big(
            e\,\cM^{\ur \us} \cM^{\ur\um}\Big)
    \eol & \quad
    - \frac{i}{24} e \, \cM^{\um \ut} \pa_\un \cM^{\ur \us}\Big( \pa_\ut \cM_{\ur \us}
   -12 \pa_\ur \cM_{\ut \us}\Big)
    + \frac{i}{2} \pa_\ur \cM^{\um \ut} \cM^{\ur \us}  \pa_\un \cM_{\ut \us }~.
\end{align}
In obtaining this result, we have used \eq{fd} and the following scalar field equation that follows from the action given in \cite{Hohm:2013uia}:
\be
(t_\alpha)_\um{}^\uk \cM_{\uk\un} \left( 
D_m\left( e D^m \cM^{\um\un}\right) +3 e \cF_{mn}{}^\um \cF^{mn\,\un} +24\, \frac{\delta (e V)}{\delta\cM_{\um\un}}\right)=0 \ ,
\ee
where
\be
V= -\frac{1}{48} \cM^{\um\un}\, \partial_\um \cM^{\ur\us}\left(  \partial_\un \cM_{\ur\us} -24\, \partial_\ur \cM_{\us\un}\right)- \left(e^{-1}\partial_\um e\right)  e^{-1} \partial_\un  \big(e\cM^{\um\un}\big)  -\frac14 \cM^{\um\un} \partial_\um g^{mn} \partial_\un g_{mn}\ .
\ee

\subsection{Supersymmetry transformation rules}
\label{S:SUSYRules}
 
 The supersymmetry transformation rules are given by
\bea
  \delta e_m{}^a &=&
  \bar\epsilon^{i}\gamma^a\psi_{m i} ~+~
  \bar\epsilon_{i}\gamma^a\psi_m{}^i \ , \nn\w2
  \delta \cV_\um{}^{ij} &=&  2\sqrt{2}\,{\cal V}_{\um kl} \, \Big(
  \bar\epsilon^{[i}\chi^{jkl]}+\frac1{24}\varepsilon^{ijklmnpq}\,
  \bar\epsilon_{m}\chi_{npq}\Big)   \ ,  \nn \\[1ex]
    \delta A_m{}^\um
    &=&
    -i\,\sqrt{2}\,\Omega^{\um\un} \cV_\un{}^{ij}\,\Big(
    \bar\epsilon^{k}\,\gamma_{\mu}\,\chi_{ijk}
    +2\sqrt{2}\, \bar\epsilon_{i}\,\psi_{\mu j}\Big) + \CC\ , 
    \nn\w2
    \Delta B_{mn\,\balpha} &=&-\frac{2}{3} \sqrt{2} \,
    (t_{\balpha})^{{\ul p}\uq}\, \Big( \cV_{{\ul p}\,ij} \cV_{\uq\,kl}\,
    \bar\epsilon^{[i}\,\gamma_{mn}\,\chi^{jkl]}
    + 2 \sqrt{2}\, \cV_{{\ul p}\,jk} \cV_{\uq}{}^{ik}\,
    \bar\epsilon_{i}\,\gamma_{[m}\,\psi_{n]}{}^{j}
    + \CC\Big)\ ,
\nn\w2
\Delta B_{mn\,\um} &=&
\frac{16}3\,
\cV^\un{}^{ij}\, \cD_\um \cV_\un{}_{jk}
 \,\bar\epsilon^k\gamma_{[m} \psi_{n]i}  
 -   \frac{4\sqrt{2}}{3} \, \cV^\un{}_{ij} \cD_\um \cV_{\un\,kl}\,
    \bar\epsilon^{[i}\,\gamma_{mn}\,\chi^{jkl]}
\nn\\
&&{}
-8i \left(
   \bar\epsilon^{i}\,\gamma_{[m} \cD_\um  \psi_{n] i}
 - \cD_\um \bar\epsilon^{i}\,\gamma_{[m}\, \psi_{n] i}\right)
+2 i \varepsilon_{mnpq}\,g^{qr}\,
\cD_\um \left( \bar\epsilon^i \gamma^p \psi_{r\,i} \right) + \CC\ ,
\nn\w2
\delta \psi_m{}^i &=& 
2 \,\cD_m \epsilon^i 
- 4i\,\cV^{\um\,ij} \widehat\nabla_\um\left(\gamma_m \epsilon_j \right)\ ,
\nn\w2
\delta \chi^{ijk} &=& -2 \sqrt{2}\, \cP_m{}^{ijkl} \gamma^{\mu} \epsilon_l
-12\sqrt{2}i \, \cV^\um{}^{[ij}\,   \widehat\nabla_\um \epsilon^{k]}\ .
\label{susy}
\eea
The coset currents $\cP_m{}^{ijkl}$ are defined as 
\be
\cD_m \cV_\un{}^{ij} \equiv  
D_m \cV_\un{}^{ij} +  \cQ_{m\,k}{}^{[i} \,\cV_\un{}^{j]k} 
 ~=~  \cP_m{}^{ijkl} \,\cV_{\un\, kl} \ ,
\label{PQ}
\ee
which also defines the composite ${\rm SU}(8)$ connection
\be
\cQ_m{}_i{}^j ~=~
\frac{2i}{3}\,\cV^{\un\,jk} \,D_m \cV_{\un\,ki}\ .
\label{defQ}
\ee

The covariant derivatives of a spinor $X_{\un i}$ with and without Christoffel connection are defined as
\bea
\cD_m X_{\un\,i} &=&
D_m X_{\un\, i} - \frac14\,\omega_m{}^{ab}\gamma_{ab} X_{\un\,i} + \frac12\cQ_{m\,i}{}^j X_{\un\,j}\ , \label{eq:cDExt}
\w2
\cD_\um X_{\un\,i}  &=&  \partial_\um  X_{\un\, i} - \frac14\,\omega_\um{}^{ab}\gamma_{ab} X_{\un\,i} +\frac12\cQ_{\um\,i}{}^j X_{\un\,j}\ , \label{eq:cDInt}
\w2
\nabla_\um X_{\un\,i} &=& \cD_\um X_{\un\,i} -\Gamma_{\um\un}{}^\ur\, X_{\ur\,i}
-\frac23 \lambda(X) \Gamma_{\underline{rm}}{}^\ur\, X_{\un\,i}\ . \label{eq:nabla}
\eea
The hatted covariant derivative $\widehat\nabla_\um$ is obtained from $\nabla_\um$  
by replacing $\omega_\um{}^{ab}$ with
\be
{\widehat\omega}_\um{}^{ab} \equiv \omega_\um{}^{ab}
+\frac1{4}\,\cM_{\um\un}\,{\cal F}_{mn}{}^\un\,e^{ma} e^{nb}\ .
\ee

Finally, let us note the simplification 
\be
\Delta B_{mn\,\um} + 2\,\Gamma_\um{}^\balpha \Delta B_{mn\balpha} = 
-8i \left(
   \bar\epsilon^{i}\,\gamma_{[m} \cD_\um  \psi_{n] i}
 - \cD_\um \bar\epsilon^{i}\,\gamma_{[m}\, \psi_{n] i}\right)
+2 i \,\varepsilon_{mnpq}\,g^{qr}\,
\cD_\um \left( \bar\epsilon^i \gamma^p \psi_{r\,i} \right) + \CC\ ,
\ee
and that writing out the covariant derivatives in the variations of the fermionic fields gives
\bea
\delta \psi_m{}^i &=& 2 \cD_m \epsilon^i 
+ \frac{1}{4} \cF_{ab}{}^{ij} \gamma^{ab} \gamma_m \epsilon_j
+ i (e_n{}^a \partial_\un e_{pa}) \cV^{\un ij} \gamma^{np}\gamma_m\epsilon_j -4i \cV^{\un ij} \partial_\un (\gamma_m\epsilon_j) 
\nn\w2
&& -2i \cV^{\um ij}\, q_{\um\, j}{}^k \gamma_m \epsilon_k -2i \cV^\um{}_{kl}\, p_\um{}^{ijkl} \gamma_m \epsilon_j\ ,
\w2
\delta \chi^{ijk} &=& -2\sqrt 2 \cP_m{}^{ijkl} \gamma^m\epsilon_l +\frac{3\sqrt 2}{4} \cF_{ab}{}^{[ij} \gamma^{ab} \epsilon^{k]} +3\sqrt 2 i (e_m{}^a \partial_\um e_{na} ) \cV^{\um[ij} \gamma^{mn} \epsilon^{k]}
\nn\w2
&& -12\sqrt 2 i \cV^{\um[ij} \partial_\um \epsilon^{k]} +6\sqrt 2 i \cV^{\um [ij}\, q_{\um l}{}^{k]} \epsilon^l 
-8\sqrt 2 i \cV^\um{}_{rs}\, p_\um{}^{ijkr} \epsilon^s
\nn\w2
&& -6\sqrt 2 i \cV^\um{}_{rs}\, p_\um{}^{rs[ij} \epsilon^{k]}\ .
\eea
where
\be \label{eq:Defqp}
q_{\um\,i}{}^j \equiv \frac23 \cV^{\un jk} \partial_\um \cV_{\un ki}\ ,\qquad p_\um{}^{ijkl} \equiv i\cV^{\un ij} \partial_\um \cV_\un{}^{kl}\ .
\ee

\section{Laying the groundwork for superspace}
The supersymmetry transformations described in the previous section
do not readily admit a lift to a conventional superspace due to a number
of obstacles. Some of these, for example, the term involving the internal
derivative of the supersymmetry parameter in the gravitino transformation \eqref{susy},
are rectified by understanding the structure of external superdiffeomorphisms in
superspace. Other issues, such as the nature of the last term in the transformation
of $B_{mn \,\um}$, require that we first make some redefinitions of fields
appearing at the component level before considering their superspace analogues.

In this section, we will elaborate upon a few modifications of the component
theory that shed light on its superspace lift. First, we describe the action
of the external diffeomorphisms on the fermions. 
Then we proceed to describe a redefinition of the two-form $B_{mn \, \um}$, which
appears necessary to make sense of its superspace analogue. The redefined two-form
turns out to have a simpler transformation under both external diffeomorphisms
and supersymmetry transformations.
Finally, we compute the algebra of external diffeomorphisms.  Afterwards, we will
describe the fate of the four generalized vielbein postulates, \eqref{vp1} -- \eqref{vp4},
in superspace. As we shall see, only the latter two of these, involving the 56-bein,
continue to play any role. We will further argue that a more democratic form of
the internal GVP \eqref{vp4}, which includes non-metricity and leaves the
generalized torsion unfixed, is more natural from a superspace perspective.
This will prove useful both for understanding some of the structure of the
supersymmetry transformations and for connecting with the superspace of gauged
supergravity after a generalized Scherk-Schwarz reduction \cite{Hohm:2014qga},
as we shall see in Section 6.

\subsection{External diffeomorphisms of the fermions}

Let us first summarize some details about external diffeomorphisms in the
supersymmetric theory that have not previously appeared in the literature.
As in the bosonic theory (see \eqref{bst}), the vierbein $e_m{}^a$ and the 56-bein
transform as tensors under external diffeomorphisms. It turns out the same is
not true of their superpartners $\psi^i$ and $\chi^{ijk}$. Rather, these fields
transform under external diffeomorphisms as
\begin{align}\label{E:ExtDiffFerms}
\delta_\xi \psi_m{}^{i}
	&= D_m \xi^n \psi_n{}^{i}
	+ \xi^n D_n \psi_m{}^i
	+ 4 i \,\cV^{\um}{}^{ij} \pa_{\um} \xi^n\, \gamma_{[n} \psi_{m]}{}_{j}
	- \tfrac{i\sqrt{2}}{2} \cV^{\um}{}_{jk} \pa_{\um} \xi^n \,\gamma_n \gamma_m \chi^{i j k} ~,\eol
\delta_\xi \chi^{i j k}
	&= \xi^n D_n \chi^{i j k}
	- 6 i \sqrt 2\, \cV^{\un\, [ij} \pa_{\un} \xi^m \psi_m{}^{k]}
	- \tfrac{i}{6} \veps^{ijk pq rst } \cV^{\un}{}_{pq} \pa_{\un} \xi^m 
		\gamma_m \chi_{rst} ~.
\end{align}
The non-tensorial terms involving internal derivatives of $\xi^m$, which we will refer to as anomalous terms,
can be justified in a few different ways. Perhaps the simplest (which we followed)
is to compute them directly in $D=11$ supergravity after reformulating it to make the local
$\rm SO(1,3) \times SU(8)$ tangent space symmetry manifest \cite{deWit:1986mz}. This corresponds
to an explicit solution of the section condition in ExFT, and so the results can be lifted
to ExFT exactly along the lines followed in
\cite{Godazgar:2013dma, Godazgar:2014sla, Godazgar:2014nqa}. Another approach, formulated
entirely within ExFT, would be to require closure of the algebra of external diffeomorphisms. It will be convenient to work out this algebra after performing a redefinition of the two-form $B_\um$ as we shall do below.

\subsection{Redefinition of the two-form and the algebra of external diffeomorphisms}
\label{sec:newB}

While, as we will see, the external diffeomorphisms of most of the fields can be directly lifted to superspace, the transformation of $B_{mn\,\um}$ -- specifically the second term in \eq{bst} -- proves to be problematic. This is
due to the presence of the inverse vierbein. Whereas we will be identifying the
component vierbein $e_m{}^a$ as the element $E_m{}^a$ of the supervielbein $E_M{}^A$, the inverse
vierbein $e_a{}^m$ has no simple interpretation in superspace, as it does
\emph{not} correspond to the element $E_a{}^m$ of the inverse supervielbein.
This is why, typically in supersymmetric theories, one can formulate supersymmetry
transformations without explicit use of the inverse vierbein.

It turns out that there is a redefinition of the two-form $B_{m n \,\um}$ 
that resolves this issue.\footnote{This redefinition is naturally related
to one recently made in the so-called ``topological phase'' of ${\rm E}_8$ ExFT
that allowed its reinterpretation as a Chern-Simons theory \cite{Hohm:2018ybo}.
There it was the constrained one-form $B_{m \um}$ that admitted a redefinition.}
We will take
\bea
B'_{m n \,\um} &=& B_{m n \,\um} - 2 i \,\veps_{m n}{}^{p q} e_p{}^a \pa_\um e_{q a}
\nn\\
&=& B_{m n \,\um} - 2 i \,\veps_{m n ab}\,\omega_\um{}^{ab}\ .
\eea
It follows that the symmetry transformations of $(A_m{}^\um, B_{mn\balpha})$ given in \eq{bst} preserve their form with $B_{mn\,\um}$ replaced by $B'_{mn\,\um}$ in ${\cal F}_{mn}{}^\um$ and $H_{mnp\,\balpha}$, provided one makes the compensating $\Xi$ and $\Omega$ transformations with parameters 
\be
\Xi_{m \,\um}= -2i \,\xi^n \varepsilon_{nm\,ab}\, \omega_\un{}^{ab}\ ,\qquad 
\Omega_{mn\,\un}{}^\um = i \Lambda^\um \varepsilon_{mnab}\,\omega_\un{}^{ab}\ .
\label{ct}
\ee
Thus we have
\bea
\delta_{\xi,\Lambda} A_m{}^\um &=& \xi^n\,{\cal F}_{nm}{}^\um(B')+ {\cal M}^{\um\un}\,g_{mn} \,\partial_\un \xi^n + D_m \Lambda^\um\ ,
\nn\w2
\Delta_{\xi,\Lambda}  B_{mn\,\balpha} &=& \xi^p\,H_{mnp\,\balpha}
+ (t_\balpha)_{\um\un}\, \Lambda^\um{\cal F}_{mn}{}^\un (B')\ .
\label{bst2}
\eea
Now noting that
\bea
\delta_\xi \omega_\um{}^{ab} &=& 
    \xi^n D_n\omega_\um{}^{ab} 
    - \partial_\um \xi^n e^{m[a} D_n e_m{}^{b]} 
    - e^{m[a} e_n{}^{b]} D_m\partial_\um \xi^n\ , \nn\w2
\varepsilon_{mn}{}^{pq} D_p(g_{qr} \partial_\um \xi^r) &=& 
    \varepsilon_{mnab}\, \left( e^{ap} e_{q}{}^b D_p \partial_\um \xi^q 
        +\partial_\um \xi^n e^{am} \cD_m e_n{}^b\right)\ ,
\eea
where we have used  $\cD_{[m} e_{n]}{}^a=0$, one finds from \eq{bst} that 
\begin{align}\label{eq:DeltaB'}
\Delta_{\xi,\Lambda} B'_{m n \um}
	&= \xi^p H_{p m n \um}(B')
	- 2 i \, \veps_{mnab}\, \pa_\um \xi^p \omega_p{}^{ab}
	+ \cF_{mn}{}^\un (B') \partial_\um \Lambda_\un - \partial_\um \cF_{mn}{}^\un (B')   \Lambda_\un\ .
\end{align}
The second term in this transformation can be readily lifted to superspace as it involves only forms. Here we interpret $\omega_p{}^{ab}$ as a one-form that can be lifted to superspace, as opposed to expressing it as a composite in terms of the vierbein and its inverse. 
In achieving this simplification, we have paid a price. The field $B_{mn\,\um}^\prime$ now transforms under local Lorentz transformations as
\be
\delta B_{mn\,\um}^\prime = -2i\, \veps_{mnab}\, \pa_\um \lambda^{ab}\ .
\ee
We will find soon find that the internal spin connection $\omega_\um{}^{ab}$ no longer appears
in any expressions and covariance under $y$-dependent Lorentz transformations is now ensured 
by the field $B'$ and the field strength $\cF(B')$ in which it appears.

Before moving on, there are a number of features of the field strength $\cF(B')$ we should
discuss. Because we have essentially redefined it as
\bea
{\cal F}_{mn\,ij}(B')
&=&
{\cal F}_{mn\,ij}(B)-i\,{\cal V}^{\um}{}_{ij}\,\varepsilon_{mnpq}\,e^{pa}\,\partial_\um e^q{}_{a}
\;,
\eea
the self-duality equation \eq{fd} now takes the form
\bea
 \widehat\cF_{mn\,ij} &=&\frac12\, e \,\varepsilon_{mnpq}\, \widehat\cF^{pq}{}_{ij}\ ,
\eea
in terms of the modified field strength
\bea
 \widehat{\cF}_{mn\,ij}
 &\equiv&
 \cF_{mn\,ij}(B')
 -2i\,{\cal V}^{\un}{}_{ij}\,e_{[m}{}^a\,\partial_\un e_{n]a}
\;,
\eea
or equivalently
\bea \label{eq:DefHatF-comp}
 \widehat{\cF}_{mn}{}^\um
 &\equiv&
 \cF_{mn}{}^\um(B')
 + 2\,\cM^{\um \un} \,e_{[m}{}^a\,\partial_\un e_{n]a}~.
\eea
The additional term in $\widehat \cF$ can be understood as the twisted
dual of the term we have added to $B$, which is necessary so that $\widehat \cF$
continues to be twisted self-dual. It transforms under Lorentz transformations as
\begin{align}
\delta \widehat \cF_{ab}{\,}_{i j}
	= - 2 i \,\cV^{\um}{}_{i j} \Big(\delta_{a}{}^{c} \delta_{b}{}^{d} + \frac{1}{2} \veps_{ab}{}^{cd} \Big)\pa_{\um} \lambda_{c d}
	= - 4i \, \cV^{\um}{}_{i j} \pa_{\um} \lambda_{ab}^+\ .
\end{align}
As a consequence of the additional term in its definition, $\widehat \cF$ satisfies a modified Bianchi identity
\bea
D_{[p} \widehat \cF_{mn]}{}^\um
	&=& -4\, (t^\balpha)^{\um\un} \partial_\un H_{mnp\, \balpha} 
-\frac16 \,\Omega^{\um\un} H_{mnp\,\un}(B')
\nn\\
&& + 2 \left( D_{[p} \cM^{\um \un}\right)  e_m{}^a \pa_\un e_{n]a}
-2 \,\cM^{\um \un} e_{[m}{}^a e_n{}^b \pa_\un \omega_{p]ab} \ ,
\label{BianchiFhat}
\eea
where we have used the vanishing of the external torsion $T_{mn}{}^p=0$.

Later on, it will be convenient to rewrite this expression in a form that is manifestly covariant under internal diffeomorphisms. To this end, we note that 
\be
\nabla_\un H_\balpha=\pa_\un H_\balpha +\Gamma_\un{}^\bbeta\,f_{\bbeta\balpha}{}^\bgamma\, H_\balpha
-\frac23 \Gamma_{\uk\un}{}^\uk\,H_\balpha\ ,
\ee
and introduce a modified three-form field strength
\bea
\scH_{\um} &=&  
{H}_{\um}(B') 
-24\,(t^\balpha)_{\um}{}^{\un}\,\Gamma_\un{}^\bbeta\,f_{\bbeta\balpha}{}^\bgamma\,H_\bgamma+16\,(t^\balpha)_{\um}{}^{\un}\,\Gamma_{\uk\un}{}^\uk\,H_\balpha
+ 2 \,\cT_\um{}^\balpha \, H_{\balpha}
\nonumber\\
&=&
{H}_{\um}(B') +2\,\Gamma_\um{}^\balpha\,H_\balpha\ .
\eea
The field strength $\scH_\um$ is a tensor under internal diffeomorphisms,
whereas $H_\um(B')$ is not.
Now the Bianchi identity \eqref{BianchiFhat} takes the form
\bea
D_{[p} \widehat \cF_{mn]}{}^\um
	&=& -4 (t^\balpha)^{\um\un} \nabla_\un H_{mnp\, \balpha} 
	+ \frac{1}{3} \, \Omega^{\um \un} \cT_\un{}^{\balpha} \, H_{mnp\, \balpha}
	-\frac16 \Omega^{\um\un} \scH_{mnp\,\un}
\nn\\
&& + 2 \left( D_{[p} \cM^{\um \un}\right)  e_m{}^a \nabla_\un e_{n]a}
-2\, \cM^{\um \un} e_{[m}{}^a e_n{}^b \nabla_\un \omega_{p]ab} \ ,
\label{bfh}
\eea
In the above expressions, we have kept explicit the generalized
torsion tensor $\cT_\um{}^\balpha$, even though it was constrained to vanish in
\cite{Godazgar:2014nqa}. We will soon see that it is convenient to relax this
requirement and allow a non-vanishing $\cT_\um{}^\balpha$.
The modified field strength $\scH_\um$ similarly appears in the covariantized variation of $B'_{mn\, \um}$, which is given by 
\begin{align}
\Delta_{\xi,\Lambda} B'_{mn\,\um} + 2 \,\Gamma_\um{}^\balpha \Delta_{\xi,\Lambda} B_{mn\,\balpha}
	&= \xi^p \scH_{p m n \, \um}
	- 2 i \, \veps_{mnab}\, \pa_\um \xi^p \omega_p{}^{ab}
	+ \cF_{mn}{}^\un (B') \,\nabla_\um \Lambda_\un 
	- \nabla_\um \cF_{mn}{}^\un (B')  \, \Lambda_\un\ .
\end{align}

While the on-shell duality equation \eqref{hd1} of $H_{mnp\, \balpha}$ is unchanged
by the redefinition of $B$, the duality equation \eqref{hd2} for $H_{mnp\, \um}$ now
reads
\begin{align}
\frac{1}{12} \varepsilon^{mnpq} H_{npq\,\um}(B') &=
    -2i \,e_a{}^m e_b{}^n \partial_\um \omega_n{}^{ab}
    - \frac13 D^m \cV^{\un ij} \partial_\um \cV_{\un ij}~.
\end{align}
Note that the duality equation for the modified field strength can be written
\begin{align}
\frac{1}{12} \varepsilon^{mnpq} \scH_{npq\,\um}(B') &=
    -2i \,e_a{}^m e_b{}^n \partial_\um \omega_n{}^{ab}
    + \frac{i}{24} \cD^m \cM^{\ul{pq}} \,\nabla_\um \cM_{\ul{pq}}~,
\end{align}
which is manifestly covariant under internal diffeomorphisms.

Turning to the supersymmetry transformations, the redefinition of the two form $B_{mn\,\um}$ clearly affects only those for $B_{mn\,\um}$, $\psi_m{}^i$, and $\chi^{ijk}$, which now take the form
\bea
\Delta B^\prime_{mn\,\um} &=&
\frac{16}3\,
\cV^\un{}^{ij}\, \cD_\um \cV_\un{}_{jk}
 \,\bar\epsilon^k\gamma_{[m} \psi_{n]i}  
 -   \frac{4\sqrt{2}}{3} \, \cV^\un{}_{ij} \cD_\um \cV_{\un\,kl}\,
    \bar\epsilon^{[i}\,\gamma_{mn}\,\chi^{jkl]}
\nn\\
&&{}
-8i \left(
   \bar\epsilon^{i}\,\gamma_{[m} \cD^\prime_\um  \psi_{n] i}
 - \cD^\prime_\um \bar\epsilon^{i}\,\gamma_{[m}\, \psi_{n] i}\right) + \CC\ ,
\nn\w2
\delta \psi_m{}^i &=& 
2 \,\cD_m \epsilon^i -4i \cV^{\um ij} \nabla^\prime_\um (\gamma_m \epsilon_j)
+ \frac14 \gamma^{np} \gamma_m \widehat\cF_{np}{}^{ij} \epsilon_j
\nn\w2
\delta \chi^{ijk} &=& -2 \sqrt{2}\, \cP_m{}^{ijkl} \gamma^{\mu} \epsilon_l
-12\sqrt{2}i \, \cV^\um{}^{[ij}\, \nabla^\prime_\um \epsilon^{k]}
+ \frac{3\sqrt2}{4} \gamma^{mn} \epsilon^{[k} \widehat\cF_{mn}{}^{ij]}\ ,
\label{susy2}
\eea
where $\cD^\prime_\um $ is obtained from $\cD_\um$ defined in \eqref{eq:cDInt} by dropping the internal connection $\omega_\um{}^{ab}$, and $\nabla^\prime_\um$ is obtained from $\nabla_\um$ defined in \eqref{eq:nabla} by replacing $\cD_\um$ with $\cD^\prime_\um$.
We also give the transformation rule
\be
\Delta B^\prime_{mn\,\um} + 2 \,\Gamma_\um{}^\balpha \Delta B_{mn\,\balpha} = 
-8i \left(
   \bar\epsilon^{i}\,\gamma_{[m} \cD^\prime_\um  \psi_{n] i}
 - \cD^\prime_\um \bar\epsilon^{i}\,\gamma_{[m}\, \psi_{n] i}\right) + \CC
\ee
Note that the last term in the supersymmetry variation of $B_{mn\,\um}$ has vanished,
and the internal connection $\omega_\um{}^{ab}$ has dropped out everywhere,
thereby making the superspace lift of these formulae possible, as we shall see later.

Finally, computing the commutators of the external diffeomorphism, with the two-form field redefinition performed, we find the following soft algebra\footnote{The 
appearance of non-trivial ${\rm SO}(1,3)$ and ${\rm SU}(8)$ parameters is less
surprising if one recalls that in the 4+7 reformulation of $D=11$ supergravity,
the external diffeomorphism corresponds to an 11D diffeomorphism plus a
local ${\rm SO}(1,10)$ transformation; the commutator of two such
transformations gives an ${\rm SO}(1,3) \times {\rm SO}(7)$ transformation.}
\begin{align}\label{E:AlgExtD}
[\delta_{\xi_1}, \delta_{\xi_2}] = \delta_{\rm ext.diff.}(\xi_{12})
	+ \delta_{\rm int.diff.}(\Lambda_{12})
	+ \delta_{\rm Lorentz}(\lambda_{12})
	+ \delta_{\rm SU(8)}(\lambda_{12})
	+ \delta (\Xi_{12}) + \delta (\Omega_{12})\ ,
\end{align}
with the composite parameters given by  
\begin{align}
\xi_{12}{}^m &= -2 \xi_{[1}{}^n D_n \xi_{2]}^m\ , 
\nn\w2
\Lambda_{12}{}^\um &= \xi_{[1}{}^n \xi_{2]}{}^p \cF_{p n}{}^\um (B')
	+ 2 \xi_{[1}{}^n g_{np} \cM^{\um \un} \pa_\un \xi_{2]}{}^p ~, 
\nn\w2
\lambda_{12}{}^{ab} &= 2 \cM^{\um \un} \pa_\um \xi_{[1}{}^n \pa_\un \xi_{2]}{}^p \,e_n{}^{[a} e_p{}^{b]}~, 
\nn\w2
\lambda_{12}{}_j{}^i &= -16 g_{mn} \cV^\um{}_{jk} \cV^{\un ik} \partial_\um \xi_{[1}{}^m \partial_\un \xi_{2]}{}^n\ ,
\nn\w2
\Xi_{m\,\balpha}^{12} &= -\xi_1^n \xi_2^p\,H_{mnp\,\balpha}\ ,
\nn\w2
\Xi^{12}_{m\,\uk} &= \xi_1^n\xi_2^{p} H_{mnp\,\uk}(B') -
	4i \,\veps_{mnab}\,\partial_\uk \xi_{[1}^p \,\xi_{2]}^n  \,\omega_p{}^{ab}\ .
\end{align} 
These can be deduced by working out the commutator algebra on $e_m{}^a, \cV_\um{}^a$ and $A_m{}^\um$. As for the composite $\Omega$-transformations, they can be computed from the closure of external diffeomorphisms on $B_{mn\,\balpha}$ and they will involve the 4-form field strengths. We shall skip the derivation of these field strengths and the resulting composite $\Omega$ parameters, as they are not needed here.

\subsection{Generalized vielbein postulates}

Now let us address the generalized vielbein postulates. Two of them,
\eqref{vp1} and \eqref{vp2}, involve the external vielbein. Neither of these
turn out to have natural superspace analogues.
A straightforward superspace generalization of the first equation \eqref{vp1}
by extending the coordinate index to a supercoordinate index, $m \rightarrow M$,
runs into the problem that its ${}_{m n}{}^a$ component differs from the
${}_{m n}{}^a$ component of the original bosonic equation \eqref{vp1} due to the
presence of a term $\Gamma_{mn}{}^{\rho} E_{\rho} {}^a$ (and its complex conjugate),
where $\rho$ ($\dot\rho$) is the 16-component index of the chiral (antichiral)
Grassmann coordinate $\theta^{\rho}$ ($\theta_{\dot\rho}$).
This is problematic because $E_{\rho}{}^a$ has no geometric meaning at the component level.\footnotemark
\footnotetext{Component fields and forms are derived from superfields and superforms
by projecting $\theta=0$ and $\rd\theta=0$. Geometrically, this is the
pullback of the inclusion map that embeds spacetime into superspace.
For the vector vielbein one-form $E^a$, the only component that survives
this projection is $e_m{}^a = E_m{}^a\vert_{\theta=0}$. The components
$E_\rho{}^a\vert_{\theta=0}$ and $E^{\dot \rho}{}^a\vert_{\theta=0}$ turn out
to be pure gauge degrees of freedom. While they can be set to zero as a Wess-Zumino
type gauge fixing condition, this is not necessary.}

One cannot circumvent this issue by setting $\Gamma_{mn}{}^{\rho}$ to zero by hand,
as this violates general supercovariance.
Conventional superspace avoids this because the affine connection is actually unnecessary for
describing supergravity; it appears in no supersymmetry transformation, nor is it
included in the gravitino kinetic term.
Instead, one uses the spin connection, which can be fixed to its usual expression
by requiring the torsion tensor $T^a = \cD e^a$ to vanish.
This condition in turn has a natural lift to superspace.
Thus, we shall abandon \eq{vp1} in superspace and instead define the vector torsion tensor
\be
T^a  :=  D E^a + E^b \wedge \Omega_b{}^a = \cD E^a = \frac{1}{2} E^B E^C T_{C B}{}^a\ ,
\ee
where $\Omega_b{}^a$ is the Lorentz-valued superconnection.
(A similar torsion tensor can be defined in terms of the gravitino one-form $E^{\alpha i}$,
but we will postpone its discussion to the next section.)
The physics originally encoded in the vanishing torsion condition will now be encoded
in constraints placed upon $T_{CB}{}^a$. We will discuss these in due course.
The point is that one avoids ever introducing an affine connection $\Gamma_{MN}{}^P$
in superspace and so there is no analogue to \eqref{vp1}.

Similar statements pertain to \eqref{vp2}, although here the situation is somewhat
different. This equation can be interpreted as a definition of a field $\pi_{\um n}{}^a$,
\begin{align}
\partial_\um e_n{}^a 
    -\frac{1}{3}\,\Gamma_{\underline{km}}{}^\uk\,e_n{}^a 
    - \omega_\um{}^{ab} e_{nb}
    =: - \pi_{\um n}{}^a ~.
\end{align}
The constraint amounts to requiring $\pi_{\um}{}^{a}{}_b := \pi_{\um n}{}^a e_b{}^{n}$
to be symmetric in $ab$, which allows one to determine the internal spin connection.
Equivalently, a choice of internal spin connection permits one to set the antisymmetric
part of $\pi_\um{}^a{}_b$ to zero.
However, this has no natural superspace lift. One would need to introduce 
$\pi_{\um N}{}^a = \pi_\um{}^a{}_B E_N{}^B$, but this involves also $\pi_{\um}{}^{a}{}_{\beta j}$ 
and $\pi_{\um}{}^a{}^{\dbeta j}$. A choice of internal spin connection leaves these
unaffected. Moreover, they cannot be set to zero without constraining the internal
derivative of $E_N{}^a$ itself. Therefore, we must dispense with \eqref{vp2} as well.

At first glance, this is problematic because it forces us to drop the
internal spin connection as there is no longer any ability to define it. But as mentioned above, the role of the internal spin connection will turn out to be
played by the constrained two-form.
The remaining two vielbein postulates \eqref{vp3} and \eqref{vp4} involve only the
56-bein and these pose no obstacles to a superspace interpretation. The external
derivative of the 56-bein \eqref{vp3} we will lift to superspace simply by
replacing $m$ with $M$. 

However, for the purely internal GVP \eqref{vp4}, we find
that it is useful to choose a more general form. It was already observed in
\cite{Godazgar:2013dma, Godazgar:2014nqa} that the internal GVP derived from
the ${\rm SU}(8)$ reformulation of $D=11$ supergravity does not take the restricted
form \eqref{vp4}, but rather includes so-called non-metricity. The most general
form of the internal GVP is
\begin{align}
\nabla_\un \cV_\um{}^{i j}
	&:= \pa_\un \cV_\um{}^{i j} 
	+ \cQ_\un{}_k{}^{[i} \cV_\um{}^{j] k}
	- \Gamma_{\un \um}{}^\up\, \cV_\up{}^{i j}
	= \cP_\un{}^{i j k l} \cV_{\um}{}_{k l} \eol
	&= \cD_\un \cV_\um{}^{i j}
	- \Gamma_{\un \um}{}^\up\, \cV_\up{}^{i j}~.
\end{align}
Here $\cQ_\um{}_i{}^j$ is the internal $\SU(8)$ connection and the non-metricity
$\cP_\um{}^{ijkl}$ is a pseudo-real expression in the $\rep{70}$ of $\SU(8)$. 
There is significant ambiguity in this expression because both the $\E7$ connection
$\Gamma$ and the set $\{\cQ, \cP\}$ are describing the same $56 \times 133$ degrees of freedom
(up to the section condition) encoded in $\pa_\un \cV_\um{}^{i j}$.
This can be clarified as follows. Using \eq{e7c},
the internal GVP can be rewritten as
\begin{align}\label{eq:IntGVPSU8}
\pa_\un \cV_\um{}^{i j} 
	+ (\cQ_\un{}_k{}^{[i} + \Gamma_\un{}^\balpha \cQ_\balpha{}_k{}^{[i}) \cV_\um{}^{j] k}
	= \Big(\cP_\un{}^{i j k l} + \Gamma_\un{}^\balpha \cP_\balpha{}^{ijkl}\Big) \cV_{\um}{}_{k l} ~,
\end{align}
where $\cQ_\balpha{}_i{}^j$ and $\cP_\balpha{}^{ijkl}$ correspond to the ``flattened''
components of the $\E7$ generator, living in the $\rep{63}$ and $\rep{70}$ of $\SU(8)$,
defined by
\begin{align}\label{eq:DefQPalpha}
(t_\balpha)_\um{}^\un \cV_\un{}^{i j}
	&= -Q_\balpha{}_k{}^{[i} \cV_\um{}^{j] k}
	+ P_\balpha{}^{i j k l} \cV_\um{}_{kl}~.
\end{align}
Evidently one can solve \eqref{eq:IntGVPSU8} either for $\cQ_\um{}_i{}^j$ and $\cP_\um{}^{ijkl}$ in terms of $\Gamma_{\um \un}{}^\up$, or $\Gamma_{\um\un}{}^\up$ in terms of $\cQ_\um{}_i{}^j$ and $\cP_\um{}^{ijkl}$. In the latter case, one finds that
\begin{align} \label{eq:GammaDef}
\Gamma_{\un \um}{}^\up
	= -i \cD_\un \cV_\um{}^{i j} \cV^{\up}{}_{i j} + i \cP_\un{}^{ijkl} \cV_\um{}_{kl} \cV^\up{}_{ij}
	+ \CC
\end{align}
There are a myriad of ways to reduce the ambiguity. One particular way is to set
$\cQ$ and $\cP$ to zero, eliminating it entirely. This is the
Weitzenb\"ock connection, and we denote it by $\mathring{}$ over the various symbols:
\begin{equation}
\text{Weitzenb\"ock connection:} \quad \large\mathring \cQ = \large\mathring \cP = 0~, \quad
\mathring \Gamma_{\um \un}{}^\up = -i \,\pa_\um \cV_\un{}^{i j} \cV_{ij}{}^\up + \CC
\end{equation}
The conventional choice in ExFT is different \cite{Godazgar:2014nqa}. It involves taking $\cP = 0$, and
then eliminating as much (but not all) of the ambiguity by imposing constraints 
on $\Gamma$, through the vanishing torsion condition \eq{t12}. 
Let's review some group theory briefly to remember how the various parts of
$\Gamma$ are usually constrained in ExFT. Decomposing under $\E7$ and then
$\SU(8)$ gives
\begin{align*}
\Gamma: \qquad \rep{56} \times \rep{133} &= \rep{56} + \rep{912} + \rep{6480}~, \\
	\rep{56} &\rightarrow \rep{28} + \brep{28}\ ,\\
	\rep{912} & \rightarrow \rep{36} + \rep{420} + \CC~, \\
	\rep{6480} &\rightarrow \rep{28} + \rep{420} 
		+ \rep{1280} + \rep{1512} + \CC~,
\end{align*}
$\cQ$ and $\cP$ decompose under $\SU(8)$ as
\begin{align*}
\cQ : \qquad (\rep{28} + \brep{28}) \times \rep{63} &= 
	\rep{28} + \rep{36} + \rep{420} + \rep{1280} + \CC\ , \\
\cP : \qquad (\rep{28} + \brep{28}) \times \rep{70} &=
	\rep{28} + \rep{420} + \rep{1512} + \CC
\end{align*}
In conventional ExFT, after setting $\cP$ to zero, $\Gamma$ and $\cQ$ have residual ambiguity:
the representations appearing in $\cQ$ are counted twice.
Killing the $\rep{912}$ part of $\Gamma$ (corresponding to torsion) removes $\rep{36}+\rep{420}$
and their conjugates. Constraining the $\rep{56}$ part of $\Gamma$ to be related to $e^{-1} \pa_\um e$ removes further ambiguity. The remaining ambiguity is the $\rep{1280}$ and its conjugate that appear in both $\cQ$ and the $\rep{6480}$ of $\Gamma$. There is no $\E7$ covariant way to eliminate this piece. However, as shown in \cite{Godazgar:2014nqa}, this undetermined piece always drops out of the SUSY transformations.

There is already a reason to reconsider this approach when generalizing to superspace.
The determinant $e = \det{e_{m}{}^a}$
is, like the inverse vielbein, an unnatural object to encounter in superspace as it violates
general supercovariance, so the constraint imposed on the $\rep{56}$ part of $\Gamma$ is
difficult to lift to superspace. The superdeterminant $E = {\rm sdet\,}{E_M{}^A}$ is more
natural, but does not reduce naturally to $e$ when returning to components. This suggests
that one should leave the $\rep{56}$ part unfixed and hope for it to drop out
of the SUSY transformations as well.
Actually, as we will demonstrate, there is no need to fix \emph{any} of the ambiguity in $\Gamma$, $\cQ$, and $\cP$.
We will allow both the non-metricity $\cP$ and the torsion tensor $\cT_{\um \un}{}^\up$, defined in \eqref{t2},
to be nonzero. This requires that the SUSY transformations be modified to include contributions
of these tensors, but in the result, all of the undetermined pieces drop out,
not just the one in the $\rep{1280}$.

Because the torsion tensor lies in the $\rep{912}$, we can employ the same representation
theory as for the embedding tensor \cite{deWit:2007kvg}.
Defining the tangent space components
$\cT_{\uc \ua}{}^\ub = \cV_\uc{}^\um \cV_\ua{}^\un \cT_{\um \un}{}^\up \cV_\up{}^\ub$
as for the embedding tensor,
\begin{equation}
\cT_{ij\,\ua}{}^\ub = 
\begin{pmatrix}
- \tfrac{2}{3} \delta_{[k}{}^{[p}\, \cT^{q]}{}_{l]ij} & 
\tfrac{1}{24} \veps_{klpq rstu} \, \cT^{rstu}{}_{ij} \cr \noalign{\vskip5mm}
\cT^{klpq}{}_{ij} &
\tfrac{2}{3} \delta_{[p}{}^{[k} \,\cT^{l]}{}_{q]ij}
\end{pmatrix}
\end{equation} 
one finds
\begin{align}
\cT^{klmn}{}_{ij} &= 
	-\tfrac{4}{3} \delta_{[i}{}^{[k} \,\cT_{j]}{}^{lmn]}\,, \eol
\cT_i{}^{jkl} &= 2 \times \Big(-\tfrac34 \,A_{2i}{}^{jkl} - \tfrac32 A_1{}^{j[k}\,\delta^{l]}{}_i\Big)~,
\end{align}
where $A_1{}^{i j} = A_1{}^{(ij)}$ is in the $\rep{36}$ and
$A_2{}_i{}^{jkl} = A_2{}_i{}^{[jkl]}$ (and traceless) is in the $\rep{420}$.
Here we have inserted an additional factor of 2 in the last relation to match the
historical conventions for the so-called $T$-tensor.\footnote{In other words, the tangent
space components of $\cT$ differ from the $T$-tensor by a factor of 2.}

Later on, it will be useful to extract the undetermined pieces in the internal connections
to ensure that they
cancel. We will do this by converting to the Weitzenb\"ock connection, and writing
expressions in terms of $\mathring \Gamma$, isolating the undetermined pieces into
$\cQ$ and $\cP$. In terms of the Weitzenb\"ock connection, one can show that
\begin{subequations}\label{E:Weitzenbock}
\begin{align}
A_1{}^{i j} &= \mathring A_1{}^{ij} - i \cV^{\um}{}^{k (i} \cQ_\um{}_k{}^{j)}~, \\
A_2{}_i{}^{jkl} &= \mathring A_2{}_i{}^{jkl} 
	+ \Big[
	4i \,\cP_{\um}{}^{j k l p} \cV^{\um}{}_{i p}
	+ 3i \,\cQ_{\um}{}_i{}^j \cV^{\um k l}
	\Big]_{\rep{420}}
	\eol
	&= \mathring A_2{}_i{}^{jkl} 
	+ 4 i \,\cP_{\um}{}^{j k l p} \cV^{\um}{}_{i p}
	- 2 i \,\delta_i{}^{[j} \cP_{\um}{}^{k l] p q} \cV^{\um}{}_{p q}
	\eol & \quad
	+ 3i \cQ_{\um}{}_i{}^{[j} \cV^{\um k l]}
	- i \,\delta_i{}^{[j} \cQ_{\um}{}_p{}^{k} \cV^{\um l] p}~, \\
\Gamma_{\un \um}{}^\up
	&= \mathring \Gamma_{\un \um}{}^\up
	+ i \cQ_\un{}_{k l}{}^{i j} \cV_\um{}^{kl} \cV^{\up}{}_{i j} 
	+ (i \cP_\un{}^{ijkl} \cV_\um{}_{kl} \cV^\up{}_{ij}
	+ \CC)
	\label{e7cc}
\end{align}
\end{subequations}
We have included above the corresponding formula for the $\E7$ connection.

\vspace{0.5cm}
\emph{From now on, unless we comment otherwise, $\nabla_\um$ will correspond to an internal covariant derivative carrying an $E_{7(7)}$
connection and SU(8) connection with arbitrary non-vanishing torsion and non-metricity.
It will not carry any internal spin connection.}

\section{$\E7$ exceptional field theory in $(4+56|32)$ superspace}

Now we turn to the construction of $\E7$ exceptional field theory in
superspace. In addition to the four external coordinates $x^m$ describing spacetime
and the $56$ internal coordinates $y^\um$ describing the exceptional structure,
there will be 32 anticommuting (Grassmann) coordinates, which we split
into chiral and antichiral coordinates $\theta^\mu$ and $\theta_{\dot\mu}$.\footnote{Here we
use the chiral (antichiral) Grassmann coordinate
$\theta^\mu$ ($\theta_{\dot\mu}$) with $\mu=1,\cdots,16$. Typically in 4D $N=8$ superspace,
one writes $\theta^{\mu I}$ with $\mu=1,2$ and $I=1,\cdots,8$ as curved analogues
of the two-component spinor and SU(8) indices. For compactness, we use $\mu$
collectively for $\mu I$.} Supersymmetry will be associated with diffeomorphisms
in the fermionic direction, in a manner to be described in due course.
The full set of coordinates are collectively denoted $Z^{\ul{M}}$,
\begin{align}
 Z^{\ul M} = \{ Z^M, y^\um\} = \{ x^m, \theta^\mu, \theta_{\dmu}, y^\um\}~.
\end{align}
We reserve $Z^M$ to denote the $(4|32)$ coordinates $(x^m, \theta^\mu, \theta_\dmu)$
parametrizing an ``external'' 4D $N=8$ superspace, with the additional 56
``internal'' coordinates $y^\um$ describing the exceptional structure.

The supervielbein on the $(4|32)$ superspace is denoted $E_M{}^A$ whose tangent
space index $A$ decomposes as $(a, \alpha \,i, \dalpha \,i)$ so that
\begin{align}
 E_M{}^A = (E_M{}^a, E_M{}^{\alpha i}, E_M{}_{\dalpha i})~.
\end{align}
We will refer to $E_M{}^a$ and $E_M{}^{\alpha i}$ as the vielbein and gravitino
super one-forms, respectively, as they are the superfield analogues of $e_m{}^a$
and $\psi_m{}^{\alpha}{}^i$. The internal exceptional space is equipped with a 56-bein,
\begin{align}
 \cV_\um{}^\ua = (\cV_\um{}^{ij}, \cV_\um{}_{ij})~.
\end{align}
While it would be natural to encode $E_M{}^A$ and $\cV_\um{}^\ua$ as components of
an even larger sehrvielbein $\cE_{\ul M}{}^{\ul A}$ (as in central charge superspace
\cite{Howe:1980th}), we will leave discussion of this
to future work. From the point of view of the external 4D $N=8$ superspace, the 56-bein
is a scalar superfield while $E_M{}^A$ are components of a super one-form
$E^A = \rd Z^M E_M{}^A$.

In addition to the external and internal vielbeins, there is a tensor hierarchy of $p$-forms.
These include the one-form vector field
\begin{align}
 A^\um = \rd Z^M\, A_M{}^\um\ ,
\end{align}
and two super two-forms
\begin{align}
B_\balpha = \frac{1}{2} \rd Z^M \rd Z^N B_{NM\, \balpha}~, \qquad
B'_\um = \frac{1}{2} \rd Z^M \rd Z^N B'_{NM\, \um}~.
\end{align}
As in components, the second two-form is a constrained tensor on its fundamental
$\E7$ index.
Just as in ExFT, there are additional 3-forms and 4-forms making up
the tensor hierarchy, but we will stop our analysis at the two-forms.
{\it Because the unprimed $B$ does not naturally occur in superspace,
henceforth, we will drop the prime.}

Finally, the superspace is also equipped with a pair of one-forms that
gauge the local tangent group ${\rm SO}(1,3) \times {\rm SU}(8)$. These are the
Lorentz connection $\Omega_a{}^b$ and the ${\rm SU}(8)$ connection $\cQ_i{}^j$,
\begin{align}
 \Omega_a{}^b = \rd Z^M \Omega_M{}_a{}^b~, \qquad
 \cQ_i{}^j = \rd Z^M \cQ_M{}_i{}^j~.
\end{align}
Constraints in superspace will be chosen so that these connections become composite,
describing no independent degrees of freedom of their own.

These superfields each have natural analogues in the component theory. The only
component field we have not mentioned in superspace yet is the spin-1/2 fermion.
And indeed, there is also a fermionic superfield, which we denote $\chi_\alpha{}^{ijk}$,
whose lowest component is the field of the same name. In conventional
4D $N=8$ superspace \cite{Brink:1979nt, Howe:1981gz}, this superfield actually appears in the curvature super-forms,
and so can be treated as a derived quantity. In exceptional superspace, it plays a
somewhat more fundamental role, as the gravitino one-form $E_M{}^{\alpha i}$ turns
out to directly transform into it under external superdiffeomorphisms.

\subsection{Symmetry transformations}

Under internal diffeomorphisms ($\Lambda^\um$) and the tensor hierarchy transformations
($\Xi_{M\, \balpha}$, $\Xi_{M\, \um}$, $\Omega_{M N}{}^\um{}_\balpha$,$\Omega_{M N}{}_\um{}^\un$)
the various superfields transform exactly as their component analogues, with
\begin{align}
\delta E_M{}^A &= \mathbb{L}_{\Lambda} E_M{}^A~, \eol
\delta \cV_\um{}^\ua &= \mathbb{L}_{\Lambda} {\cal V}_\um{}^\ua~, \eol
\delta \chi_\alpha{}^{ijk} &= \mathbb{L}_{\Lambda} \chi_\alpha{}^{ijk}~, \eol
\delta A_M{}^\um &=
    D_M \Lambda^\um  
   +  12\,(t^\balpha)^{\um\un} \partial_\un \Xi_{M\,\balpha}
   +\frac{1}{2}\,\Omega^{\um\un}\,\Xi_{M\,\un} \ ,\eol
\Delta  B_{NM\,\balpha} &=
    (t_\balpha)_{\um\un}\, \Lambda^\um{\cal F}_{NM}{}^\un 
    + 2\,{ D}_{[N}\Xi_{M]\balpha}
    + \partial_\um \Omega_{NM}{}^\um{}_\balpha + (t_\balpha)_\um{}^\un \Omega_{NM}{}_\un{}^\um\ ,\eol
\Delta B_{NM\,\um} &=
    \cF_{NM}{}^\un \partial_\um \Lambda_\un 
    - \partial_\um \cF_{NM}{}^\un \Lambda_\un
    + 2 D_{[N}\Xi_{M]\um} 
    + 48\,(t^\balpha)_\un{}^\ur  \left(\partial_\ur\partial_\um A_{[N}{}^\un\right) \Xi_{M]\balpha}
    \eol & \quad
    -  \partial_\um \Omega_{NM}{}_\un{}^\un - 2 \,\partial_\un \Omega_{NM}{}_\um{}^\un\ ,
\end{align}
where
\begin{align}
\Delta  B_{NM\,\balpha} &:=
    \delta  B_{NM\,\balpha} + (t_\balpha)_{\um\un}\,A_{[N}{}^\um\, \delta A_{M]}{}^\un \ , \eol
\Delta B_{NM\,\um} &:=
    \delta B_{NM\,\um} 
    + A_{[N}{}^\un \partial_\um \delta A_{M]}{}_\un - \partial_\um A_{[N}{}^\un \delta A_{M]}{}_\un~.
\end{align}
Here the $\lambda$ weights of the various superfields match their component cousins.
The curvature two-form $\cF_{NM}{}^\um$ (and later, the three-forms $H_{PNM\, \balpha}$
and $H_{PNM\, \um}$) are defined exactly as in components, replacing $m \rightarrow M$. 
We collect their superspace definitions in the next subsection. 

There is a minor technical subtlety that the superindices $M$ and $A$ come equipped
with a $\mathbb Z_2$ grading, which causes certain signs to appear when their 
relative ordering changes. For example $[NM]$ above should be understood as a graded
commutator. This is common in superspace and we briefly review it in
Appendix \ref{app:conv}. To keep formulae as legible as possible, we suppress such
grading factors.

The transformations under external diffeomorphisms are somewhat more involved.
We list first the ones whose transformations can be directly compared to \eqref{bst}:
\begin{align}
\delta E_M{}^a &=
	D_M \xi^N\, E_N{}^a + \xi^N D_N E_M{}^a\ ,
\nn\w2
\delta \cV_\um{}^\ua &=  \xi^M D_M \cV_\um{}^\ua \ ,
\nn\w2
\delta A_M{}^{\um} &= \xi^N \cF_{NM}{}^{\um} + \cM^{\underline{mn}} \pa_\un \xi^N\, G_{NM}\ ,
\qquad\qquad G_{NM} := E_N{}^a E_{M a}
\nn\w2
\Delta B_{NM \,\balpha} &= \xi^P H_{PNM \,\balpha}\ .
\end{align}
A relevant feature is, as in components, the appearance of an anomalous term
in $\delta A_M{}^\um$ involving $\pa_\um \xi^N$.
The constrained two-form $B_{NM \,\um}$ transforms not only with an additional explicit spin
connection term (matching the redefined component two-form \eqref{eq:DeltaB'}), but with a few
additional terms involving the gravitino one-forms,
\begin{align}
\Delta B_{NM\, \um} &= 
        \xi^P H_{PNM \um}
        - 2 i \, E_{N}{}^a E_{M}{}^b \veps_{abcd}\, \pa_\um \xi^P \Omega_P{}^{cd}
        \eol & \quad
	- 16 i \,\pa_\um \xi^P E_{P \dalpha i} \,E_{[N}{}^{\alpha i} E_{M]}{}^a (\gamma_a)_\alpha{}^\dalpha
	+ 16 i \,\pa_\um \xi^P E_P{}^{\alpha i} \,E_{[N \dalpha i} E_{M]}{}^a (\gamma_a)_\alpha{}^\dalpha~.
\end{align}
The gravitino one-form $E_M{}^{\alpha i}$ and the spin-1/2 fermion $\chi^{\alpha\, ijk}$
have even more involved anomalous terms,
\begin{align}
\delta E_M{}^{\alpha i}
	&= D_M \xi^N E_N{}^{\alpha i} + \xi^N D_{N} E_{M}{}^{\alpha i} 
	+ 2 i \,\cV^{\um}{}^{ij} \pa_{\um} \xi^N 
	\Big( E_N{}_{\dbeta j} E_M{}^c - E_N{}^c E_M{}_{\dbeta j}
	\Big) (\gamma_c)^{\dbeta \alpha}
	\eol & \quad
	- \frac{i}{2\sqrt 2} \cV^{\um}{}_{jk} \pa_{\um} \xi^N E_N{}^d E_M{}^c
		\chi^{\beta j k i} (\gamma_c \gamma_d)_\beta{}^{\alpha}
	~, \eol
\delta \chi^{\alpha ijk}
	&=  \xi^N D_N \chi^{\alpha ijk}
 	- 12 i \sqrt 2\, \cV^{\un\, [ij} \pa_\un \xi^N E_N{}^{\alpha k]}
        + \frac{i}{6} \veps^{ijk pq rst } \cV^\un{}_{pq} \pa_{\un} \xi^N E_N{}^b
		\bar \chi_{\dbeta rst} (\gamma_b)^{\dbeta \alpha}\ .
\end{align}
While the exact relation between  the supervielbein and the component fields has not been specified yet, a natural definition, which will be provided in Section 5, motivates the above form of the external superdiffeomorphisms in view of the supersymmetry transformation rules \eq{E:ExtDiffFerms}. Note also that, as already mentioned above, because the gravitino transforms directly into $\chi$, we are led to treat $\chi$ on the same level as the gravitino and the other fundamental superfields rather than as a derived curvature superfield.

Finally, we should mention that just as in components, the constrained two-form
$B_{NM \um}$ possesses an anomalous Lorentz transformation,
\begin{align}\label{eq:deltaB.Lor}
\delta B_{NM\,\um} = -2i\, E_N{}^a E_M{}^b \veps_{abcd}\, \pa_\um \lambda^{cd}~.
\end{align}

\subsection{Covariant external superdiffeomorphisms and modified curvature tensors}

The curvature super-forms of the tensor hierarchy have already appeared above in
the symmetry transformations of the tensor hierarchy fields. They are defined
as at the component level\footnote{The reader is cautioned that we employ
superspace conventions for differential forms, see e.g. \cite{Wess:1992cp} and Appendix \ref{app:conv}.}
\begin{align}
\cF^\um &=
    \rd A^\um
    + \frac{1}{2} [A, A]^\um_{\rm E}
    - 12 \,  (t^\balpha)^{\um\un} \,\partial_\un B_{\balpha}
    - \frac12\,\Omega^{\um\un}\,B_{\un}~, \\
H_\balpha &= D B_\balpha 
        - \frac{1}{2} (t_{\balpha})_{\um\un} A^\um \rd A^\un
	+ \frac{1}{12} (t_{\balpha})_{\um\un} A^\um [A,A]_E^\un
	- \pa_\um C^\um{}_\balpha - (t_\balpha)_\um{}^\un C_\un{}^\um
	\label{h133a} ~, \\
H_\um &=
	D B_\um + 24\, (t_\balpha)_\ur{}^\us\, \pa_\um \pa_\us A^\ur \, B_\balpha
	- \frac{1}{2} \Omega_{\ur\us}\, \pa_\um A^\ur \, \rd A^\us
	+ \frac{1}{2} \Omega_{\ur\us}\, A^\ur \, \pa_\um \rd A^\us
    \eol & \quad
    - \frac{1}{3} \Omega_{\ur\us}\, \Big( A^\ur\, \pa_\um [A,A]_E^\us
	+ [A,A]_E^\us \,\pa_\um A^\ur \Big)
	+ \pa_\um C_\un{}^\un + 2\, \pa_\un C_\um{}^\un\ . 
\end{align}
We have included the super 3-forms $C^\um{}_\balpha$ and $C_\um{}^\un$ 
in their definitions for completeness, but they will not play a major
role in the subsequent discussion. We emphasize that because $B_\um$
transforms anomalously under Lorentz transformations, the same is true
of its curvature $H_\um$. In fact, the curvature $H_\um$ is not even
a tensor under internal diffeomorphisms, a fact that we will return to soon.

In addition to the tensor hierarchy curvatures, there are curvature super-forms
associated with the supervielbein $E_M{}^A$ and the Lorentz and SU(8) connections,
$\Omega$ and $\cQ$. The former define the super torsion tensor,
\begin{align}\label{eq:DefT}
T^A := \cD E^A = D E^A + E^B \Omega_B{}^A + E^B \cQ_B{}^A
\end{align}
where
\begin{align}
\Omega_B{}^A =
\begin{pmatrix}
 \Omega_b{}^a & 0 & 0 \\
 0 & \tfrac{1}{4} \Omega^{cd} (\gamma_{cd})_\beta{}^\alpha & 0 \\
 0 & 0 & \tfrac{1}{4} \Omega^{cd} (\gamma_{cd})^\dbeta{}_\dalpha
\end{pmatrix}~, \qquad
\cQ_B{}^A = 
\begin{pmatrix}
 0 & 0 & 0 \\
 0 & -\tfrac{1}{2} \cQ_j{}^i \delta_\beta{}^\alpha & 0 \\
 0 & 0 & \tfrac{1}{2} \cQ_i{}^j \delta^\dbeta{}_\dalpha
\end{pmatrix}~,
\end{align}
as well as the Lorentz and SU(8) curvatures,
\begin{align}
R_a{}^b := D \Omega_a{}^b + \Omega_a{}^c \Omega_c{}^b~, \qquad
R_i{}^j := D \cQ_i{}^j - \frac{1}{2} \cQ_i{}^k \cQ_k{}^j~.
\end{align}
Typically, the superspace Bianchi identities determine the latter curvatures in
terms of the torsion tensor. One then finds that imposing suitable constraints on
the tangent space components $T_{CB}{}^A$ of the torsion tensor prove to define
the supergeometry. However, the situation is more subtle in exceptional superspace.
The main reason is that the curvature tensors we have introduced
above are not actually the natural curvature tensors from the point of view of
superspace. By this, we mean that some of them do not possess natural expansions
in terms of the superspace frame $E^A$.

It turns out to be more illuminating to first consider the curvature two-form $\cF^\um$.
Recall that under external diffeomorphisms
\begin{align*}
 \delta A_M{}^\um 
    = \xi^N \cF_{N M}{}^\um + \cM^{\um \un} \nabla_\un \xi^N E_N{}^a E_{M a}~,
\end{align*}
where we have used $\pa_\un \xi^N = \nabla_\un \xi^N$.
We would like to rewrite this expression as a covariant external diffeomorphism.
A covariant external diffeomorphism is defined in terms of the tangent space parameter
$\xi^A = \xi^M E_M{}^A$ as
\begin{align}\label{eq:covxi}
 \delta_{\rm cov}(\xi^A) = \delta(\xi^M) 
    - \delta_{\rm Lorentz}(\xi^N \Omega_N{}^{ab})
    - \delta_{\rm SU(8)}(\xi^N \cQ_N{}_i{}^j)~.
\end{align}
For the vector fields, the additional transformations do not contribute, but nevertheless
the transformation rule takes a different form when rewritten in terms of
$\xi^A$:
\begin{align}\label{eq:dcovA}
\delta_{\rm cov}(\xi) A_M{}^\um 
	&= \xi^N \widehat\cF_{NM}{}^{\um}
	+ \cM^{\um \un} \nabla_{\un} \xi^a  E_{M a}
	- \cM^{\um \un} \xi^a\, \nabla_{\un} E_{M a}\ , 
\end{align}
where we have defined
\begin{align} \label{eq:DefHatF}
\widehat{\cF}_{NM}{}^\um &:= \cF_{NM}{}^\um + 2 \cM^{\um \un} \,E_{[N}{}^a\,\nabla_\un E_{M]a}~.
\end{align}
This expression for $\widehat \cF^\um$ proves to be the superspace analogue of the
component modified field strength introduced in \eqref{eq:DefHatF-comp}.\footnote{One
can trade $\pa_\un$ for $\nabla_\un$ in \eqref{eq:DefHatF} as the connection terms drop out.}
Recall that the motivation for introducing this modified field strength in components was that, upon
redefining the constrained two-form, it was this modified field strength that possessed
the twisted self-duality relation. The analogous statement in superspace is that
$\widehat\cF^\um$ will be the tensor that is constrained in order to define
the supergeometry. That is, it will be chosen to possess a sensible expansion\footnotemark
\begin{align}\label{eq:FTanDecomp}
 \widehat \cF_{NM}{}^\um = E_N{}^B E_M{}^A \widehat\cF_{B A}{}^\um\ ,
\end{align}
where (as we will discuss in the next section)
the tangent space components $\widehat\cF_{B A}{}^\um$ 
are set equal to other covariant superfields (such as $\chi^{\alpha ijk}$) or are constrained in
some other way, e.g. twisted self-duality (up to fermions) in the case of $\widehat \cF_{ba}{}^\um$.
\footnotetext{Of course, because the supervielbein is assumed to be invertible, one can always
define $\cF_{BA}{}^\um := E_B{}^N E_A{}^M \cF_{N M}{}^\um$. The problem is that one finds
a contribution to $\cF_{Ba}{}^\um$ of the form
$\cM^{\um \un} E_B{}^N \nabla_{\un} E_{N a}$ that is difficult to make sense of upon reducing
to components. No such contribution to $\widehat \cF_{BA}{}^\um$ occurs.}
As at the component level, the term we have added in \eqref{eq:DefHatF} is not
Lorentz invariant because the internal derivative carries no internal spin connection.
But $\cF^\um$ is itself not Lorentz invariant due to the anomalous Lorentz transformation
\eqref{eq:deltaB.Lor} of the constrained two-form. This leads (only)
the top component $\widehat \cF_{ab}{}^\um$ in \eqref{eq:FTanDecomp} to transform.
As in components, we find, for the inhomogeneous part of the Lorentz transformation,
\begin{align}\label{eq:deltaF.Lor}
\delta_{\rm anom} \widehat \cF_{ab}{}^{i j}
	= 2 i \,\cV^{\um i j} \Big(\delta_{a}{}^c \delta_b{}^{d} - \frac{1}{2} \veps_{ab}{}^{cd} \Big)\pa_{\um} \lambda_{c d}
	= 4 i \,\cV^{\um i j} \pa_{\um} \lambda^-_{ab}~.
\end{align}

What about the supervielbein?
For the vierbein one-form $E_M{}^a$, it turns out that a covariant external diffeomorphism
leads to the usual expression
\begin{align}\label{eq:dcovEa}
 \delta_{\rm cov}(\xi) E_M{}^a = \cD_M \xi^a + \xi^N T_{N M}{}^a\ ,
\end{align}
which suggests that the vector torsion tensor possesses a sensible tangent space expansion
\begin{align}
T_{NM}{}^a = E_N{}^C E_M{}^B T_{CB}{}^a \ ,
\end{align}
without modification.
For the gravitino $E_M{}^{\alpha i}$, the situation is more subtle. We find
\begin{align}\label{eq:dcovEalpha.0}
\delta_{\rm cov}(\xi) E_M{}^{\alpha i}
	&= \cD_M \xi^{\alpha i} + \xi^N T_{NM}{}^{\alpha i} 
	+ 2 i \,\cV^{\um}{}^{ij} \nabla_{\um} \xi^N 
	\Big( E_N{}_{\dbeta j} E_M{}^c - E_N{}^c E_M{}_{\dbeta j}
	\Big) (\gamma_c)^{\dbeta \alpha}
	\eol & \quad
	- \frac{i}{2\sqrt 2} \cV^{\um}{}_{jk} \nabla_{\um} \xi^N E_N{}^d E_M{}^c
		\chi^{\beta j k i} (\gamma_c \gamma_d)_\beta{}^{\alpha}~.
\end{align}
This can be rewritten as
\begin{align}\label{eq:dcovEalpha}
\delta_{\rm cov}(\xi) E_M{}^{\alpha i}
	&= \cD_M \xi^{\alpha i}
	+ \xi^N \widehat T_{NM}{}^{\alpha i}
	+ 2i \,\cV^{\um i j} \nabla_{\um} \Big(\bar \xi_{\dbeta j} E_M{}^c (\gamma_c)^{\dbeta \alpha}\Big)
	- 2i \,\cV^{\um i j} \nabla_\um \Big(\xi^c E_{M \dbeta j} (\gamma_c)^{\dbeta \alpha}\Big)
    \eol & \quad
    -\frac{i}{2\sqrt 2} \left( \nabla_\um \xi^d\,E_M{}^c -\xi^c \nabla_\um E_M{}^d \right) \cV^\um{}_{jk} \chi^{\beta jki} (\gamma_c\gamma_d)_\beta{}^\alpha~,
\end{align}
where we have exchanged $\xi^M$ in the additional terms for $\xi^A = \xi^M E_M{}^A$.
The modified gravitino torsion tensor in this expression is
\begin{align}\label{eq:DefGravT}
 \widehat T^{\alpha i} &= T^{\alpha i}
	+ 2i \,\cV^{\um ij} \nabla_{\um} (E_{\dbeta j} \wedge E^c) (\gamma_c)^{\dbeta \alpha}
	- \frac{i}{2\sqrt 2} \cV^\um{}_{jk}\, \chi^{\beta jki} (\gamma_c \gamma_d)_\beta{}^\alpha
	\nabla_{\um} E^d \wedge E^c\ ~.
\end{align}
As we will see, this leads to a sensible tangent space expansion
$\widehat T_{NM}{}^{\alpha i} = E_N{}^C E_M{}^B \widehat T_{CB}{}^{\alpha i}$.

The internal counterparts to the supervielbein are the two superfields
$\cV_\um{}^{ij}$ and $\chi^{\alpha i j k}$. While these are not gauge superfields,
they also have curvatures naturally associated with them:
their covariant derivatives. Because the 56-bein transforms under external
covariant diffeomorphisms as
\begin{align}
\delta_{\rm cov}(\xi) \cV_\um{}^{ij} = \xi^M \cD_M \cV_\um{}^{ij}\ ,
\end{align}
its covariant derivative should possess a sensible tangent space expansion.
As in components we take
\begin{align}
\cD \cV_\um{}^{ij}= \cP^{ijkl} \cV_{\um}{}_{kl}~, \qquad \cP^{ijkl} = E^A \cP_A{}^{ijkl}\ ,
\label{eq:dV}
\end{align}
to both define the one-form $\cP^{ijkl}$ valued in the $\rep{70}$ of SU(8) and
the SU(8) connection $\cQ_i{}^j$.
However, the fermion superfield $\chi^{\alpha ijk}$ has additional terms in its
transformation rule, suggesting we define the one-form 
\begin{align}\label{eq:Deftau}
\widehat\tau_M{}^{\alpha i j k} &:=
	\cD_M \chi^{\alpha i j k}
	+ 12 i \sqrt{2} \,\cV^{\un [i j} \nabla_{\un} E_M{}^{\alpha k]}
	- \frac{i}{6} \veps^{i j k l p q r s} \,\cV^{\un}{}_{lp} \nabla_{\un} E_{M b}
		\chi_{\dbeta q r s} (\gamma_b)^{\dbeta \alpha}\ ,
\end{align}
so that
\begin{align}\label{eq:dcovchi}
\delta_{\rm cov}(\xi) \chi^{\alpha ijk} &=
    \xi^M \widehat\tau_M{}^{\alpha i j k}
    - 12 i \sqrt{2} \,\cV^{\un [i j} \nabla_{\un} \xi^{\alpha k]}
    + \frac{i}{6} \veps^{i j k l p q r s} \,\cV^{\un}{}_{lp} \nabla_{\un} \xi^{b}
	\chi_{\dbeta q r s} (\gamma_b)^{\dbeta \alpha}~.
\end{align}
This suggests that the one-form $\widehat\tau$ possesses a natural tangent
space expansion, $\widehat \tau_M{}^{\alpha i j k} = E_M{}^B \widehat\tau_B{}^{\alpha i j k}$.

For the two-forms, the situation is again somewhat subtle. 
The transformation for $B_{NM\,\balpha}$ is unchanged,
\begin{align} \label{eq:dcovBalpha}
\Delta_{\rm cov}(\xi) B_{NM\, \balpha} &= \xi^P H_{PNM \,\balpha}\ ,
\end{align}
suggesting that $H_{PNM\, \balpha}$ possesses a sensible tangent space expansion. The covariant transformation of $B_{NM\,\um}$ is more involved. Keeping in mind the anomalous Lorentz transformation \eqref{eq:deltaB.Lor} and the Lorentz connection contribution to covariant diffeomorphisms \eqref{eq:covxi}, one finds that
\begin{align}
\Delta_{\rm cov}(\xi) B_{NM\, \um} &= 
        \xi^P H_{PNM \um}
        + 2 i \, E_{N}{}^a E_{M}{}^b \veps_{abcd}\, \xi^P \nabla_\um \Omega_P{}^{cd}
        \eol & \quad
	- 16 i \,\nabla_\um \xi^P E_{P \dalpha i} \,E_{[N}{}^{\alpha i} E_{M]}{}^a (\gamma_a)_\alpha{}^\dalpha
	+ 16 i \,\nabla_\um \xi^P E_P{}^{\alpha i} \,E_{[N \dalpha i} E_{M]}{}^a (\gamma_a)_\alpha{}^\dalpha~.
\end{align}
In light of the comments in Section \ref{sec:newB}, a more natural form for the covariant variation of $B_{NM\, \um}$ is
\begin{align}\label{eq:dcovBm.v1}
\Delta_{\rm cov} B_{NM\,\um} + 2 \,\Gamma_\um{}^\balpha \Delta_{\rm cov} B_{NM\,\balpha}
	&= \xi^P \scH_{P N M \, \um}
        + 2 i \, E_{N}{}^a E_{M}{}^b \veps_{abcd}\, \xi^P \nabla_\um \Omega_P{}^{cd}
	\eol & \quad
	- 16 i \,\nabla_\um \xi^P E_{P \dalpha i} \,E_{[N}{}^{\alpha i} E_{M]}{}^a (\gamma_a)_\alpha{}^\dalpha
	\eol & \quad
	+ 16 i \,\nabla_\um \xi^P E_P{}^{\alpha i} \,E_{[N \dalpha i} E_{M]}{}^a (\gamma_a)_\alpha{}^\dalpha~,
\end{align}
where $\scH_\um$ is the modification of the field strength $H_\um$,
\begin{align} \label{hhd}
\scH_\um &:=  {H}_\um +2\,\Gamma_\um{}^\balpha\,H_\balpha~,
\end{align}
which transforms covariantly under internal diffeomorphisms, in contrast to
$H_\um$ itself. The form of \eqref{eq:dcovBm.v1} suggests the definition
\begin{align}\label{eq:defscH}
\widehat \scH_\um &:= \scH_\um
    - i E^a E^b \,\veps_{abcd} \nabla_\um  \Omega^{cd}
    + 8 i \,E_a \, \nabla_\um E^{\alpha i} E_{\dalpha i} (\gamma^{a})_\alpha{}^{\dalpha} 
    - 8 i \,E_a \, \nabla_\um E_{\dalpha i} E^{\alpha i} (\gamma^{a})_\alpha{}^{\dalpha}\ ,
\end{align}
so that
\begin{align}\label{eq:dcovBm}
\Delta_{\rm cov} B_{NM\,\um} + 2 \,\Gamma_\um{}^\balpha \Delta_{\rm cov} B_{NM\,\balpha}
	&= \xi^P \widehat\scH_{P N M \, \um}
	\eol & \quad
        - 16 i \,\xi^{\alpha i} \,\overset{\leftrightarrow}{\nabla}_\um E_{[N \dalpha i}\,  E_{M]}{}^a (\gamma_a)_\alpha{}^\dalpha
	+ 16 i \,\xi_{\dalpha i} \,\overset{\leftrightarrow}{\nabla}_\um E_{[N}{}^{\alpha i}\,  E_{M]}{}^a (\gamma_a)_\alpha{}^\dalpha
	\eol & \quad
	- 4i \,\xi^a \Big(
	\veps_{abcd} \,E_{[N}{}^b \nabla_\um \Omega_{M]}{}^{cd}
        + 4\,E_{[N}{}^{\alpha i} \, \overset{\leftrightarrow}{\nabla}_\um E_{M]\dalpha i} (\gamma_a)_\alpha{}^\dalpha
	\Big)~.
\end{align}
Again, the suggestion is that $\widehat\scH_{P N M \, \um}$ should possess a sensible
tangent space expansion.

\subsection{Superspace Bianchi identities}
Having now some idea of the relevant superspace curvatures and what combinations of
them should involve sensible tangent space expansions, we turn to a brief
discussion of the Bianchi identities that need to be solved.

We begin with the fields of the $p$-form hierarchy. As at the component level, the
field strength $\cF^\um$ must obey the Bianchi identity
\begin{align}
\cD \cF^\um
    &= -12 \,(t^\balpha)^{\um\un} \partial_\un H_\balpha 
    - \frac{1}{2} \Omega^{\um\un} H_\un \eol
    &= -12 \,(t^\balpha)^{\um\un} \nabla_\un H_\balpha 
    + \Omega^{\um\un} \,\cT_\un{}^\balpha H_\balpha
    - \frac12 \Omega^{\um\un} \scH_\un\ ,
\end{align}
where we have used the definition \eqref{eq:defscH} for $\scH_\um$. The above form
of the Bianchi makes it apparent that $H_\um$ cannot be covariant under internal
diffeomorphisms (because $\pa_\un H_\balpha$ is not), whereas $\scH_\um$ is.
Keeping in mind that it is $\widehat \cF^\um$ rather than
$\cF^\um$ that will possess a conventional tangent space expansion, one can rewrite
this Bianchi identity as in \eqref{bfh}
\begin{align} \label{bfh_super}
\cD \widehat \cF^\um
    &= -12 (t^\balpha)^{\um\un} \nabla_\un H_\balpha 
    + \Omega^{\um\un} \,\cT_\un{}^\balpha H_\balpha
    - \frac12 \Omega^{\um\un} \scH_\un
    -\cM^{\um\un} E^a \nabla_\un T_a + \cM^{\um\un} T^a \nabla_\un E_a
    \eol & \quad
    - \left( D\cM^{\um \un}\right) E^a \nabla_\un E_a
    + \cM^{\um\un} E^a E^b \nabla_\un \Omega_{ab} \ .
\end{align}

The 3-form field strength $H_\balpha$ in turn obeys the Bianchi identity
\bea
\cD{H_{\balpha}} &=&
	- \frac12 (t_\balpha)_{\um\un} \,\cF^\um \cF^\un 
	- \pa_\um{G^\um{}_{\balpha}} - (t_\balpha)_\um{}^\un G_\un{}^\um 
\nn\w2
	&=&
	- \frac12 (t_\balpha)_{\um\un} \,\cF^\um \cF^\un 
	- \nabla_\um{G^\um{}_{\balpha}} 
	- \frac13 (t_\balpha)_\un{}^\um \,\cT_\um{}^\bbeta G^\un{}_\bbeta 
	- (t_\balpha)_\um{}^\un\, \scG_\un{}^\um \ .
\label{eq:Halpha.Bianchi}
\eea
In the second line, we have introduced
\begin{align}
\scG_\um{}^\un := G_\um{}^\un - \Gamma_\um{}^\balpha G^\un{}_\balpha~,
\end{align}
which unlike $G_\um{}^\un$ is a tensor under internal diffeomorphisms.
These 4-form curvatures are further discussed in Section \ref{app:gc}.

The constrained 3-form $H_\um$ obeys the Bianchi identity
\begin{align}
\cD H_\um &=
	\Omega_{\ur\us} \cF^\ur \pa_\um \cF^\us
	+ 24 (t^\balpha)_\ur{}^\un \pa_\um \pa_\un A^\ur\, H_\balpha
	+ \pa_\um G_\un{}^\un + 2 \,\pa_\un G_\um{}^\un\ .
\end{align}
Its covariant version $\scH_\um$ in turn obeys
\bea
\cD \scH_\um  &=&
    \Omega_{\up\uq} \cF^\up \nabla_\um \cF^\uq - 2 \,R_\um{}^\balpha\, H_{\balpha}
    + \nabla_\um \scG_\un{}^\un + 2 \,\nabla_\un \scG_\um{}^\un
    - 2 \,\mathbf T_{\um\un}{}^\uk \scG_\uk{}^\un
\nn\\
    && + \frac17  \nabla_\um G^\un{}_{\balpha} \, \cT_\un{}^\balpha
    + \frac37 G^\un{}_{\balpha} \, \nabla_\um \cT_\un{}^\balpha
    - 2 R_{\um\un}{}^\balpha G^\un{}_{\balpha} ~.
    \label{eq:Hm.Bianchi}
\eea
The tensors $\mathbf T_{\um \un}{}^\uk$, $R_\um{}^\balpha = \rd Z^N R_{N \um}{}^\balpha$,
and $R_{\um \un}{}^\balpha$ correspond to objects that appear in the
commutators between internal covariant derivatives. They are collected in Appendix \ref{app:Derivs}.

The Bianchi identities for the supervielbein are a bit more complicated.
From the definition \eqref{eq:DefT}, one concludes that
\begin{align}
\cD T^a &= E^b R_b{}^a - \mathbb L_{\cF} E^a~, \\
\cD T^{\alpha i} &= 
    \frac{1}{4} E^{\beta i} R^{cd} (\gamma_{cd})_\beta{}^\alpha
    - \frac{1}{2} E^{\alpha j} R_j{}^i
    - \mathbb L_{\cF} E^{\alpha i}~. \label{eq:Bianchi.psi}
\end{align}
Typically in superspace, the torsion Bianchi identity allows one to determine
the curvature tensors $R_b{}^a$ and $R_j{}^i$ in terms of the torsion tensor.
This is somewhat more subtle in exceptional superspace because these curvature tensors
may now involve terms with internal derivatives of the supervielbein.
This is apparent when considering the Bianchi identities for the 56-bein,
which read (using $R_{kl}{}^{ij} = \delta_{[k}{}^{[i} R_{l]}{}^{j]}$)
\begin{align}
\cD^2 \cV_\um{}^{i j} = - R_{kl}{}^{i j} \cV_\um{}^{kl} - \mathbb L_\cF \cV_\um{}^{i j}
\quad \implies \quad
\begin{cases}
\cD \cP^{i j k l}
	= -i \cV^{\um k l} \mathbb L_\cF \cV_\um{}^{i j} \\
R_{kl}{}^{ij}
	= i \cV^\um{}_{kl} \mathbb L_\cF \cV_\um{}^{ij} - \cP^{i j r s} \cP_{r s k l}
\end{cases}~.
\end{align}
The second identity defines the SU(8) curvature and involves terms
with internal derivatives on the field strength tensor $\cF$ (which itself
involves internal derivatives of the vielbein). Finally, we mention the Bianchi
identities for the $\chi$ curvature, which we leave in the form
\begin{align}\label{eq:Bianchi.chi}
\cD^2 \chi^{\alpha i j k}
    = \frac{1}{4} \chi^{\beta i j k} R^{cd} (\gamma_{cd})_\beta{}^\alpha
    - \frac{3}{2} \chi^{\alpha l [i j} R_l{}^{k]}
    - \mathbb L_{\cF} \chi^{\alpha i j k}~.
\end{align}

\subsection{Constraints and solution of the Bianchi identities}
We present here the set of constraints on the various curvatures that
provide the solution to the Bianchi identities. While we have not explicitly
checked the higher dimension components of \eqref{eq:Bianchi.psi} or 
\eqref{eq:Bianchi.chi}, which provide the explicit form of the Riemann
tensor (and the superspace version of Einstein's equation), the other
Bianchi identities are sufficient to determine the other curvatures.
We leave its full characterization to future work,
where a unified exceptional geometry would be expected
to shed light on some of the structure encountered. While the identities that
we need to solve are a good bit more involved than in conventional superspace,
luckily, most of the relations correspond exactly to results
expected from $N=8$ superspace \cite{Brink:1979nt, Howe:1981gz,Bandos:2015ila}.
We summarize them below. Most of the computations were achieved using \textit{Cadabra}
\cite{Peeters.Cadabra1, Peeters.Cadabra2}.

\paragraph{The 56-bein curvature $\cP^{ijkl}$.}

We impose the following constraints on the supercovariant derivative of the
56-bein \eqref{eq:dV}:
\begin{align}\label{eq:Pijkl.Constraints}
\cP_{\alpha k}{}^{ij pq} = 2 \sqrt 2 \, \delta_k^{[i} \chi_\alpha{}^{jpq]}~, \qquad
\cP^{\dalpha k}{\,}^{ijpq} = \frac{\sqrt 2}{12}\, \veps^{ijpqkrst} \chi^{\dalpha}{}_{rst}\ . \end{align}
%

\paragraph{The vector torsion tensor.} 

The vector torsion tensor is constrained so that its non-vanishing components in tangent space are
\begin{align}\label{eq:Ta.Constraints}
T_{\alpha i}{}^{\dbeta j}{}^c &= 2 \,\delta_i{}^j\, (\gamma^c)_\alpha{}^\dbeta~, \qquad
T_{a b c} = \frac{1}{12} \,\veps_{abcd} \chi^{\alpha ijk}\, (\gamma^d)_{\alpha \dalpha} \, \chi^\dalpha{}_{ijk}\ .
\end{align}
The choice of $T_{ab}{}^c$ is a matter of convention and can be altered by a covariant redefinition of the spin connection. The choice we expect here is to match the convention used in 4D gauged supergravity \cite{deWit:2007kvg}, although it is easy to change this. Therefore, the full constraint on the covariant derivative of $E^a$ can be written as
\begin{align}\label{eq:ExtTorsion.Full}
T^a := \cD E^a
	= -\frac{1}{24} \, \veps^{abcd} \,E_b \wedge E_c \,\chi^{\alpha ijk}\, 
	(\gamma_d)_{\alpha\dalpha} \, \chi^\dalpha{}_{ijk} 
	+ 2\,E^{\alpha i} \wedge E_\dbeta\,(\gamma^a)_\alpha{}^\dbeta\ .
\end{align}

\paragraph{The two-form curvature $\cF^\um$.}

The two-form curvature $\cF^\um$ is constrained through the modified field strength
$\widehat \cF^\um$ given in \eqref{eq:DefHatF}.
The lower dimension parts of $\widehat \cF_{BA}{}^\um$ are constrained as
\begin{subequations}\label{eq:F.Constraints}
\begin{alignat}{2}
\widehat\cF_{\beta j\, \alpha i}{}^{\um} &=
	-8 i\,\cV^{\um}{}_{ji} \,\eps_{\beta\alpha}~, &\qquad
\widehat\cF^{\dbeta j\, \dalpha i}{}^{\um} &=
	+ 8 i \,\cV^{\um}{}^{ji} \,\eps^{\dbeta\dalpha}~, \\
\widehat\cF_{\beta j \,a}{}^{\um}
	&= - \sqrt 2\,  i \, \cV^{\um kl} (\gamma_a)_{\beta \dbeta} \chi^\dbeta{}_{jkl}~, &\quad
\widehat\cF^{\dbeta j}{}_{a}{}^{\um}
	&= \sqrt 2\, i \, \cV^{\um}{}_{kl} (\gamma_a)^{\dbeta \beta} \chi_\beta{}^{jkl}~.
\end{alignat}
\end{subequations}
The vector-vector component $\widehat \cF_{ab}{}^\um$ is also constrained so that
\begin{align}\label{eq:SelfDuality}
\widehat \cF^+_{ab}{}^{i j}  := \widehat \cF^+_{ab}{}^{\um} \cV_{\um}{}^{ij}
	= \frac{1}{144} \veps^{ijklpqrs} \bar \chi_{klp} \gamma_{ab} \chi_{qrs}\ ,
\end{align}
as in gauged supergravity. This is the twisted self-duality constraint in ExFT.
Note that the self-dual part of $\widehat \cF_{ab}{}^{ij}$ is actually
Lorentz covariant, whereas the anti-self-dual part transforms as \eqref{eq:deltaF.Lor}.

\paragraph{The 3-forms $H_\balpha$ and $H_\um$.}

Analyzing the Bianchi identity \eq{bfh_super},
and comparing terms with explicit internal derivatives, one determines the
tangent space components of $H_\balpha = \tfrac{1}{3!} E^A E^B E^C H_{C B A \, \balpha}$
to be
\begin{subequations} \label{eq:Halpha.Constraints}
\begin{align}
H^{\dgamma k}{}_{\beta j\, a\, \balpha} &= 
	\frac{8}{3} (t_\balpha)^{\um \un} \cV_{\um j l} \cV_\un{}^{kl} (\gamma_a)^\dgamma{}_\beta~,\\
H_{\gamma i\, b a\, \balpha} &=
	-\frac{2}{3} \sqrt{2} \,(t_\balpha)^{\um \un} \cV_{\um i j} \cV_{\un kl}
		(\gamma_{b a})_\gamma{}^\beta \chi_\beta{}^{jkl}~, \\
H^{\dgamma i}{}_{b a\, \balpha} &=
	-\frac{2}{3} \sqrt{2} \,(t_\balpha)^{\um \un} \cV_{\um}{}^{i j} \cV_{\un}{}^{kl}
		(\gamma_{b a})^\dgamma{}_\dbeta \chi^\dbeta{}_{jkl}~, \\
H_{c b a\, \balpha} &=
	- \frac{1}{3} \veps_{c b a d} \,(t_{\balpha})^{\um \un} \Big(
	\frac{1}{2} \cV_{\um}{}^{i j} \cV_{\un}{}^{k l} \cP^{d}{}_{i j k l}
	- \frac{1}{2} \cV_{\um}{}_{i j} \cV_{\un}{}_{k l} \cP^{d}{}^{i j k l}
	+ \cV_{\um}{}^{i k} \cV_{\un}{}_{j k} \,
	\chi^{\alpha j p q} (\gamma^{d})_{\alpha}{}^{\dalpha} \chi_{\dalpha i p q}
	\Big)\ ,
\label{eq:Halpha.Constraints.d}
\end{align}
\end{subequations}
with all other components vanishing. From the component perspective, the last
equality is an equation of motion on the three-form field strength and corresponds
to its on-shell duality condition. These expressions agree with those from 
ungauged 4D $N=8$ superspace \cite{Bandos:2015ila}.

Determining the constrained 3-form field strength is somewhat more involved.
From the Bianchi identity \eqref{bfh_super}, one can directly show that
\begin{align}\label{eq:scHm}
\scH_\um &=
    i E^a E^b \,\veps_{abcd} \nabla_\um  \Omega^{cd} 
    - 8 i \,E_a \, \nabla_\um E^{\alpha i} E_{\dalpha i} (\gamma^{a})_\alpha{}^{\dalpha} 
    + 8 i \,E_a \, \nabla_\um E_{\dalpha i} E^{\alpha i} (\gamma^{a})_\alpha{}^{\dalpha} 
    \eol & \quad
    - \tfrac{2i}{3} \sqrt{2} E^a E^b E^{\beta i} (\gamma_{ab})_{\beta\alpha}
        \chi^{\alpha jkl}  \cP_{\um ijkl} 
    + \tfrac{2i}{3} \sqrt{2} E^a E^b E_{\dbeta i} (\gamma_{ab})^{\dbeta\dalpha} 
        \chi_{\dalpha jkl} \cP_\um{}^{ijkl} 
    \eol & \quad
    - \frac{i}{9} \, E^a E^b E^c \,\veps_{abcd} \,(
        \cP_\um{}^{ijkl} \cP^d{}_{ijkl}
        + \tfrac{1}{2} \nabla_\um \chi^{\alpha ijk} 
            \chi_{\dalpha ijk} (\gamma^d)_\alpha{}^{\dalpha} 
	+ \tfrac{1}{2} \nabla_\um {\chi_{\dalpha ijk}} 
            \chi^{\alpha ijk} (\gamma^d)_\alpha{}^{\dalpha} 
	)~.
\end{align}
This explicit expression for the field strength $\scH_\um$ is covariant under internal diffeomorphisms as required by the Bianchi identity \eqref{bfh_super}. However, the presence of the explicit spin connection means it cannot be covariant under Lorentz transformations, which is as expected. Note that the definition \eqref{eq:defscH} for $\widehat \scH_\um$,
which we motivated in the hope of it having a conventional tangent space expansion,
indeed leads to such a result:
\begin{align}\label{eq:Hm.Constraints}
\widehat \scH_\um &:= \frac{1}{3!} E^A E^B E^C \widehat \scH_{CBA\, \um} = 
    - \tfrac{2i}{3} \sqrt{2} E^a E^b E^{\beta i} (\gamma_{ab})_{\beta\alpha}
        \chi^{\alpha jkl}  \cP_{\um ijkl} 
    + \tfrac{2i}{3} \sqrt{2} E^a E^b E_{\dbeta i} (\gamma_{ab})^{\dbeta\dalpha} 
        \chi_{\dalpha jkl} \cP_\um{}^{ijkl} 
    \eol & \qquad \qquad
    - \frac{i}{9} \, E^a E^b E^c \,\veps_{abcd} \,(
        \cP_\um{}^{ijkl} \cP^d{}_{ijkl}
        + \tfrac{1}{2} \nabla_\um \chi^{\alpha ijk} 
            \chi_{\dalpha ijk} (\gamma^d)_\alpha{}^{\dalpha} 
	+ \tfrac{1}{2} \nabla_\um {\chi_{\dalpha ijk}} 
            \chi^{\alpha ijk} (\gamma^d)_\alpha{}^{\dalpha} 
	)~.
\end{align}
From the component perspective, the constraint on $\widehat\scH_{cba\, \um}$ corresponds to
an on-shell duality condition.

The actual (non-covariant) field strength $H_\um$ can be found by inverting \eqref{hhd}:
\begin{align}\label{eq:Hm}
H_\um &=
    i E^a E^b \,\veps_{abcd} \pa_\um \Omega^{c d}
    - 8 i \,E^a \, \Big(\pa_\um E^{\alpha i} - \tfrac12 q_\um{}_j{}^i E^{\alpha j} \Big) 
	E_{\dalpha i} (\gamma_{a})_\alpha{}^{\dalpha} 
    + 8 i \,E^a \, \Big(\pa_\um E_{\dalpha i} + \tfrac12 q_\um{}_i{}^j E_{\dalpha j} \Big) 
	E^{\alpha i} (\gamma_{a})_\alpha{}^{\dalpha} 
    \eol & \quad
    - \tfrac{2i}{3} \sqrt{2} E^a E^b E^{\beta i} (\gamma_{ab})_{\beta\alpha}
        \chi^{\alpha jkl}  p_{\um ijkl} 
    + \tfrac{2i}{3} \sqrt{2} E^a E^b E_{\dbeta i} (\gamma_{ab})^{\dbeta\dalpha} 
        \chi_{\dalpha jkl} p_\um{}^{ijkl} 
    \eol & \quad
    -\tfrac{i}{9} \, E^a E^b E^c \,\veps_{abcd} \,\Big(
	p_\um{}^{ijkl} \cP^d{}_{ijkl}
	+ \tfrac{1}{2} \pa_\um{\chi^{\alpha ijk}} \chi_{\dalpha ijk} (\gamma^d)_\alpha{}^{\dalpha} 
	+ \tfrac{1}{2} \pa_\um{\chi_{\dalpha ijk}} \chi^{\alpha ijk} (\gamma^d)_\alpha{}^{\dalpha} 
	\eol & \qquad \qquad
	- \tfrac32 q_\um{}_l{}^i \chi^{\alpha ljk} \chi_{\dalpha ijk} (\gamma^d)_\alpha{}^{\dalpha} 
    \Big)\ ,
\end{align}
where $q_\um{}_i{}^j$ and $p_\um{}^{ijkl}$ are given by \eqref{eq:Defqp}.

There are several consistency checks which the expressions for $H_\um$ and $\scH_\um$ satisfy:
\begin{itemize}
\item The definition of the covariant $\scH_\um$ \eqref{hhd} involves the 
$\E7$ connection, which as we have discussed contains undetermined pieces.
These drop out from the Bianchi identity \eqref{bfh_super} when the explicit
expressions for $\scH_\um$ and $H_\balpha$ are used, as well as the expression
\eqref{eq:GammaDef} for the $\E7$ connection.

\item $H_\um$ is a constrained tensor on its $\um$ index, obeying the conditions \eqref{eq:ConstrainedTensor}.
However, $\scH_\um$ does not unless the $\E7$ connection is also constrained. As this involves undetermined pieces, this may or may not be the case.

\item Because of the underlying non-Lorentz invariance of the two-form field strength,
the curvature $H_\um$ has the appropriate anomalous Lorentz transformation, consistent
with \eqref{eq:deltaB.Lor},
\begin{align}
\delta H_\um = D \Big(
    i \veps_{abcd} E^a E^b \pa_\um \lambda^{cd}
\Big)~.
\end{align}
\end{itemize}

\paragraph{The gravitino torsion tensor.}

The modified gravitino torsion tensor $\widehat T^{\alpha i}$, defined in \eqref{eq:DefGravT},
is constrained so that its lower tangent space components are
\begin{subequations} \label{eq:Talpha.Constraints}
\begin{align}
\widehat T_{\gamma k\, \beta j}{}^{\alpha i} &= 0\ , \\
\widehat T^{\dgamma k}{}_{\beta j}{}^{\alpha i} &= 0~, \\
\widehat T^{\dgamma k\, \dbeta j\, \alpha i} &= \sqrt 2\, \eps^{\dgamma\dbeta} \chi^{\alpha k j i}~,\\
\widehat T_{\beta j}{}_c{}^{\alpha i}
	&= \tfrac{1}{8} (\bar \chi^{ikl} \gamma^a \chi_{jkl}) (\gamma_c \gamma_a)_\beta{}^\alpha~, \\
\widehat T^{\dbeta j}{}_c{}^{\alpha i}
	&= \tfrac{1}{8} (\gamma_c \gamma^{a b})^{\dbeta \alpha} \widehat\cF_{ab}{}^{i j} 
	- \tfrac{1}{1152} \veps^{ijklpqrs} 
	(\bar \chi_{klp} \gamma^{ab} \chi_{q r s})  (\gamma_{ab} \gamma_c)^{\dbeta \alpha}
	+ i (\gamma_c)^{\dbeta \alpha} \cV^\um{}_{kl} \cP_\um{}^{ijkl}
	- (\gamma_c)^{\dbeta \alpha} A_1{}^{i j} \eol
	&= \tfrac{1}{8} (\gamma_c \gamma^{a b})^{\dbeta \alpha} \,\widehat\cF_{ab}{}^{i j} 
	- \tfrac{1}{8} (\gamma^{ab} \gamma_c)^{\dbeta \alpha} \, \widehat \cF_{ab}{}^{i j}
	+ i (\gamma_c)^{\dbeta \alpha} \cV^\um{}_{kl} \cP_\um{}^{ijkl} 
	- (\gamma_c)^{\dbeta \alpha} A_1{}^{i j} ~.
\end{align}
\end{subequations}
Because the internal covariant derivative $\nabla_\um$ does not carry any internal
spin connection, the modified tensor $\widehat T^{\alpha i}$ has an anomalous Lorentz
transformation; this is reproduced by the constraints above due to the field strength
$\widehat \cF_{ab}{}^{ij}$.
Also, because $\nabla_\um$ \emph{does} depend on the internal $\E7$ connection,
$\widehat T^{\alpha i}$ depends on the precise choice of internal connections even
though $T^{\alpha i} := \cD E^{\alpha i}$ does not.
This is apparent above in the appearance of both
$\cP_\um{}^{ijkl}$ and component $A_1{}^{ij}$ of the $\E7$ torsion tensor.
However, one can check that the undetermined pieces of the internal GVP
drop out of $T^{\alpha i}$ itself. This is the superspace version of
the observation in \cite{Godazgar:2014nqa} that the $\rep{1280}$ component
of the $\SU(8)$ connection drops out of the SUSY transformation of the gravitino.

As a consequence of the Bianchi identities,
in particular the $\cF^\um$ Bianchi identity \eqref{bfh_super},
the top component $\widehat T_{c b}{}^{\alpha i}$ itself obeys several constraints.
Its self-dual component is fixed as
\begin{align}
\widehat T^+_{a b}{}^{\alpha i} &=
\frac{\sqrt{2}}{12} \cP_{[a}{}^{i j k l} (\gamma_{b]})^{\dalpha \alpha} \chi_{\dalpha j k l} 
+ \frac{\sqrt{2}}{24} \veps_{a b c d} \,\cP^{c}{}^{i j k l} \chi_{\dalpha j k l} (\gamma^{d})^{\dalpha \alpha}
+ \frac{\sqrt{2}}{8} \chi^{\alpha i j k} \widehat\cF^+_{ab \,j k}\ ,
\end{align}
whereas the spin-1/2 part of its anti-self-dual component is
\begin{align}
\widehat T^-_{a b}{}^{\alpha i} (\gamma^{b})_{\alpha \dalpha}
    &= 
    \tfrac{\sqrt{2}}{12} A_2{}^{i}{}_{j k l} \chi^{\alpha j k l} (\gamma_{a})_{\alpha \dalpha} 
    + \tfrac{\sqrt{2}}{24} \cP^{b}{}^{i j k l} \chi_{\dbeta j k l} (\gamma_b \gamma_a)^\dbeta{}_\dalpha
    + i \tfrac{\sqrt{2}}{2} \cV^{\um}{}_{j k} \, \nabla_{\um}{\chi^{\alpha i j k}} \,
        (\gamma_{a})_{\alpha \dalpha} 
    \eol & \quad
    - i \tfrac{\sqrt{2} }{3} \cV^{\um i j} \cP_{\um j k l p} \chi^{\alpha k l p} (\gamma_{a})_{\alpha \dalpha} 
    + i \tfrac{\sqrt{2}}{4} \cV^{\um j k} \cP_{\um j k l p} \chi^{\alpha i l p} (\gamma_{a})_{\alpha \dalpha} 
    \eol & \quad
    + \tfrac{\sqrt{2}}{1728} \bar \chi^{i j k} \chi^{l p q} \, 
        \chi^{\alpha r s t} (\gamma_{a})_{\alpha\dalpha} \, \veps_{j k l p q r s t}~.
\end{align}
These correspond to the gravitino equations of motion in the underlying component theory.
Note that $\widehat T_{ab}{}^{\alpha i}$ itself is not Lorentz covariant, although
$T_{nm}{}^{\alpha i}$ is.
This is because it is $T_{nm}{}^{\alpha i}$ itself that is directly related
to the gravitino equations of motion.

\paragraph{The $\chi$ curvature.}

The curvature  $\widehat \tau_M{}^{\alpha i}$ is defined in \eqref{eq:Deftau},
and it takes values in tangent space as
\begin{subequations}\label{eq:tau.Constraints}
\begin{align} 
\widehat\tau_{\beta l}{}^{\alpha}{}^{ijk}
	&= -\frac{3}{4} \sqrt{2} \,\delta_{l}^{[i} (\gamma_{a b})_{\beta}{}^\alpha \widehat \cF_{a b}{}^{j k]}
	- \frac{\sqrt{2}}{24} \delta_\beta{}^\alpha \veps^{ijk pqr st} \bar\chi_{pqr} \chi_{st l}
	\eol &\quad
	- 6 i \sqrt 2\,\delta_\beta{}^\alpha \cV^\un{}_{pq} \cP_\un{}^{p q [i j} \delta_l{}^{k]}
	- 8i \sqrt 2 \, \delta_\beta{}^\alpha  \cV^\un{}_{pl} \cP_\un{}^{ijk p} 
	- 2 \sqrt{2}\, \delta_\beta{}^\alpha A_2{}_l{}^{ijk}
	~, \\
\widehat\tau^{\dbeta l\, \alpha i j k}
	&= 2 \sqrt{2} (\gamma_a)^{\dbeta \alpha} \cP_{a}{}^{i j k l}~.
\end{align}
\end{subequations}
The spin-1/2 part of its top component $\widehat \tau_a{}^{\alpha i j k}$ is constrained as
\begin{align}
\widehat\tau_a{}^{\alpha i j k} (\gamma_a)_{\alpha \dalpha} &=
    \tfrac{1}{6} A_2{}^{[i}{}_{l p q} \,\veps^{j k] l p q r s t} \, \chi_{\dalpha r s t} 
    + \tfrac{i}{3} \veps^{i j k l p q r s} \cV^{\um}{}_{r s} \nabla_{\um}{\chi_{\dalpha l p q}} 
    \eol & \quad
    - 2i \cV^{\um r s} \cP_{\um}{}^{i j k t}  \chi_{\dalpha r s t}
    - 6i \cV^{\um r [i} \cP_{\um}{}^{j k] s t}  \chi_{\dalpha r s t}
    + 6i \cV^{\um [i j} \cP_{\um}{}^{k] r s t}  \chi_{\dalpha r s t}
    \eol & \quad
    + \tfrac{1}{4} \bar\chi^{i j k} \chi^{r s t} \,\chi_{\dalpha r s t} 
    - \tfrac{3}{4} \bar\chi^{r [i j} \chi^{k] s t} \,\chi_{\dalpha r s t} 
    + \tfrac{1}{48} \veps^{i j k l p q r s} \widehat \cF_{a b}{}_{r s}\,
        \chi_{\dbeta l p q} (\gamma_{a b})^{\dbeta}{}_{\dalpha}  
\end{align}
corresponding to the $\chi$ equation of motion. As with the gravitino torsion
components, these constraints arise most directly by analyzing the $\cF^\um$ Bianchi
identity \eqref{bfh_super}.
Here as well the specific choice of connection terms in $\nabla_\un$
influences the $\chi$ curvature. The absence of an internal Lorentz connection 
is reflected in the appearance of $\widehat \cF_{ab}{}^{ij}$, and the dependence
on the precise $\E7$ connection is reflected by the appearance of
$\cP_\um{}^{ijkl}$ and the generalized torsion component $A_2{}_l{}^{ijk}$.
As with the gravitino curvature, one can check that the undetermined pieces
of the internal GVP drop out of $\cD \chi^{\alpha i j k}$.

\subsection{The $G$ curvatures}\label{app:gc}

For the sake of completeness, we record here a number of results related to the 4-form
field strengths $G^\um{}_\balpha$ and $G_\um{}^\un$ in superspace. These arise by solving
the Bianchi identities \eqref{eq:Halpha.Bianchi} and \eqref{eq:Hm.Bianchi}, which provided
for us a consistency check on our solutions for $H_\balpha$ and $\scH_\um$.
As when one solves for the $H$ field strengths using the $\cF$ Bianchi identities, there is ambiguity
in these solutions having to do with the kernel of the projector appearing on the right-hand
side of the Bianchi identity. To put it more simply, to solve for the 4-form curvatures $G$,
we must implicitly make a choice for the 4-form potentials of the tensor hierarchy, as these
have not yet appeared in any curvatures. It is interesting that the superspace versions
of the $G$ curvatures that we will give below possess on-shell duality conditions
that \emph{do not} reduce to the ones given in \eqref{eq:G912duality} and
\eqref{eq:Gmn.duality}, and thus must correspond to a redefinition of one or more
of the 4-form potentials.

The superspace curvature $G^\um{}_\balpha$, which is in the $\rep{912}$, 
enjoys like $H_\balpha$ a standard tangent space expansion with components
\begin{subequations}\label{eq:G912.Constraints}
\begin{align}
G_{\beta j\, \alpha i\, b a}{}^{\um}{}_{\balpha}
	&= - \frac{32i}{3} (\gamma_{a b})_{\alpha\beta}\,
	\cV^{\um}{}_{i k} \, \cV^{\un}{}_{j l} \cV^{\up}{}^{k l} \, (t_{\balpha})_{\ul{np}}~, \qquad
G^{\dbeta j}{}_{\alpha i\, b a}{}^{\um}{}_{\balpha}	= 0~, \\
G_{\alpha i\, c b a}{}^\um{}_\balpha &= 
\frac{14i}{3} \sqrt{2}\, \veps_{a b c d} (\gamma_{d})_{\alpha}{}^{\dalpha}  \,
	(\chi_{\dalpha j k l} \,\cV^{\un}{}_{i q}
	\,\cV_{\ur}{}^{q j} \,\cV_{\us}{}^{k l})\,
    t_{\bbeta}{}^{\ur \us} \, \mathbb P_{912}{}_{\un}{}^{\bbeta}{}_{\balpha}{}^{\um}~, \\
G_{d c b a}{}^\um{}_\balpha &=
	\frac{7i}{3}
	\veps_{a b c d}
	\Big(
	\chi^{\alpha i j k} \eps_{\alpha \beta} \chi^{\beta r s t} \cV_{\ur}{}_{k r} \cV_{\us}{}_{s t} \cV^{\un}{}_{i j}
	+ \chi_{\dalpha i j k} \eps^{\dalpha \dbeta} \chi_{\dbeta r s t} \cV_{\ur}{}^{k r} \cV_{\us}{}^{s t} \cV^{\un}{}^{i j}
	\Big)\, t_{\bbeta}{}^{\ur \us} \, \mathbb P_{912}{}_{\un}{}^{\bbeta}{}_{\balpha}{}^{\um}
	\eol & \quad
	+ 24 i \,\veps_{a b c d} \,\cZ^{\um}{}_\balpha
\end{align}
\end{subequations}
where $\cZ^{\um}{}_\balpha$ is a purely scalar expression determined only by derivatives
of the coset fields. In terms of the Weitzenb\"ock connection, it can be written most
simply as
\begin{align}\label{eq:G912Z}
\cZ^{\um}{}_{\balpha}
	&= \frac{1}{288} \Big( t_{\bbeta}{}_{\uk}{}^{\ur} \mathring\cT_{\ul{pq}}{}^{\us} \cM^{\ul{np}} \cM^{\ul{kq}} \cM_{\ur \us}
		+ 7 \,t_{\bbeta}{}_{\uk}{}^{\uq} \mathring\cT_{\ul{pq}}{}^{\uk} \cM^{\ul{np}} \Big) 
	\mathbb P_{912}{}_{\un}{}^{\bbeta}{}_{\balpha}{}^{\um}~,
\end{align}
entirely in terms of the Weitzenb\"ock torsion and the internal metric.
Note that this \emph{does not} coincide with the bosonic expression \eqref{eq:G912duality}
given in Section \ref{S:Sec2}.
This suggests that these two bosonic results for $G^\um{}_\balpha$ must differ by a
redefinition of a 4-form potential. We will show this below. The fact that this
form of the expression seems to more naturally arise in superspace is quite
remarkable for the following reason.
In a Scherk-Schwarz reduction of the type we will discuss in Section \ref{S:SS}, the Weitzenb\"ock torsion
is replaced by the embedding tensor, and the above result is then proportional to the variation
of the scalar potential of gauged supergravity with respect to the embedding tensor.
It is expected that the $D$-form field strengths of gauged supergravities
should be equal to this quantity, see e.g. the $D=3$ discussion of \cite{deWit:2008ta}.

The compact expression \eqref{eq:G912Z} can be rewritten as
\begin{align}
\cZ^{\um}{}_{\balpha}
	&= \Big[
	\cQ_{\bbeta i}{}^{j} \Big(
	\frac{i}{8} \cV^{\un k i} A_1{}_{j k} 
	+ \frac{i}{8} \cV^{\un}{}_{k j} A_1{}^{i k} 
	+ \frac{1}{8} \cV^{\un k i} \cV^{\um}{}_{j l} \cQ_{\um k}{}^{l} 
	+ \frac{1}{8} \cV^{\un}{}_{k j} \cV^{\um i l} \cQ_{\um l}{}^{k})
	\eol & \quad
	+ \cP_{\bbeta i j k l} \Big( 
	- \frac{7i}{144} \cV^{\un l p} A_2{}_{p}{}^{i j k} 
	- \frac{7}{36} \cV^{\un l p} \cV^{\um}{}_{p q} \cP_{\um}{}^{i j k q} 
	- \frac{7}{48} \cV^{\un l p} \cV^{\um i j} \cQ_{\um}{}_{p}{}^{k} \Big)
	\eol & \quad
	+ \cP_{\bbeta}{}^{i j k l} \Big(
	\frac{7i}{144} \cV^{\un}{}_{l p} A_2{}^{p}{}_{i j k} 
	- \frac{7}{36} \cV^{\un}{}_{l p} \cV^{\um p q} \cP_{\um i j k q} 
	+ \frac{7}{48} \cV^{\un}{}_{l p} \cV^{\um}{}_{i j} \cQ_{\um k}{}^{p} \Big)
	\Big] \,\mathbb P_{912}{}_{\un}{}^{\bbeta}{}_{\balpha}{}^{\um}~,
	\label{eq:G912Z2}
\end{align}
where $\cQ_\balpha$ and $\cP_\balpha$ are the $\SU(8)$ projections of
$(t_\balpha)_\um{}^\un$, see \eqref{eq:DefQPalpha}.
In the latter expression, we have done two things. First, we have exchanged $\mathring \cT$
built from the Weitzenb\"ock connection for $\cT$ built from the generic $\E7$ connection
$\Gamma$. Remember this carries undetermined pieces, corresponding to freedom to redefine
$\cQ_\um$ and $\cP_\um$, which now appear explicitly. We have subsequently rewritten $\cT$ in
terms of the $\SU(8)$ tensors $A_1$ and $A_2$.

The reason for rewriting $Z^\um{}_\balpha$ in this way is to emphasize that it
is not $\SU(8)$ invariant, with the internal connection $\cQ_\um{}_i{}^j$ appearing
explicitly. The $\SU(8)$ transformation of $Z^\um{}_\balpha$
leads to a transformation of $G^\um{}_\balpha$ itself,
\begin{align}\label{eq:G912.SU8}
\delta G^\um{}_\balpha
	&= i E^{a} E^{b} E^{c} E^{d} \veps_{a b c d} \Big[
	\tfrac{1}{36} \cP_{\balpha i j k l} \cV^{\um i j} \cV^{\un k p} \pa_{\un}{\lambda_{p}{}^{l}} 
	+ \tfrac{1}{12} \cP_{\balpha i j k l} \cV^{\um i p} \cV^{\un j k} \pa_{\un}{\lambda_{p}{}^{l}} 
	\eol & \quad
	- \cQ_{\balpha i}{}^{j} (
	\tfrac{1}{18} \cV^{\um}{}_{j k} \cV^{\un i l} \pa_{\un}{\lambda_{l}{}^{k}} 
	+ \tfrac{5}{72} \cV^{\um}{}_{j k} \cV^{\un k l} \pa_{\un}{\lambda_{l}{}^{i}} 
	+ \tfrac{1}{24} \cV^{\um}{}_{k l} \cV^{\un i k} \pa_{\un}{\lambda_{j}{}^{l}} 
	+ \tfrac{1}{48} \cV^{\um}{}_{k l} \cV^{\un k l} \pa_{\un}{\lambda_{j}{}^{i}} 
	)
	+ \CC
	\Big]
\end{align}
Because there is no internal derivative on a frame field, this can only arise from
a non-covariant $\SU(8)$ transformation of one of the constrained 4-form potentials
$D_{QPNM}{}_\um{}^{\up \uq}$. Extrapolating from the pattern of the 2-form and 3-form potentials,
this field should obey the section condition on $\um$ and the upper pair of indices
$\up\uq$ should live in some particular representation of $\E7$ in the product
$\rep{56} \times \rep{56} = \rep{1} + \rep{133} + \rep{1463} + \rep{1539}$.
It is not hard to show that all but the $\rep{1463}$ are in principle present, meaning
that they are projected out from the right-hand side of the $H$ Bianchi identities.
(It may be that they are not actually required to ensure gauge invariance of the
field strengths.) Writing these three fields as
$D_\um$, $D_\um{}^\balpha$, and $D_\um{}^{\ul{np}}$,
their full contributions to the field strengths $G^\um{}_\balpha$ and $G_\um{}^\un$ are
\begin{subequations}
\begin{align}
G^{\um}{}_\balpha
    &= \cdots + 7 \,\mathbb P_{912}{}^\bbeta{}_\un{}^\um{}_\balpha \Big(
    \kappa_{\bbeta \bgamma} \, \Omega^{\ul{np}} \,D_\up{}^\bgamma
    + D_\up{}^{\ul{qn}} \, (t_{\bbeta})_\uq{}^\up
    \Big)~, \\
G_{\um}{}^\un
    &= \cdots
    + \Omega^{\ul{np}} (\pa_\um D_\up - \pa_\up D_\um)
    + 12 \,\pa_\up D_\um{}^\bbeta \, (t_{\bbeta})^{\ul{pn}}
    + 4 \,\pa_\um D_\up{}^\bbeta \, (t_{\bbeta})^{\ul{pn}}
    \eol & \quad
    + 2 \,\pa_\up D_\um{}^{\ul{pn}}
    + \tfrac{2}{3} \pa_\um D_\up{}^{\ul{pn}}
    ~.
\end{align}
\end{subequations}
In order to generate the anomalous $\SU(8)$ transformation \eqref{eq:G912.SU8}, one
should assign the following anomalous transformation to $D_\um{}^{\ul{np}}$,
\begin{align}
\delta D_{\un}{}^{\ul{pq}} =
	\frac{i}{4!} E^{a} E^{b} E^{c} E^{d}\, \veps_{a b c d} \,(
-  \pa_{\un}{\lambda_{i}{}^{j}} \cV^{\up}{}^{i k} \cV^{\uq}{}_{k j} 
+ \pa_{\un}{\lambda_{j}{}^{i}} \cV^{\up}{}_{i k} \cV^{\uq}{}^{k j} )~.
\end{align}
This is an intriguing result, because it seems very similar to what we found
for the constrained 2-form, where it seemed necessary to assign an anomalous
Lorentz transformation \eqref{eq:deltaB.Lor}.

This anomalous $\SU(8)$ behavior seems to be at the root of the difference between
the bosonic part \eqref{eq:G912Z} of the superform $G^\um{}_\balpha$
and the purely bosonic expression \eqref{eq:G912duality} for $G^\um{}_\balpha$.
Namely, there seems to be a tension between maintaining $\SU(8)$ invariance
and maintaining covariance under internal diffeomorphisms.
We derived the expression \eqref{eq:G912duality} in a formulation with only an
internal metric and no explicit 56-bein; it was not possible to violate $\SU(8)$
invariance, and we were led to an expression that transforms anomalously
under internal diffeomorphisms. In deriving the superform $G^\um{}_\balpha$
above, manifest $\E7$ diffeomorphism covariance was assumed everywhere and led to an expression
that violates $\SU(8)$ invariance. As it turns out, one can write down the bosonic
shift in $D_\um{}^{\ul{pq}}$ that exchanges \eqref{eq:G912duality} for \eqref{eq:G912Z}:
\begin{align}\label{eq:Delta4-form}
\Delta D_{\um}{}^{\ul{pq}} = \frac{i}{4!} E^a E^b E^c E^d \, \veps_{abcd} \Big(
    \pa_{\um} \cV^{[\up}{}^{i j} \cV^{\uq]}{}_{i j}
    + \pa_{\um}{\cV^{[\up}{}_{i j}} \cV^{\uq]}{}^{i j}
    + \tfrac{1}{56} \Omega^{\up \uq} (
        \pa_{\um}{\cV^{\un}{}^{i j}} \cV_{\un}{}_{i j}
        +\pa_{\um}{\cV^{\un}{}_{i j}} \cV_{\un}{}^{i j}
    )
\Big)~.
\end{align}
It is easy to see that this induces the anomalous $\SU(8)$ transformation discussed above.

For completeness, we also give the rather complicated expression for
$\scG_\um{}^\un$ that we found in superspace. As with $\scH_\um$, it
is useful to separate out a part $\widehat\scG_\um{}^\un$
that possesses a conventional tangent
space expression from the rest:
\begin{align}
\scG_\um{}^\un - \widehat \scG_\um{}^\un
        &= 
	8 \,E^{a} E^{b} E^{\alpha i} \nabla_{\um}{E^{\beta j}} (\gamma_{a b})_{\alpha\beta} \cV^{\un}{}_{i j}
	- 4 \,E^{\alpha i} E^{\beta j} E^{a} \nabla_{\um}{E_{a}} \, \eps_{\alpha\beta} \cV^{\un}{}_{i j}
	\eol & \quad
	+ \tfrac{\sqrt 2}{3} E^{a} E^{b} \nabla_{\um}{E^{c}} E^{\alpha i} 
	\veps_{a b c d} (\gamma^{d})_{\alpha}{}^{\dalpha}  \chi_{\dalpha i j k} \cV^{\un}{}^{j k}
	+ \sqrt{2} \,E^{a} E^{b} \nabla_{\um}{E_{b}} E^{\alpha i} 
	(\gamma_{a})_{\alpha}{}^{\dalpha}  \chi_{\dalpha i j k} \cV^{\un}{}^{j k}
	\eol & \quad
	- \tfrac{5 \sqrt 2}{9} E^{a} E^{b} E^{c} \nabla_{\um} E^{\alpha i} 
	\veps_{a b c d} (\gamma^{d})_{\alpha}{}^{\dalpha} \chi_{\dalpha i j k} \cV^{\un}{}^{j k}
	\eol & \quad
	+ E^{a} E^{b} E^{c} \nabla_{\um}{E_{c}} \Big(
        \tfrac{i}{8}  \veps_{a b d e} \widehat \cF^{d e}{}^{\un}
        - \tfrac{1}{144} (\gamma_{a b})_{\alpha \beta} \chi^{\alpha i j k} \chi^{\beta p q r} \veps_{i j k p q r s t} \cV^{\un}{}^{s t}
	\Big)
	\eol & \quad
	+ \tfrac{i}{6} \veps_{a b c d} \,E^{a} E^{b} E^{c} \nabla_{\um}{\nabla_{\up}{E^{d}}} \,\cM^{\ul{pn}} 
	\eol & \quad
	+ \tfrac{i}{9} \veps_{a b c d} E^{a} E^{b} E^{c} \nabla_{\um}{E^{d}}  \Big(
	\cP_{\up}{}^{i j k l} \cV^{\up}{}_{i j} \cV^{\un}{}_{k l} 
	+ \cQ_{\up i}{}^{j} \cV^{\up i k} \cV^{\un}{}_{j k} 
	\Big)
	\eol & \quad
	+ \tfrac{i}{3} \veps_{a b c d} E^{a} E^{b} E^{c} \nabla_{\up}{E^{d}}  \Big(
	\cP_{\um}{}^{i j k l} \cV^{\up}{}_{i j} \cV^{\un}{}_{k l} 
	+ \cQ_{\um i}{}^{j} \cV^{\up i k} \cV^{\un}{}_{j k} 
	\Big)
	\eol & \quad
	+ \CC
\end{align}
The conventional part
$\widehat\scG_\um{}^\un = \frac{1}{4!} E^A E^B E^C E^D \widehat\cG_{DCBA}{\,}_\um{}^\un$ 
has non-vanishing pieces
\begin{subequations}
\begin{align}
\widehat\scG_{c b a \, \alpha i}{\,}_\um{}^\un
    &= 6\sqrt{2}\, \veps_{a b c d} (\gamma_{d})_{\alpha}{}^{\dalpha} 
	\Big(
	\tfrac{2}{9} \chi_{\dalpha i j k} \cP_{\um}{}^{j k l p} \cV^{\un}{}_{l p}
	- \tfrac{4}{9} \chi_{\dalpha j k l} \cP_{\um}{}^{j k l p} \cV^{\un}{}_{i p}
	- \tfrac{2}{18} \nabla_{\um} \chi_{\dalpha i j k}  \,\cV^{\un}{}^{j k} \Big)
	~, \\
\widehat\scG_{c b a}{}^{\dalpha i}{\,}_\um{}^\un 
    &= - 6 \sqrt{2}\, \veps_{abcd} (\gamma_{d})_{\alpha}{}^{\dalpha} 
	\Big(
	\tfrac{2}{9} \chi^{\alpha i j k} \cP_{\um}{}_{j k l p} \cV^{\un}{}^{l p}
	- \tfrac{4}{9} \chi^{\alpha j k l} \cP_{\um}{}_{j k l p} \cV^{\un}{}^{i p}
	- \tfrac{2}{18} \nabla_{\um} \chi^{\alpha i j k}  \,\cV^{\un}{}_{j k} \Big)
	~, \\
\widehat\scG_{dcba}{\,}_\um{}^\un &=
    24 i \,\veps_{a b c d} \Big(
            \tfrac{i}{18} \chi^{\alpha i j k} \eps_{\alpha\beta} \chi^{\beta l r s} \,\cP_{\um}{}_{i j k l}\,\cV^{\un}{}_{r s} 
            - \tfrac{i}{432} \chi^{\alpha i j k} \eps_{\alpha\beta} \nabla_{\um}{\chi^{\beta l p q}} \veps_{i j k l p q r s} \cV^{\un}{}^{r s}
            + \CC
    \Big)
    \eol & \quad
    + 24 i \,\veps_{a b c d}\, \cZ_\um{}^\un~.
\end{align}
\end{subequations}
The last term $\cZ_\um{}^\un$ gives the purely bosonic part of $\widehat\scG_\um{}^\un$ and involves
the rather unwieldy expression\footnote{The $\nabla_\um$ in $\nabla_\um \cQ_\un{}_i{}^j$ is to
be understood to carry the same $\SU(8)$ connection as if $\cQ$ were a tensor.}
\begin{align}
\cZ_\um{}^\un &=
	- \tfrac{1}{96} \cQ_{\um i}{}^{j} \cQ_{\up j}{}^{i} \cM^{\ul{pn}}
	+ \tfrac{1}{9} \cV^{\un}{}_{i j} \cV^{\up i k} \nabla_{\um}{\cQ_{\up k}{}^{j}} 
	+ \tfrac{1}{9} \cV^{\un}{}_{i j} \cV^{\up}{}_{k l} \nabla_{\um}{\cP_{\up}{}^{i j k l}} 
	\eol & \quad
	+ \tfrac{i}{8} A_1{}_{i j} \cQ_{\um k}{}^{i} \cV^{\un j k} 
	- \tfrac{i}{48} A_2{}_{i}{}^{j k l} \cQ_{\um j}{}^{i} \cV^{\un}{}_{k l} 
	+ \tfrac{i}{36} A_2{}_{i}{}^{j k l} \cP_{\um j k l p} \cV^{\un i p} 
	\eol & \quad
	+  \cP_{\um i j k l} \Big(\tfrac{1}{9} \cP_{\up}{}^{i j k p} \cV^{\up}{}_{p q} \cV^{\un l q} 
	+ \tfrac{1}{9} \cP_{\up}{}^{i j p q} \cV^{\un}{}_{p q} \cV^{\up k l} 
	- \tfrac{1}{18}\cP_{\up}{}^{i j p q} \cV^{\up}{}_{p q} \cV^{\un k l}
        \eol & \qquad \qquad
        - \tfrac{1}{36} \cQ_{\up p}{}^{i} \cV^{\un j k} \cV^{\up l p} 
	+ \tfrac{1}{36} \cQ_{\up p}{}^{i} \cV^{\un j p} \cV^{\up k l} 
	\Big)
	\eol & \quad
	+ \tfrac{1}{12} \cQ_{\um p}{}^{i} \cP_{\up i j k l} \Big(\cV^{\un j p} \cV^{\up k l} 
	- \cV^{\un j k} \cV^{\up l p} \Big)
	+ \cQ_{\um i}{}^{j} \cQ_{\up j}{}^{k} \Big(
	\tfrac{1}{24} \cV^{\un}{}_{k l} \cV^{\up i l} 
	+ \tfrac{1}{8} \cV^{\up}{}_{k l} \cV^{\un i l} 
	\Big)
	\eol & \quad
	+ \CC
\end{align}
As with $\scH_\um$, much of the structure is determined by requiring that the
undetermined parts of the various connections cancel when one computes
$G_\um{}^\un$ from $\scG_\um{}^\un$. The expression for $G_\um{}^\un$ can be recovered
by setting $\Gamma=0$, $\cQ=q$, and $\cP = p$ in the expression for $\scG_\um{}^\un$.
We should also add that the expression for $\mathring\scG_\um{}^\un$ (that is,
with the Weitzenb\"ock connection) is quite simple as $\mathring\cQ_\um$ and
$\mathring\cP_\um$ both vanish.

The bosonic part of $G_\um{}^\un$ must coincide with the one given in
\eqref{eq:Gmn.duality} after some redefinition of the 4-forms. We have already
seen for $G^\um{}_\balpha$ that the redefinition is restricted to the constrained
4-forms, in particular \eqref{eq:Delta4-form} for $D_\um{}^{\ul{np}}$.
No redefinition was needed for $D_\um{}^{\balpha}$, but we have not checked if one
is needed for $D_\um$. (The latter constrained 4-form is absent in
$G^\um{}_\balpha$.)

\section{Component results from superspace}

Here we verify that the use of the proposed generalized superdiffeomorphisms and 
constraints on torsion and curvatures produce the component results. 

\subsection{Component fields and supersymmetry transformations}
First, we must identify the component fields in terms of the various superfields.
For the component one-forms, the correct procedure is to identify them as the
$\theta= \rd\theta=0$ part of the superspace one-form. Formally, this corresponds
to the pullback of the inclusion map embedding spacetime into superspace.
For the vierbein, this amounts to
\begin{align}
e^a := E^a \vert_{\theta = \rd\theta=0} \quad \implies \quad e^a =  \rd x^m e_m{}^a~, \quad
e_m{}^a(x,y) := E_m{}^a(Z) \vert_{\theta=0} ~.
\end{align}
For the gravitino, it is conventional to include an additional factor of $2$,
\begin{align}
\psi_\alpha{}^{i} := 2 \, E_\alpha{}^{i} \vert_{\theta=\rd\theta=0} \quad \implies \quad
\psi_m{}_\alpha{}^i(x,y) = 2\, E_m{}_\alpha{}^i(Z)\vert_{\theta=0}~.
\end{align}
For all other one-forms, we make the analogous choices, i.e.
\begin{align}
A_m{}^\um(x,y) := A_m{}^\um(Z)\vert_{\theta=0}~, \quad
B_{m n}{\,}_\balpha(x,y) := B_{m n}{\,}_{\balpha}(Z)\vert_{\theta=0}~, \quad
B_{m n}{\,}_\um(x,y) := B_{m n}{\,}_{\um}(Z)\vert_{\theta=0}~,
\end{align}
for the fundamental one-forms, and similarly for the composite one-forms,
\begin{align}
\omega_m{}^{ab}(x,y) := \Omega_m{}^{ab}(Z) \vert_{\theta=0}~, \qquad
\cQ_m{}_i{}^j(x,y) := \cQ_m{}_i{}^j(Z) \vert_{\theta=0}~.
\end{align}
All other component fields correspond to $\theta=0$ parts of identically named
superfields. For example, $\chi_\alpha{}^{ijk}(x,y) := \chi_\alpha{}^{ijk}(Z)\vert_{\theta=0}$,
and so forth.

To derive their symmetry transformations, we must compute their transformations
under \emph{covariant} external diffeomorphisms where the diffeomorphism parameter,
written in tangent space, takes the form
$\xi^A := \xi^M E_M{}^A = (0, \eps^{\alpha i}, \eps_{\dalpha i})$. 
Let's discuss first how this works with the vierbein and gravitino. 
From \eqref{eq:dcovEa} and \eqref{eq:Ta.Constraints}, we find for the vierbein
\begin{align}
\delta E_M{}^a 
&= 
	2 \, E_M{}^{\beta j} \eps_{\dgamma k} (\gamma^a)_{\beta}{}^\dgamma \delta_j{}^k
	+ 2 \, E_M{}_{\dbeta j} \eps^{\gamma k} (\gamma^a)_{\gamma}{}^\dbeta \delta_k{}^j\ ,
\end{align}
which reduces to the component result, rewritten in four-component notation,
\begin{align}
\delta e_m{}^a &= 
	\bar \eps_j \gamma^a \psi_m{}^j
	+ \bar \eps^j \gamma^a \psi_m{}_j~.
\end{align}
From \eqref{eq:dcovEalpha} and \eqref{eq:Talpha.Constraints}, we find
\begin{align}
\delta E_M{}^{\alpha i}
	&=
	\cD_M \eps^{\alpha i}
	+ 2i \cV^{\um i j} \nabla_{\um} \Big(\eps_{\dbeta j} E_M{}^c (\gamma_c)^{\dbeta \alpha}\Big)
	\eol & \quad
	+ E_M{}_{\dbeta j} \eps_{\dgamma k} \sqrt 2\, \eps^{\dgamma\dbeta} \chi^{\alpha k j i}
	+ \tfrac{1}{8} E_M{}^c \eps^{\beta j} (\bar \chi^{ikl} \gamma^a \chi_{jkl}) (\gamma_c \gamma_a)_\beta{}^\alpha
	\eol & \quad
	+ \tfrac{1}{8} E_M{}^c \eps_{\dbeta j} (\gamma_c \gamma^{a b})^{\dbeta \alpha} \,\widehat\cF_{ab}{}^{i j} 
	- \tfrac{1}{8} E_M{}^c \eps_{\dbeta j} (\gamma^{ab} \gamma_c)^{\dbeta \alpha} \, \widehat \cF_{ab}{}^{i j}
	\eol & \quad
	+ i E_M{}^c \eps_{\dbeta j} (\gamma_c)^{\dbeta \alpha} \cV^\um{}_{kl} \cP_\um{}^{ijkl} 
	- E_M{}^c \eps_{\dbeta j} (\gamma_c)^{\dbeta \alpha} A_1{}^{i j}~.
\end{align}
Lowering and suppressing the spinor index, and then reducing to the $\theta=\rd\theta=0$
part gives the gravitino supersymmetry transformation
\begin{align}\label{eq:delta.psi}
\delta \psi_m{}^{i}
	&=
	2 \cD_m \eps^{i}
	- 4i \,\cV^{\um i j} \nabla_{\um} (\gamma_m \eps_j)
	- \sqrt 2 \,\bar\eps_{j} \psi_m{}_{k} \, \chi^{i j k}
	+ \tfrac{1}{4} (\bar \chi^{ikl} \gamma^a \chi_{jkl}) \,\gamma_a \gamma_m \eps^{j}
	\eol & \quad
	+ \tfrac{1}{4}  \gamma^{ab} \gamma_m \eps_{j} \,\widehat\cF_{ab}{}^{i j} 
	- \tfrac{1}{4} \gamma_m \gamma^{ab}  \eps_{j}  \, \widehat \cF_{ab}{}^{i j}
	+ 2 \gamma_m\eps_{j} \Big(A_1{}^{i j} + i \cV^\um{}_{kl} \cP_\um{}^{ijkl}\Big)~.
\end{align}

For the 56-bein, \eqref{eq:dV} and \eqref{eq:Pijkl.Constraints} lead to
\begin{align}
\delta \cV_\um{}^{ij} &= 2 \sqrt{2} \,\bar \eps^{i} \chi^{jkl} \cV_{\um kl}
	+ \frac{\sqrt{2}}{12} \,\veps^{ijklpqrs} \,\bar\eps_p \chi_{qrs} \cV_{\um kl}
\end{align}
whereas for $\chi$ \eqref{eq:dcovchi} and \eqref{eq:tau.Constraints} imply
\begin{align}
\delta \chi^{ijk}
	&= 
	- 2 \sqrt{2} \,\cP_{a}{}^{i j k l}\,\gamma^a \eps_{l} 
	- 12 i \sqrt{2} \,\cV^{\un [i j} \nabla_{\un} \eps^{k]}
	+ \tfrac{3\sqrt{2}}{4}  \,\widehat \cF_{a b}{}^{[i j} \,\gamma_{a b} \eps^{k]} 
	- \tfrac{\sqrt{2}}{24} \veps^{ijk pqr st} \bar\chi_{pqr} \chi_{st l}\, \eps^{l} 
	\eol & \quad
	- 2 \sqrt{2}\, A_2{}_l{}^{ijk}\, \eps^{l} 
	- 6 i \sqrt 2 \, \cV^\un{}_{pq} \cP_\un{}^{p q [i j}\, \eps^{k]} 
	- 8 i \sqrt 2 \, \cV^\un{}_{pq} \cP_\un{}^{ijk p} \, \eps^{q}~.
\end{align}

For $A_m{}^\um$, we combine \eqref{eq:dcovEalpha} with \eqref{eq:F.Constraints} to
recover
\begin{align}
\delta A_m{}^\um
	&= 4 i\,\cV^{\ul M}{}_{ij} \,\bar\eps^{i} \psi_m{}^{j}  
	- i \sqrt 2\, \cV^{\ul M i j} \bar\eps^{k} \gamma_m \chi_{ijk}
	+ \CC
\end{align}
For the adjoint-valued two-forms, \eqref{eq:dcovBalpha} and \eqref{eq:Halpha.Constraints}
lead to
\begin{align}
\Delta B_{m n\, \balpha} &=
	- \tfrac{2}{3} \sqrt{2} \,(t_\balpha)^{\um \un} \cV_{\um i j} \cV_{\un kl} \,\bar\eps^i \gamma_{mn} \chi^{jkl}
	- \tfrac{8}{3} \,(t_\balpha)^{\um \un} \cV_{\um j l} \cV_\un{}^{kl} \, \bar\eps_{k}\, \gamma_{[m} \psi_{n]}{}^{j} 
	+ \CC
\end{align}
For the constrained two-forms, \eqref{eq:dcovBm} and \eqref{eq:Hm.Constraints} lead to
\begin{align}
\Delta B_{mn\,\um} + 2 \,\Gamma_\um{}^\balpha \Delta B_{mn\,\balpha}
	&= 
	- 8i \,\bar \eps^{i} \,\gamma_{[m}  \nabla_\um \psi_{n]}{}_{i}
	+ 8i \,\nabla_\um \bar \eps^{i} \,\gamma_{[m} \psi_{n]}{}_{i}
	+ 8i \,\bar\eps_{i} \,\gamma_{[m} \nabla_\um \psi_{n]}{}^{i}
	- 8i \,\nabla_\um \bar\eps_{i} \,\gamma_{[m} \psi_{n]}{}^{i}
	\eol & \quad
	- \frac{4i}{3} \sqrt{2}\, \bar\eps^{i} \gamma_{mn} \chi^{j k l} \cP_{\un i j k l}
	+ \frac{4i}{3} \sqrt{2}\, \bar\eps_{i} \gamma_{mn} \chi_{j k l} \cP_{\un}{}^{i j k l}~.
\end{align}

Above, we have recovered the SUSY transformations of $e_m{}^a$, $\cV_\um{}^{ij}$, and
$B_{mn\,\balpha}$ in the form of \eqref{susy}. For the gravitino, $\chi^{ijk}$, and
$B_{mn\, \um}$, one must keep in mind that the above rules involve the redefined field
$B'_{mn\,\um}$ (which we have denoted in Section 4 and onward without a prime)
and so one should compare instead with \eqref{susy2}. Aside from higher
fermionic corrections, some deviations arise in these transformations having to do
with allowing the internal GVP to take a more general form. For example,
in comparing with the gravitino transformation \eqref{eq:delta.psi},
one finds in addition to \eqref{susy2},
two explicit higher-order fermion terms,
one implicit higher-order fermion term (the second $\widehat\cF_{ab}{}^{ij}$ term, which
is on-shell related to a fermion bilinear via \eqref{eq:SelfDuality}),
and the last two terms involving the $A_1$ component of the $\E7$ torsion tensor
and the non-metricity $\cP_\um{}^{ijkl}$, which vanish under the internal GVP assumptions
made in Sections 2 and 3.

We emphasize that as in \cite{Godazgar:2014nqa}
one can confirm that all of the undetermined components of the internal connections drop out from
the above transformations. This is most easily seen by using \eqref{E:Weitzenbock}
to rewrite $\Gamma$ in terms of the Weitzenb\"ock connection, isolating the undetermined pieces
in the fields $\cQ_\um{}_i{}^j$ and $\cP_\um{}^{ijkl}$. The latter two fields then cancel out
of all equations.

\subsection{Composite connections and supercovariant curvatures}
The supersymmetry transformations discussed above involve several composite
quantities -- the spin connection, the $\SU(8)$ connection, the covariant field
strength $\widehat \cF_{ab}{}^{ij}$ -- and their component definitions need to
be given for the component SUSY transformations to be fully realized.

From the constraints on the torsion two-form, one determines the component external
spin connection by projecting \eqref{eq:ExtTorsion.Full} to spacetime,
\begin{align}
2 \,\cD_{[m} e_{n]}{}^a &= 2 \,D_{[m} e_{n]}{}^a + 2 \,\omega_{m n}{}^a =
    \bar \psi_{[m}{}^i \gamma^a \psi_{n]}{}_i 
    + \tfrac{1}{12} \veps_{mn}{}^{ab} \,\bar \chi^{ijk} \gamma_b \chi_{ijk}~,
\end{align}
and solving for $\omega_m{}^{ab}$ in the usual way.
One similarly obtains the component SU(8) from \eqref{eq:dV},
\begin{align}
\cD_m \cV_\un{}^{ij} = D_m \cV_\un{}^{ij} + \cQ_m{}_k{}^{[i} \cV_\un{}^{j]k} = 
    \Big(
    e_m{}^a \cP_{a}{}^{ijkl}\, 
    + \sqrt2\, \bar\psi_m{}^{[i} \chi^{jkl]}
    + \tfrac{\sqrt 2}{24} \veps^{ijkl pqrs} \, \bar\psi_m{}_p \chi_{qrs} \Big) \cV_{\un \, kl}~,
\end{align}
and inverting the relation to solve for $\cQ_m{}_i{}^j$. Both expressions for $\omega_m{}^{ab}$ and $\cQ_\um{}_i{}^j$ match those of ungauged $N=8$ supergravity upon replacing $\pa_m \rightarrow D_m$. Note that this expression defines $\cP_a{}^{ijkl}$ to coincide with the supercovariant one-form of ungauged $N=8$ supergravity, where it is usually denoted $\hat \cP_a{}^{ijkl}$.

The supercovariant field strength for the vector fields arises by projecting
\eqref{eq:DefHatF} to components, using the constraints \eqref{eq:F.Constraints},
and solving for $\widehat \cF_{ab}{}^{\um}$ as
\begin{align}
\widehat \cF_{ab}{}^\um &=
    e_a{}^m e_b{}^n \cF_{m n}{}^\um
    + i \,\cV^{\um\, i j} \Big(
    4 \,\bar \psi_{a i} \psi_{b j}
    + \sqrt{2} \bar\psi_{[a}{}^k \gamma_{b]} \chi_{i j k}
    \Big)
    - i \,\cV^{\um}{}_{i j} \Big(
    4 \,\bar \psi_{a}{}^i \psi_{b}{}^j
    + \sqrt{2} \bar\psi_{[a}{}_k \gamma_{b]} \chi^{i j k}
    \Big)
    \eol & \quad
    - \cM^{\um \un} e_{[a}{}^m \nabla_\un e_{m b]}~.
\end{align}
It is this quantity that obeys the twisted self-duality condition \eqref{eq:SelfDuality},
equivalently written
\begin{align}
\widehat \cF_{ab}{}^\um &= \frac{i}{2}\,\varepsilon_{abcd}\, \Omega^{\um\un}\,\cM_{\un\uk}\,
    \widehat\cF^{cd\,\uk}~.
\end{align}

Although we don't need them to realize the SUSY transformations, it is worth giving the
explicit formulae for the supercovariant 3-form field strengths of the 2-form fields.
The supercovariant form of $H_{mnp\, \balpha}$ corresponds to the lowest component of the
superspace tensor $H_{abc\, \balpha}$, which is
\begin{align}
H_{a b c \, \balpha}
    &= e_a{}^m e_b{}^n e_c{}^p H_{mnp\, \balpha}
    + \sqrt{2}\, (t_\balpha)^{\um \un} \Big(
        \cV_\um{}_{i j} \cV_\un{}_{kl} \, \bar\psi_{[a}{}^i \gamma_{bc]} \chi^{jkl} + \CC \Big)
    + 4 \,(t_\balpha{})^{\um \un} \cV_\um{}_{jk} \cV_\un{}^{i k}\,\bar\psi_{[a}{}^i \gamma_b \psi_{c]}{}_j ~.
\end{align}
The constraint \eqref{eq:Halpha.Constraints.d} corresponds to the on-shell duality equation
\begin{align}
H_{a b c\, \balpha} &=
- \frac{1}{6} \veps_{a b c d} \,(t_{\balpha})^{\um \un} \Big(
	\cV_{\um}{}^{i j} \cV_{\un}{}^{k l} \cP^{d}{}_{i j k l}
	- \cV_{\um}{}_{i j} \cV_{\un}{}_{k l} \cP^{d}{}^{i j k l}
	- 2\,\cV_{\um}{}^{i k} \cV_{\un}{}_{j k} \,
	\bar \chi^{j p q} \gamma^{d}\chi_{i p q}
	\Big)
\end{align}
which is a natural generalization of the bosonic result \eqref{hd1}.

For the constrained two-form, the supercovariant form of its field strength corresponds to
the lowest component of $\widehat\scH_{abc\, \um}$, which is a component of the
superspace tensor $\widehat\scH_{\um}$ defined in \eqref{eq:defscH}. Using the constraints
\eqref{eq:Hm.Constraints}, we find
\begin{align}
\widehat\scH_{abc\, \um} &=
    e_{[a}{}^m e_b{}^n e_{c]}{}^p \Big[
    H_{mnp\,\um} + 2 \, \Gamma_\um{}^\balpha H_{mnp\, \balpha}
    + 6 i \, \veps_{mn \, ef} \nabla_\um \omega_p{}^{ef}
    \eol & \qquad \qquad
    + 12i\, \bar\psi_m{}^i \gamma_n \nabla_\um \psi_{p i}
    - 12i\, \bar\psi_m{}_i \gamma_n \nabla_\um \psi_{p}{}^{i}
    \eol & \qquad \qquad
    + 2 i \sqrt{2}\, \bar\psi_m{}^i \gamma_{np} \chi^{jkl}\, \cP_{\um\, ijkl}
    - 2 i \sqrt{2}\, \bar\psi_m{}_i \gamma_{np} \chi_{jkl}\, \cP_{\um}{}^{ijkl}
    \Big]\ ,
\end{align}
where the on-shell duality equation is given by
\begin{align}
\widehat\scH_{a b c\, \um} &=
    \frac{2i}{3} \veps_{abcd} \,(
        \cP_\um{}^{ijkl} \cP^d{}_{ijkl}
	+ \tfrac{1}{2} \bar\chi^{ijk} \gamma^d \overset{\leftrightarrow}{\nabla}_\um \chi_{ijk}
	)~.
\end{align}
This generalizes the bosonic result \eqref{hd2}, where one must take care to
note that the terms involving the internal and external spin connections,
$\pa_\um \omega_n{}^{ab} - D_n \omega_\um{}^{ab}$,
corresponding to a mixed internal/external Riemann tensor $R_{\um n}{}^{ab}$,
have been eliminated in different ways: the former by absorption into the
definition of $\scH_{abc\,\um}$, and the latter by redefining $B$ to $B'$.

\section{Consistent Scherk-Schwarz reductions in superspace}
\label{S:SS}

It has already been shown in \cite{Hohm:2014qga} that the (bosonic) $\E7$ ExFT
admits a consistent Scherk-Schwarz reduction to gauged supergravity with an
embedding tensor related to the twist matrices associated with the reduction,
provided the twist matrices themselves obey the section condition.
It is no surprise that a similar statement can be made connecting 
$\E7$ ExFT superspace with $N=8$ superspace with an arbitrary embedding tensor.
We sketch the construction here for two reasons. First, with the more generic
internal GVP we have advocated, the connection between ExFT and gauged supergravity
becomes completely transparent. Second, to our knowledge, the corresponding
$N=8$ superspace with generic embedding tensor has not actually appeared explicitly
in the literature, although it is by no means difficult to construct it directly from
the component results \cite{deWit:2007kvg}.

In complete analogy to \cite{Hohm:2014qga}, a generalized Scherk-Schwarz reduction
in superspace arises by assuming that the $y$-dependence of any superfield is
sequestered into two special fields, a so-called twist matrix $U_\um{}^\tM(y)$ and
a scale factor $\rho(y)$; hereafter we refer to these collectively as twist matrices.
We employ $\tM,\tN,\tP,\cdots$ to denote the ``flat''
$\E7$ 56-plet indices of gauged supergravity.\footnote{These ``flat'' indices should not
be confused with the $\SU(8)$ tangent space indices ${}^{ij}$ and ${}_{ij}$
which arise when one contracts with $\cV_\um{}^{ij}$ or $\cV_\um{}_{ij}$.}
For a superfield $\Phi_\um$ of weight $\lambda$,
carrying a single fundamental $\E7$ index, we call a \emph{covariant twist} one for which
\begin{align}\label{eq:CovTwist}
\Phi_\um(x,\theta,y) =  \rho^{-2 \lambda} \, U_\um{}^\tM \,\Phi_\tM(x,\theta)~,
\end{align}
with a straightforward generalization to different $\E7$ representations.
Nearly every superfield is covariantly twisted, e.g.
\begin{subequations}
\begin{align}
E_M{}^a(x,\theta,y) &= \rho^{-1}\, E_M{}^a(x,\theta)~, \\
E_M{}^{\alpha i}(x,\theta,y) &= \rho^{-1/2}\, E_M{}^{\alpha i}(x,\theta)~, \\
\cV_\um{}^{ij}(x,\theta,y) &= U_\um{}^\tM \,\cV_\tM{}^{ij}(x,\theta)~, \\
\chi_\alpha{}^{i j k}(x,\theta,y) &= \rho^{1/4}\, \chi_\alpha{}^{ijk}(x,\theta)~, \\
A_M{}^\um(x,\theta,y) &= \rho^{-1}\, (U^{-1})_\tM{}^\um\, A_M{}^\tM(x,\theta) \, ~, \\
B_{NM\,\balpha}(x,\theta,y) &= \rho^{-2}\, U_\balpha{}^\bbeta\, B_{N M\,\bbeta}(x,\theta)~.
\end{align}
\end{subequations}
The exception is for the constrained $p$-form fields, e.g. $B_{NM\, \um}$, which we
will describe shortly.

The twist matrices cannot be chosen arbitrarily. Rather, they must obey the following
two conditions
\begin{align}
7\, \Big[ (U^{-1})_\tM{}^\um (U^{-1})_\tN{}^\un 
	\pa_\um U_\un{}^\tP \Big]_{\rep{912}} &= \rho\, X_{\tM \tN}{}^\tP
	\equiv \rho\, \Theta_\tM{}^\balpha (t_\balpha)_\tN{}^\tP~, \\
\pa_\un (U^{-1})_\tM{}^\un - 3 \rho^{-1} \pa_\un \rho\,  (U^{-1})_\tM{}^\un
	&= 2\, \rho\, \vartheta_\tM~,
\end{align}
where $X_{\tM\tN}{}^\tP$ 
and $\vartheta_\tM$ are constant matrices. These correspond
to the two components of the embedding tensor of gauged supergravity,
with $X_{\tM\tN}{}^\tP = \Theta_\tM{}^\balpha (t_\balpha)_\tN{}^\tP$
corresponding to the $\rep{912}$ component \cite{deWit:2007kvg} and
$\vartheta_\tM$ corresponding to the $\rep{56}$ component associated only to
trombone gaugings \cite{LeDiffon:2008sh}. Provided one can choose twist matrices
in this way, one can show that the two pieces of the embedding tensor,
$X_{\tM \tN}{}^\tP$ and $\vartheta_\tM$, obey the quadratic constraints \cite{Hohm:2014qga}.

In order to convert the various ExFT formulae, it is useful to eliminate the ambiguity
inherent in the internal GVP. The easiest way to do this is to choose the Weitzenb\"ock
connection where $\mathring \Gamma_\um{}^\balpha$ is determined entirely in terms of the derivative
of the 56-bein with $\mathring \cQ_\um{}_i{}^j$ and $\mathring \cP_\um{}^{ijkl}$ both vanishing.
Because of the ansatz made for the coset fields, it is easy to see that
the $\E7$ connection is given purely by the twist matrices
\begin{align}
\mathring \Gamma_{\um \un}{}^\up = \pa_\um U_\un{}^\tP (U^{-1})_\tP{}^\up~.
\end{align}
In particular, it follows that the torsion tensor is
\begin{align}
\mathring \cT_{\um \un}{}^\up = \rho \, U_\um{}^\tM\, U_\un{}^\tN X_{\tM \tN}{}^\tP \, (U^{-1})_\tP{}^\up~.
\end{align}
Considering the torsion tensor as an internal tensor of weight $\lambda=-1/2$, 
we identify its ``flattened'' version as the embedding tensor. 
It is also straightforward to show that the covariant derivative of any superfield
obeying the covariant twist ansatz \eqref{eq:CovTwist} is
\begin{align}
\mathring \nabla_\un \Phi_\um
	= \rho^{-2 (\lambda-1)}\, U_\un{}^\tN \, U_\um{}^\tM \Big( \frac{4}{3} \lambda\, \vartheta_\tN \Phi_\tM \Big)~.
\end{align}
This generalizes easily to any other $\E7$ representation carried by $\Phi$.
Thus, covariant derivatives of covariantly twisted objects just map
to the trombone part of the embedding tensor, multiplied by a factor of $\tfrac{4}{3}\lambda$.

The notable exception to the covariant twist ansatz is the constrained two-form
$B_{NM \,\um}$ (and the higher constrained $p$-form fields).
The appropriate ansatz, given in component form in \cite{Hohm:2014qga},
can be motivated by considering a covariantized version of $B_{NM\, \um}$,
\begin{align}
\scB_{N M\um} &= B_{N M \, \um} + 2 \,\Gamma_{\um}{}^{\balpha} B_{N M\, \balpha}~.
\end{align}
This redefined 2-form is the natural potential associated with the 3-form field
strength $\scH_{PNM\, \um}$ that we have been employing. For example, the field
strength superform $\cF^\um$ can be rewritten
\begin{align}
\cF^{\um}
&= \rd{A^{\um}} + A^{\un} \nabla_{\un}{A^{\um}}
+ \frac{1}{4} (
    24 \,t^{\balpha \,\um \un} t_{\balpha \,\ur \us}
    - \Omega^{\um \un} \Omega_{\ur \us}
  ) A^{\ur} \nabla_{\un}{A^{\us}}
  - \frac{1}{2} A^\un A^\up \cT_{\ul{np}}{}^\um
  \eol & \quad
- 12\, t^{\balpha}{}^{\um \un} \nabla_\un B_\balpha
	+ \Omega^{\um \un} \,T_\un{}^\balpha \,B_\balpha
	- \frac{1}{2} \Omega^{\um \un} \scB_\un~,
\end{align}
in terms of $\scB$. Above, we have converted all internal derivatives to covariant
ones. Here we are using a generic $\E7$ connection, but now we will specialize
to the Weitzenb\"ock connection.
The reduction ansatz for $\scB$, when the Weitzenb\"ock connection is chosen, can
be simply written as
\begin{align}\label{eq:Bm.ansatz}
\mathring \scB_{N M \um}(x,\theta,y) = 0~.
\end{align}

It is straightforward now to apply the reduction ansatz to all of the various curvature superforms.
For example, the field strength superfield $\cF^\tM$ becomes
\begin{align}\label{eq:GaugedSugra.DefF}
\cF^\tM &= \rd{A^{\tM}} 
	+ \frac{2}{3} \vartheta_\tN A^{\tN} A^{\tM}
	- \frac{1}{6} \Omega^{\tM \tN} \vartheta_\tN \, \Omega_{\tR \tS} \,A^{\tR} A^{\tS} 
	- \frac{1}{2} A^\tN A^\tP X_{\tN \tP}{}^\tM
  \eol & \quad
	- 16 \,t^{\balpha}{}^{\tM \tN} \vartheta_\tN B_\balpha
	+ \Omega^{\tM \tN} \Theta_\tN{}^\balpha B_\balpha~,
\end{align}
as expected for gauged supergravity \cite{LeDiffon:2008sh}. Note that there is no longer any difference
between $\cF^\tM$ and $\widehat \cF^\tM$. Now the superspace constraints on $\cF^\tM$ are just
given by \eqref{eq:F.Constraints} and \eqref{eq:SelfDuality}, with the index $\um$ replaced by $\tM$.
Similar considerations apply to the higher $p$-form field strengths in the tensor hierarchy (with the
exception of the constrained field strengths discussed below).
For example, the field strength $H_{\balpha}$ of gauged supergravity will obey the same constraints
\eqref{eq:Halpha.Constraints}, although its explicit form in terms of the potentials will now involve the embedding tensor as in \eqref{eq:GaugedSugra.DefF}.

For quantities that are covariant under internal diffeomorphisms, it is useful to first
fully covariantize any internal derivatives. In particular, the external covariant derivative $\cD$ of any superfield that transforms
covariantly under internal diffeomorphisms is altered as follows. For the prototypical
superfield $\Phi_\um$ of weight $\lambda$ discussed above,
\begin{align}
\cD_N \Phi_\um := \pa_N \Phi_\um - \mathbb L_{A_N} \Phi_\um = \pa_N \Phi_\um 
    - \mathbb L^\nabla_{A_N} \Phi_\um - A_N{}^\un \cT_{\un \um}{}^\up \Phi_\up~.
\end{align}
Now covariantly twisting quantities and specializing to the Weitzenb\"ock connection,
this becomes
\begin{align}
\cD_N \Phi_\tM = \pa_N \Phi_\um
    - 2 \lambda\, \vartheta_\tK A_N{}^\tK \Phi_\tM
    - 8 \,\vartheta_\tK A_N{}^\tL \mathbb P^\tK{}_\tL{}^\tN{}_\tM \Phi_\tN
    - A_N{}^\tN X_{\tN \tM}{}^\tP \Phi_\tP~.
\end{align}
The last term is the usual embedding tensor contribution, whereas the middle two
terms correspond to trombone contributions.

For the vierbein $E_M{}^a$, the new torsion tensor $T^a = \cD E^a$ is unchanged.
Similarly, the constraints on $\cD \cV_\tM{}^{i j} = \cP^{i j k l} \cV_{\tM k l}$
exactly match the superspace ExFT results. For the gravitino $E_M{}^{\alpha i}$, we define 
$T^{\alpha i} = \cD E^{\alpha i}$, and using the definition
\eqref{eq:DefGravT} of $\widehat T^{\alpha i}$ with its ExFT constraints
\eqref{eq:Talpha.Constraints}, leads to the gauged supergravity constraints
\begin{subequations} 
\begin{align}
T_{\gamma k\, \beta j}{}^{\alpha i} &= 0\ , \\
T^{\dgamma k}{}_{\beta j}{}^{\alpha i} &= 0~, \\
T^{\dgamma k\, \dbeta j\, \alpha i} &= \sqrt 2\, \eps^{\dgamma\dbeta} \chi^{\alpha k j i}~,\\
T_{\beta j}{}_c{}^{\alpha i}
	&= \tfrac{1}{8} (\bar \chi^{ikl} \gamma^a \chi_{jkl}) (\gamma_c \gamma_a)_\beta{}^\alpha~, \\
T^{\dbeta j}{}_c{}^{\alpha i}
	&= \tfrac{1}{8} (\gamma_c \gamma^{a b})^{\dbeta \alpha} \cF_{ab}{}^{i j} 
	- \tfrac{1}{1152} \veps^{ijklpqrs} 
	(\bar \chi_{klp} \gamma^{ab} \chi_{q r s})  (\gamma_{ab} \gamma_c)^{\dbeta \alpha}
	\eol & \quad
	+ (\gamma_c)^{\dbeta \alpha} (
        2 B^{i j}
	- A_1{}^{i j})	
\end{align}
\end{subequations}
It helps to recall here that $A_1{}^{ij}$ in \eqref{eq:Talpha.Constraints} corresponded
to a specific component of the $\E7$ torsion tensor. Adopting the Weitzenb\"ock
connection and making the reduction ansatz converts this to the corresponding
component of the embedding tensor. The trombone contribution 
$B^{i j} := i \cV^{\tM i j} \vartheta_\tM$ arises from the
second term in the definition of $\widehat T^{\alpha i}$. We also emphasize that
taking the Weitzenb\"ock connection has eliminated all factors of $\cP$ and $\cQ$.
For the one-form $\chi$ curvature, $\tau^{\alpha i j k} := \cD\chi^{\alpha i j k}$,
we recover the constraints
\begin{align}
\tau_{\beta l}{}^{\alpha}{}^{ijk}
	&= -\frac{3}{4} \sqrt{2} \,\delta_{l}^{[i} (\gamma_{a b})_{\beta}{}^\alpha \cF_{a b}{}^{j k]}
	- \frac{\sqrt{2}}{24} \delta_\beta{}^\alpha \veps^{ijk pqr st} \bar\chi_{pqr} \chi_{st l}
	\eol &\quad
	- 2 \sqrt{2}\, \delta_\beta{}^\alpha A_2{}_l{}^{ijk}
	- 4 \sqrt{2}\, \delta_\beta{}^\alpha \delta_l{}^{[i} B^{j k]}
	~, \eol
\tau^{\dbeta l\, \alpha i j k}
	&= 2 \sqrt{2} (\gamma_a)^{\dbeta \alpha} \cP_{a}{}^{i j k l}~.
\end{align}
Using these constraints, one can recover the expected SUSY transformations
of the component gravitino and $\chi$ field.

Because of the structure of the generalized Scherk-Schwarz reduction, where
group-valued twist matrices govern the entirety of the $y$-dependence, consistency
of the reduced theory is straightforward. The only meaningful check is to ensure
that the trivial ansatz \eqref{eq:Bm.ansatz} for $\scB_{N M \, \um}$ is consistent
with the other ans\"atze, where a general $(x,\theta)$ dependent piece remained.
This amounts to checking that the curvature associated with $\scB_{NM\, \um}$ actually
vanishes. Although we have not discussed this explicitly, it is relatively straightforward
to show that $\scH_\um$ can be defined directly in terms of $\scB_\um$ and corresponds
to the covariantization of its exterior derivative. Now upon specializing to the
Weitzenb\"ock connection and making the ans\"atze discussed above, one can
see that $\mathring \scH_\um$ does indeed vanish.\footnote{This is slightly more
subtle than we have described, because one must ensure that the full definition
of $\scH_\um$ in terms of $\scB_\um$ consistently vanishes. This expression involves
$R_{[P \um}{}^\balpha B_{N M]\,\balpha}$, the first factor being the mixed
external/internal $\E7$ curvature, and indeed one can show this vanishes in the Weitzenb\"ock
connection under the ans\"atze.} (The same is true for $\mathring\scG_\um{}^\un$,
ensuring that the constrained 3-forms drop out as well.)

In summary, the constraints discussed above characterize the structure of
gauged supergravity in superspace.

\section{Conclusions}

In this paper we have provided the superspace formulation for $\E7$ exceptional field theory. We have shown how the external diffeomorphisms and local supersymmetry transformations can be understood in a unified fashion as superdiffeomorphisms. In doing so, we have found that a redefinition of constrained 2-form potential is necessary, and it provides a geometrical framework in which the internal Lorentz connection is removed everywhere. Interestingly, a similar field redefinition in $\E8$ exceptional field theory, this time involving a constrained 1-form, allowed a reinterpretation of the theory as a Chern-Simons theory \cite{Hohm:2018ybo}. As an application of our superspace $\E7$ ExFT, we have performed a generalized Scherk-Schwarz reduction to obtain the superspace formulation of maximal gauged supergravities parametrized by an embedding tensor. 

The ideas of this paper are expected to be applicable to all other exceptional field theories. Two challenging future directions are as follows. The first is an application of our results to the construction of actions for particle, string and brane actions as suitable sigma models in which the target space manifold is the superspace we have constructed here.  The second is to aim for a further unification. Although we have combined supersymmetry and external diffeomorphisms, they remain distinct from internal diffeomorphisms. A master formulation should exist where these emerge as different parts of a single set of generalized (super)diffeomorphisms. In such a formulation, including fermions and their local supersymmetry transformations, it would be interesting to understand better the reason for the redefinitions we have encountered and whether
 it is indeed essential for some of the constrained $p$-form fields to adopt anomalous $R$-symmetry transformations.

\section*{Acknowledgments}

We would like to thank Guillaume Bossard, Olaf Hohm, Axel Kleinschmidt, and
William Linch for helpful discussions.
The work of ES is supported in part by NSF grants PHY-1521099 and PHY-1803875.
The work of DB is supported in part by NSF grants PHY-1521099 and
the Mitchell Institute for Fundamental Physics and Astronomy.

\begin{appendix}

\section{Conventions}\label{app:conv}
The Lorentz metric is $\eta_{ab} = \textrm{diag}(-1,1,1,1)$ and the
antisymmetric tensor $\veps_{abcd}$ is imaginary, with $\veps_{0123} = -i$.
We employ the pseudotensor
$\veps_{mnpq} := e_m{}^a e_n{}^b e_p{}^c e_q{}^d \, \veps_{abcd}$,
which introduces some factors of $e = \det e_m{}^a$ versus corresponding
formulae in \cite{Godazgar:2014nqa}.

\subsection{Spinor conventions}
We employ both four-component and two-component conventions.
Our two-component conventions follow mainly Wess and Bagger \cite{Wess:1992cp}.
Left-handed spinors are denoted with two-component
Greek indices $\alpha,\beta,\gamma,\cdots$, while right-handed spinors are
denoted with dotted indices $\dalpha, \dbeta, \dgamma, \cdots$.
Spinor indices are raised and lowered using the antisymmetric tensor
$\eps_{\alpha \beta}$,
\begin{gather}\label{eq:epsLorentz}
\psi^\beta = \eps^{\beta \alpha} \psi_\alpha~, \qquad \psi_\alpha = \eps_{\alpha \beta} \psi^\beta~,\qquad
\eps_{\alpha \beta} \eps^{\beta \gamma} = \delta_\alpha^\gamma~, \qquad
\eps^{12} = \eps_{21} = 1~,
\end{gather}
and similarly with dotted indices.
A complex four-component Dirac spinor $\Psi$ decomposes into left-handed and
right-handed spinors $\psi_\alpha$ and $\chi^\dalpha$. Its charge conjugate
$\Psi^c$ decomposes into $\chi_\alpha = (\chi_\dalpha)^*$ and
$\psi^\dalpha = (\psi^\alpha)^*$ so that $\Psi$ and its Dirac conjugate $\bar\Psi$
are given by
\begin{align}
\Psi =
\begin{pmatrix}
\psi_\alpha \\
\chi^\dalpha
\end{pmatrix}~, \qquad
\bar\Psi =
\begin{pmatrix}
\chi^\alpha~, & \psi_\dalpha
\end{pmatrix}~.
\end{align}
For a Majorana spinor, $\chi = \psi$ above.
Our 4D gamma matrices obey
\begin{align}
\{\gamma^a, \gamma^b\} = 2 \eta^{ab}~, \qquad
(\gamma^a)^\dag = \gamma_a~, \qquad
\gamma_5 = -\ri \gamma^0 \gamma^1 \gamma^2 \gamma^3~.
\end{align}
They decompose as
\begin{gather}
\gamma^a =
\left(\begin{array}{cc}
0 & (\gamma^a)_{\alpha \dbeta} \\
(\gamma^a)^{\dalpha \beta} & 0
\end{array}\right)~, \qquad 
\gamma_5 = \begin{pmatrix}
\delta_\alpha{}^\beta & 0 \\
0 & -\delta^\dalpha{}_\dbeta
\end{pmatrix}~.
\end{gather}
The two-component matrices $(\gamma^a)_{\alpha\dalpha}$ are formally identical to
$i (\sigma^a)_{\alpha\dalpha}$ where $\sigma^a$ obey the same relations as in \cite{Wess:1992cp}, i.e.
\begin{align}
(\gamma^a)^{\dalpha \alpha} := \eps^{\dalpha \dbeta} \eps^{\alpha \beta} (\gamma^a)_{\beta \dbeta}~.
\end{align}
Antisymmetric combinations of $\gamma$ matrices are
\begin{align}
\gamma^{ab} = \gamma^{[a} \gamma^{b]} = 
\begin{pmatrix}
(\gamma^{ab})_\alpha{}^\beta & 0 \\
0 & (\gamma^{ab})^\dalpha{}_\dbeta 
\end{pmatrix}\ ,
\end{align}
where $(\gamma^{ab})_{\alpha\beta} = \eps_{\beta \gamma} (\gamma^{ab})_\alpha{}^\gamma$
is \emph{symmetric} in its spinor indices and similarly for
$(\gamma^{ab})_{\dalpha\dbeta} = \eps_{\dalpha \dgamma} (\gamma^{ab})^\dgamma{}_\dbeta$.
These obey the duality properties
\begin{align}
\tfrac{1}{2} \veps_{abcd} \gamma^{cd} = -\gamma_5 \gamma_{ab}~, \qquad
\tfrac{1}{2} \veps_{abcd} (\gamma^{cd})_\alpha{}^\beta = -(\gamma_{ab})_\alpha{}^\beta~, \qquad
\tfrac{1}{2} \veps_{abcd} (\gamma^{cd})^\dalpha{}_\dbeta = +(\gamma_{ab})^\dalpha{}_\dbeta~.
\end{align}

\subsection{SO(1,3) and SU(8) transformations and connections }
Lorentz transformations act as
\begin{align}
\delta e_m{}^a &= \lambda^{a}{}_b \,e_m{}^b = -e_m{}^b\, \lambda_b{}^a~, \eol
\delta \psi_{m\, \alpha}{}^i &= \lambda_\alpha{}^\beta \psi_{m\,\beta i}
    = \frac{1}{4} \lambda^{cd} (\gamma_{cd})_\alpha{}^\beta \psi_{m \, \beta i}~, \eol
\delta \psi_{m}{}^{\alpha i} &= - \psi_{m}{}^{\beta i} \lambda_\beta{}^\alpha
    = - \frac{1}{4} \lambda^{cd}  \, \psi_{m}{}^{\beta i} (\gamma_{cd})_\beta{}^\alpha ~,
\end{align}
Our conventions for $\SU(8)$ indices follow \cite{deWit:2007kvg}. In particular,
the $\SU(8)$ transformations of the $\rep{8}$ and $\brep{8}$ involve factors of
$\tfrac{1}{2}$ as
\begin{align}
\delta V_i = - \frac{1}{2} \lambda_i{}^j V_j~, \qquad
\delta V^i = + \frac{1}{2} \lambda_j{}^i V^j \quad \implies \quad
\delta \cV_\um{}^{ij} = \lambda_{kl}{}^{ij} \cV_{\um}{}^{kl} 
    = \delta_{k}{}^{[i} \lambda_{l}{}^{j]}\, \cV_\um{}^{kl}~.
\end{align}
The corresponding connections appear in the covariant derivative with a minus sign
so that
\begin{align}
\cD_m \psi_n{}^i
    = \pa_m \psi_n{}^i
    - \frac{1}{4} \omega_m{}^{cd} \gamma_{cd} \psi_n{}^i
    - \frac{1}{2} \cQ_m{}_j{}^i \psi_n{}^j
     .
\end{align}

\subsection{Differential forms}
Our conventions for differential forms follow the usual superspace conventions.
For a $p$-form $\Omega$, we write
\begin{align}
\Omega = \frac{1}{p!} \rd Z^{M_1} \cdots \rd Z^{M_p} \Omega_{M_p \cdots M_1}
    = \frac{1}{p!} E^{A_1} \cdots E^{A_p} \Omega_{A_p \cdots A_1}~.
\end{align}
Differential forms and interior products act from the right, so that
\begin{align}
\rd \Omega &= \frac{1}{p!} \rd Z^{M_1} \cdots \rd Z^{M_p} \rd Z^N \pa_N \Omega_{M_p \cdots M_1}~, \\
\imath_V \Omega &= \frac{1}{(p-1)!} \rd Z^{M_1} \cdots V^{M_p} \Omega_{M_p \cdots M_1}~.
\end{align}
Whenever superindices $M$ and $N$ are antisymmetrized, this carries a
usual grading so that
\begin{align}
2 \,V_{[M} W_{N]} = V_M W_N - (-1)^{\eps(M) \eps(N)} V_N W_M
\end{align}
where $\eps(M) = 0$ or $1$ depending on if $M$ is a bosonic or fermionic index.
The grading can be understood as arising because the indices $M$ and $N$ have been
interchanged from their ordering on the left-hand side. In a similar way,
gradings appear in expressions like \eqref{eq:FTanDecomp}, which should actually be read as
\begin{align}
 \widehat \cF_{NM}{}^\um = E_N{}^B E_M{}^A \widehat\cF_{B A}{}^\um\, 
    (-1)^{\eps(B) (\eps(M)+\eps(A))}~.
\end{align}
Gradings also arise from pushing super-indices past other fermionic indices.
For example, \eqref{eq:dcovEalpha.0} should be read as
\begin{align}
\delta_{\rm cov}(\xi) E_M{}^{\alpha i}
	&= \cD_M \xi^{\alpha i} + \xi^N T_{NM}{}^{\alpha i} 
	+ 2 i \,\cV^{\um}{}^{ij} \nabla_{\um} \xi^N 
	\Big( (-1)^{\eps(M)} E_N{}_{\dbeta j} E_M{}^c - E_N{}^c E_M{}_{\dbeta j}
	\Big) (\gamma_c)^{\dbeta \alpha}
	\eol & \quad
	- \frac{i}{2\sqrt 2} \cV^{\um}{}_{jk} \nabla_{\um} \xi^N E_N{}^d E_M{}^c
		\chi^{\beta j k i} (\gamma_c \gamma_d)_\beta{}^{\alpha}~,
\end{align}
as the $M$ index must be pushed all the way to the left
and picks up a sign when passing $\dbeta j$.

\section{Algebra of external and internal derivatives}
\label{app:Derivs}

In analyzing the superspace Bianchi identities, it is useful to employ
covariant external and internal derivatives to maintain manifest internal
diffeomorphism covariance. In this appendix, we summarize the commutation
relations of these covariant derivatives. A number of these formulae have
appeared elsewhere (see e.g. \cite{Godazgar:2014sla}), but we present
them here in a unified way in our conventions.

\paragraph{External derivative algebra.}
Defining the exterior (external) covariant differential
$D := \rd - \mathbb L_A$, we have as usual $D^2 = - \mathbb L_\cF$.
Because $\cF^{\um}$ has weight $\tfrac{1}{2}$, it follows that
for arbitrary tensor $V^\um$,
\begin{align}
D^2 V^\um = - \mathbb L_\cF V^\um
	= - \mathbb L_\cF^\nabla V^\um + \cF^\un \cT_{\ul{np}}{}^\um V^\up~.
\end{align}
The last equation is useful because it allows us to maintain manifest internal
diffeomorphism covariance.

\paragraph{Mixed external/internal derivative algebra.}
On an internal vector $V^\um$ of weight $\lambda$, one can show
\begin{align}
[D_M, \nabla_\um] V^\un &= 
	R_{M \ul{mp}}{}^\un V^\up
	- \frac{2}{3} \lambda \,R_{M \uk\, \um}{}^\uk V^\un\ ,
\end{align}
where the mixed $\E7$ curvature
\begin{align}
R_{M \ul{mp}}{}^\un
	&:= D_M \Gamma_{\ul{mp}}{}^\un 
	- 12 \,\mathbb P^\un{}_\up{}^\uk{}_{\ul l} \,\pa_\um \pa_\uk A_M{}^{\ul l}~.
\end{align}
Provided the $\E7$ connection $\Gamma$ transforms as a proper affine
connection, the mixed $\E7$ curvature transforms covariantly, i.e. as a proper
curvature.

\paragraph{Internal derivative algebra.}
The commutator of internal derivatives $V^\um$ can be written
\begin{align}\label{E:IntComm}
[\nabla_\um, \nabla_\un] V^\up
	&= - \mathbf T_{\um \un}{}^\ur \nabla_\ur V^\up
	+ R_{\ul{mnq}}{}^\up \,V^\uq
	- \frac{2}{3} \lambda R_{\um \un}\, V^\up~.
\end{align}
While the full right-hand side is a covariant expression by construction,
the individual terms are not. The ``torsion tensor'' defined by
\begin{align}
\mathbf T_{\um \un}{}^\up := 2\, \Gamma_{[\um \un]}{}^\up - \frac{2}{3} \Gamma_{[\um} \delta_{\un]}{}^\up~,\qquad
\Gamma_\um := \Gamma_{\uk \um}{}^\uk\ ,
\end{align}
is only a tensor if it is contracted with a constrained vector on the $\up$ index,
as one finds a non-covariant part to its transformation,
\begin{align}
\delta_{\rm nc} \mathbf T_{\um \un}{}^\up 
	&= Y_{\un {\ul l}}{}^{\ul{pk}}\, \pa_\um \pa_\uk \Lambda^{\ul l}
	- Y_{\um {\ul l}}{}^{\ul{pk}} \,\pa_\un \pa_\uk \Lambda^{\ul l}~, \eol
Y_{\um \un}{}^{\uk {\ul l}} &:= 12 (t_{\balpha})_{\um \un} (t^{\balpha})^{\uk {\ul l}}
	- \frac{1}{2} \Omega_{\um \un} \Omega^{\uk {\ul l}}\ .
\end{align}
It cannot generically be chosen to vanish for this reason, although one can
have a situation where it vanishes always upon contraction with a constrained
vector, as in \cite{Godazgar:2014sla}.
Note that the internal covariant derivative $\nabla_\um$ is not necessarily a
constrained object.

Some observations are in order. The object $\mathbf T$ defined above and the
actual $\E7$ torsion tensor $\cT$ are related by
\begin{align}
\cT_{\um \un }{}^\up
	&= \mathbf T_{\um \un}{}^\up
	- Y_{\un \ur}{}^{\ul{ps}} \Gamma_{\us \um}{}^\ur
	+ \tfrac{1}{3}\, Y_{\un \um}{}^{\up \uq} \Gamma_{\ul{rp}}{}^\ur\ .
\end{align}
These happen to coincide when using the Weitzenb\"ock connection and contracting
with a constrained vector, i.e.
$\mathring \cT_{\um \un}{}^\up \pa_\up = \mathring {\mathbf T}_{[\um \un]}{}^\up \pa_\up$.

The $\E7$ ``curvature''
\begin{align}
R_{\ul{mnq}}{}^\up
	:= \pa_\um \Gamma_{\ul{nq}}{}^\up
	- \pa_\un \Gamma_{\ul{mq}}{}^\up
	- \Gamma_{\ul{mq}}{}^\ur \Gamma_{\un \ur}{}^\up
	+ \Gamma_{\ul{nq}}{}^\ur \Gamma_{\um \ur}{}^\up\ ,
\end{align}
is also non-covariant, transforming as
\begin{align}
\delta_{\rm nc} R_{\ul{mnq}}{}^\ur =
    \delta_{\rm nc} \mathbf T_{\um \un}{}^\ur \,\Gamma_{\ul{rq}}{}^\up~.
\end{align}
The scale curvature is
\begin{align}
R_{\um \un} := \pa_{\um} \Gamma_{\uk\un}{}^\uk - \pa_{\un} \Gamma_{\uk\um}{}^\uk~,
\end{align}
which is indeed covariant. This is not given
by a simple contraction of $R_{\ul{mnq}}{}^\ur$, but instead by
\begin{align}
R_{\ul{mpn}}{}^\up - R_{\ul{npm}}{}^\up
	= \frac{4}{3} R_{\um \un} - \nabla_\up \mathbf T_{\um \un}{}^\up\ .
\end{align}

\paragraph{Covariant derivatives, their connections, and curvatures.}
It is straightforward to modify the definitions of $D_M$ and $\nabla_\um$ so that
they carry spin and $\SU(8)$ connections. That is, we take
\begin{align}
\cD_M &:= D_M - \frac{1}{2} \Omega_M{}^{a b} M_{a b} - \frac{1}{2} \cQ_M{}_j{}^i I_i{}^j~, \\
\nabla_\um' &:= \nabla_\um - \frac{1}{2} \Omega_\um{}^{a b} M_{a b} - \frac{1}{2} \cQ_\um{}_j{}^i I_i{}^j~,
\end{align}
where $M_{ab}$ and $I_i{}^j$ are the Lorentz and $\SU(8)$ generators, which act
on a spinor $X_{\um \,i}$ as $M_{ab} X_{\um \, i} = \tfrac{1}{2} \gamma_{ab} X_{\um \, i}$
and $I_k{}^l X_{\um\, i} = -\delta^l_i X_{\um \, k} + \tfrac{1}{8} \delta^l_k X_{\um\,i}$.
Henceforth, we drop the prime on $\nabla_\um$. It is now easy to show that
\begin{align}
[\cD_M, \nabla_\um] X_{\un \,i}&= 
	- \frac{1}{4} R_{M \um}{}^{ab} \gamma_{a b}\, X_{\un \,i}
	+ \frac{1}{2} R_{M \um}{}_i{}^j X_{\un \,j}
	- R_{M \um \un}{}^\up X_{\up \, i}
	- \frac{2}{3} \lambda \,R_{M \uk\, \um}{}^\uk X_{\un \,i}\ ,
\end{align}
involving the $R_{M \un\, ab}$ and $R_{M \un}{}_i{}^j$, with their obvious
definitions.
In the body of the paper, we have taken the internal part of the spin connection
to vanish, so that $R_{M \un}{}^{ab } = -\nabla_\un \Omega_{M}{}^{ab}$.

\paragraph{Curvature relations.}
As a consequence of the external and internal GVPs,
\begin{align}
\cD_M \cV_\up{}^{i j} = \cP_{M}{}^{i j k l} \cV_{\up\, kl}~, \qquad
\nabla_\um \cV_\up{}^{i j} = \cP_{\um}{}^{i j k l} \cV_{\up\, kl}~,
\end{align}
one can show that
\begin{align}
R_{M \un\, \ul{pq}}
	&= 
	2i \,\Big(R_{M \un}{\,}{}_{k l}{}^{i j} 
	+ \cP_\un{}^{i j r s} \cP_M{}_{r s k l}
	- \cP_M{}^{i j r s} \cP_\un{}_{r s k l}
	\Big) \cV_{(\up}{}^{kl} \cV_{\uq)}{}_{i j}
	\eol & \quad
	+ i (\cD_M \cP_\un{}^{i j k l} - \nabla_\un \cP_M{}^{i j k l}) \cV_{\up}{}_{ij} \cV_{\uq}{}_{kl}
	- i (\cD_M \cP_\un{}_{i j k l} - \nabla_\un \cP_M{}_{i j k l}) \cV_{\up}{}^{ij} \cV_{\uq}{}^{kl}~.
\end{align}
This condition allows one to determine the $\E7$ curvature $R_{M \un}{}_\up{}^\uq$
from the $\SU(8)$ curvature 
$R_{M \un}{\,}{}_{k l}{}^{i j} = \delta_{[k}{}^{[i} R_{M \un}{}_{l]}{}^{j]}$ and the mixed curl of $\cP^{ijkl}$,
or vice-versa, reflecting the ambiguity in the internal GVP.
Similarly, using
\begin{align}
\cD^2 \cV_\um{}^{i j} = R_{k}{}^{[i}  \cV_\um{}^{j] k} - \mathbb L_\cF \cV_\um{}^{i j}\ ,
\end{align}
one can determine the two-form $\cD \cP^{ijkl}$ and the external $\SU(8)$ curvature
$R_i{}^j$ as
\begin{align}
\cD \cP^{i j k l}
	&= -i \cV^{\um k l} \mathbb L_\cF \cV_\um{}^{i j}~, \qquad
R_{kl}{}^{i j}
	= i \cV^\um{}_{kl} \mathbb L_\cF \cV_\um{}^{ij} - \cP^{i j r s} \cP_{r s k l}~.
\end{align}
For these last relations, it is helpful to use
\begin{align}
\mathbb L_\cF \cV_\um{}^{i j} = 
	\mathbb L^\nabla_\cF \cV_\um{}^{i j}
	+ \cF^\un \cT_{\un \um}{}^\up \cV_\up{}^{i j}~.
\end{align}
Then one can show for example that
\begin{align}
R_{kl}{}^{ij} &= - \cP^{i j r s} \wedge \cP_{r s k l}
    + 12 i \cV^\um{}_{kl} \cV_\un{}^{ij} \mathbb P^\un{}_\um{}^\uk{}_{\ul l} \nabla_\uk \cF^{{\ul l}}
    + i \cF^\un \cT_{\un \uk}{}^{\ul l} \cV_{{\ul l}}{}_{kl} \cV^\uk{}^{i j} \eol
    &= - \cP^{i j r s} \wedge \cP_{r s k l}
    - \Big(
    12 \nabla_\up \cF^{\ul q} \, (t^{\balpha})_{\ul q}{}^\up
    + \cF^\un \cT_\un{}^\balpha
    \Big) \cQ_\balpha{}_{k l}{}^{i j}\ .
\end{align}
Finally, we mention for the purely internal curvatures that
\begin{align}
R_{\ul{mnp}}{}^{\uq} &=
	\Big(2i \,\nabla_{[\um}{\cP_{\un]}{}^{i j k l}}
		+ i \,\mathbf T_{\um \un}{}^{\ur} \cP_{\ur}{}^{i j k l} \Big) \cV_{\up}{}_{k l} \cV^{\uq}{}_{i j}
	+ \Big(
	i \,R_{\um \un}{}_{k l}{}^{i j}
	- 2i \,\cP_{[\um}{}^{i j r s} \cP_{\un]}{}_{r s k l}
	\Big) \cV_{\up}{}^{k l} \cV^{\uq}{}_{i j}
	+ \CC
\end{align}

\paragraph{External derivatives of the $\E7$ torsion.}
One final set of relations prove useful: the external derivative of the
$\E7$ torsion tensor. This can be written
\begin{align}
\cD_M \cT_{\un \um}{}^{\up} = 
	R_{M \un \um}{}^{\up}
	-12 \,\mathbb P^\up{}_\um{}^{\ur}{}_\us\, R_{M \ur \un}{}^\us
	+ 4 \, \mathbb P^\up{}_\um{}^{\ur}{}_\un\, R_{M \us \ur}{}^\us~.
\end{align}
From this equation, one can determine the covariant exterior derivatives of
$A_1{}^{i j}$ and $A_2{}_i{}^{jkl}$,
\begin{subequations}
\begin{align}
\cD_M A_1{}^{i j} &=
	\frac{1}{3} A_2{}^{(i}{}_{k l p} \cP_M{}^{j) k l p} 
	+ i R_{M \um k}{}^{(i} \cV^{\um j) k}
	- \frac{4i}{3} \cP_{\um k l p q} \cV^{\um q (i} \cP_M{}^{j) k l p}~, \\
\cD_M A_2{}_i{}^{jkl} &=
	\Big[
	2 A_1{}_{i p} \cP_M{}^{p j k l} 
	+ 3 A_2{}^{j}{}_{i p q} \cP_M{}^{p q k l} 
	+ 3i \,R_{M \um \,i}{}^{j} \cV^{\um}{}^{k l}
	\eol & \quad
	+ 4i \, \cV^{\um}{}_{i p}\, (\cD_M \cP_{\um}{}^{j k l p} - \cD_{\um}{\cP_M{}^{j k l p}})
	- 4i \,\cP_{\um i p q r} \,\cP_M{}^{j p q r} \,\cV^{\um}{}^{k l}
	\Big]_{\rep{420}}\ ,
\end{align}
\end{subequations}
where a projection onto the $\rep{420}$ of $\SU(8)$ is implied in the last equality.

\end{appendix}


\providecommand{\href}[2]{#2}\begingroup\raggedright\endgroup

\end{document}

%% file: sugracom_v3.tex
\newcommand {\cD}{{\cal D}}
\newcommand {\cE}{{\cal E}}
\newcommand {\cF}{{\cal F}}
\newcommand {\cG}{{\cal G}}

\newcommand {\cM}{{\cal M}}

\newcommand {\cP}{{\cal P}}
\newcommand {\cQ}{{\cal Q}}

\newcommand {\cT}{{\cal T}}

\newcommand {\cV}{{\cal V}}

\newcommand {\cZ}{{\cal Z}}

\newcommand{\rd}{{\mathrm d}}
\newcommand{\ri}{{\mathrm i}}

\newcommand{\dalpha}{{\dot{\alpha}}}
\newcommand{\dbeta}{{\dot{\beta}}}
\newcommand{\dgamma}{{\dot{\gamma}}}

\newcommand{\dmu}{{\dot{\mu}}}

\newcommand{\veps}{\varepsilon}
\newcommand{\eps}{{\epsilon}}
\newcommand{\eol}{\notag \\}

\newcommand{\ul}{\underline}
\newcommand{\pa}{\partial}
